%% file: main.tex
\documentclass[lettersize,journal]{IEEEtran}
\usepackage{amsmath,amsfonts}
\usepackage{latexsym}
\usepackage{algorithmic}
\usepackage{algorithm}
\usepackage{array}
\usepackage[caption=false,font=normalsize,labelfont=sf,textfont=sf]{subfig}
\usepackage{textcomp}
\usepackage{stfloats}
\usepackage{url}
\usepackage{verbatim}
\usepackage{graphicx}
\usepackage{multirow}
\usepackage[binary-units,detect-all]{siunitx}
\DeclareSIUnit{\bps}{bps}
\DeclareSIUnit{\dBm}{dBm}
\DeclareUnicodeCharacter{2009}{\,}
\usepackage{pifont}
\usepackage{cite}
\usepackage{comment}
\usepackage{circuitikz}
\usepackage[nolist,nohyperlinks]{acronym} 
\input{config.tex} 

\usepackage{soul} 
\soulregister{\si}{1}
\graphicspath{{Figures/}}

\begin{document}

\title{Terahertz Communications and Sensing for 6G and Beyond: A Comprehensive Review}
{\author{Wei~Jiang,~\IEEEmembership{Senior~Member,~IEEE,}~Qiuheng~Zhou,~Jiguang~He,~\IEEEmembership{Senior~Member,~IEEE,}~Mohammad~Asif~Habibi,~\IEEEmembership{Member,~IEEE,}~Sergiy~Melnyk,~Mohammed~El-Absi,~\IEEEmembership{Member,~IEEE,}~Bin~Han,~\IEEEmembership{Senior~Member, IEEE,}~Marco~Di~Renzo,~\IEEEmembership{Fellow,~IEEE,}~Hans~Dieter~Schotten,~Fa-Long~Luo,~\IEEEmembership{Fellow,~IEEE,}~Tarek~S.~El-Bawab,~\IEEEmembership{Fellow,~IEEE,}~Markku~Juntti,~\IEEEmembership{Fellow,~IEEE,}~M\'erouane~Debbah,~\IEEEmembership{Fellow,~IEEE,}~Victor~C.~M.~Leung,~\IEEEmembership{Life~Fellow,~IEEE}
\thanks{Manuscript received June 16, 2023; revised November 13, 2023; accepted December 21, 2023. The work of W. Jiang, Q. Zhou, S. Melnyk, and H. D. Schotten was supported in part by German Federal Ministry of Education and Research (BMBF) through \emph{Open6G-Hub} (Grant \emph{16KISK003K}), \textit{6G NeXt} (Grant \emph{16KISK177}), and \emph{AI-NET PROTECT} (Grant \emph{16KIS1283}), and in part by the European Commission (EC) H2020 Framework through \textit{AI@EDGE} (Grant \emph{101015922}). The work of M. A. Habibi and H. D. Schotten was supported by the BMBF through \textit{Open6G-Hub} (Grant \emph{16KISK004}). The work of M. El-Absi was supported in part by BMBF through \textit{6GEM} (Grant \emph{16KISK038}) and in part by the State of Northrhine Westphalia, Germany through \textit{Netzwerke 2021}. The work of M. Di Renzo was supported in part by the EC through \textit{COVER} (Grant \emph{101086228}), \textit{UNITE} (Grant \emph{101129618}),  and \textit{INSTINCT} (Grant \emph{101139161}), and in part by the Agence Nationale de la Recherche (ANR) through \textit{France 2030} (Grant \emph{NF-YACARI 22-PEFT-0005} and \emph{NF-SYSTERA 22-PEFT-0006}) and \textit{PASSIONATE} (Grant \emph{CHIST-ERA-22-WAI-04 through ANR-23-CHR4-0003-01}). M. Juntti's work was supported by the Research Council of Finland through the 6G Flagship Programme (Grant \emph{346208}). The work of V. C. M. Leung was supported in part by Guangdong Pearl River Talent Program (Grant \emph{2019ZT08X603} and \emph{2019JC01X235}) and in part by Canadian Natural Sciences and Engineering Council (Grant \emph{RGPIN-2019-06348}). The editor to coordinate the review of this survey is Prof. Dusit (Tao) Niyato. (\textit{Corresponding author: Wei Jiang.})}
\thanks{W. Jiang, Q. Zhou, S. Melnyk, and H. D. Schotten are with German Research Center for Artificial Intelligence (DFKI),  67663 Germany  (e-mail: \{wei.jiang, qiuheng.zhou, sergiy.melnyk, hans.schotten\}@dfki.de).}
\thanks{J. He is with the Technology Innovation Institute, Abu Dhabi, UAE, and also with University of Oulu, 90014 Oulu, Finland (e-mail: jiguang.he@tii.ae).}
\thanks{M. A. Habibi, B. Han, and H. D. Schotten are with University of Kaiserslautern (RPTU), 67663 Kaiserslautern, Germany (e-mail: \{m.asif, bin.han, schotten\}@rptu.de). }
\thanks{M. El-Absi is with University of Duisburg-Essen, 47057 Duisburg, Germany (e-mail: mohammed.el-absi@uni-due.de).}
\thanks{M. Di Renzo is with Universit\'e Paris-Saclay, CNRS, CentraleSup\'elec, Laboratoire des Signaux et Syst\`emes, 3 Rue Joliot-Curie, 91192 Gif-sur-Yvette, France (email: marco.di-renzo@universite-paris-saclay.fr).} 
\thanks{F. L. Luo is with the Electrical and Computer Engineering Department, University of Washington, Seattle, WA 98195, USA (email: falong@uw.edu).}
\thanks{T. S. El-Bawab is with the School of Engineering at the American University of Nigeria (AUN), 640101 Nigeria (email: telbawab@ieee.org).}
\thanks{M. Juntti is with the Center for Wireless Communications (CWC) at the University of Oulu, 90014 Oulu, Finland (email: markku.juntti@oulu.fi).}
\thanks{ M. Debbah is with KU 6G Research Center, Khalifa University of Science and Technology, 127788 Abu Dhabi, UAE, and also with CentraleSupelec, University Paris-Saclay, 91192, France (email: merouane.debbah@ku.ac.ae). } 
\thanks{Victor C. M. Leung is with Shenzhen University, 518060 Shenzhen, China, and is also with the University of British Columbia (UBC), BC V6T 1Z4, Vancouver, Canada (email: vleung@ieee.org).}
}

\maketitle

\begin{abstract}
Next-generation cellular technologies, commonly referred to as the \ac{6G}, are envisioned to support a higher system capacity, better performance, and network sensing capabilities. The \ac{THz} band is one potential enabler to this end due to the large unused frequency bands and the high spatial resolution enabled by the short signal wavelength and large bandwidth. Different from earlier surveys, this paper presents a comprehensive treatment and technology survey on \ac{THz} communications and sensing in terms of advantages, applications, propagation characterization, channel modeling, measurement campaigns, antennas, transceiver devices, beamforming, networking, the integration of communications and sensing, and experimental testbeds. Starting from the motivation and use cases, we survey the development and historical perspective of \ac{THz} communications and sensing with the anticipated \ac{6G} requirements. We explore the radio propagation, channel modeling, and measurement for the \ac{THz} band. The transceiver requirements, architectures, technological challenges, and state-of-the-art approaches to compensate for the high propagation losses, including appropriate antenna design and beamforming solutions. We overview several related technologies that either are required by or are beneficial for \ac{THz} systems and networks. The synergistic design of sensing and communications is explored in depth. Practical trials, demonstrations, and experiments are also summarized. The paper gives a holistic view of the current state of the art and highlights the open research challenges towards \ac{6G} and beyond. 

\end{abstract}
 

\section{Introduction}

\IEEEPARstart{T}{oday}, the \ac{5G} of mobile networks are being deployed \cite{Ref_dahlman20215gNR}. Meanwhile, both academia and industry have shifted their research focus to the next generation of communications technologies, which are commonly referred to as the \acf{6G} and are officially named by the \ac{ITU-R} as International Mobile Telecommunications for 2030 (IMT-2030) \cite{Ref_IMT2030requirements}. Several research groups, standardization bodies, regulatory organizations, and government agencies \cite{Ref_jiang2021kickoff} have initiated a variety of programs to discuss the \ac{6G} vision \cite{Ref_saad2020vision} and to develop key technologies \cite{Ref_huang2019survey}, as we will elaborate in Sec. \ref{sec:potential}. To support disruptive applications, such as virtual and augmented reality \cite{Ref_huawei2018cloud}, Internet of Things \cite{Ref_Han2017Optimal}, Industry 4.0, connected and autonomous vehicles \cite{Ref_Nasimi2019Platoon},  and yet-to-be-conceived  use cases like Metaverse, holographic-type telepresence \cite{Ref_jiang2021road},  Tactile Internet \cite{Ref_fettweis2014tactile}, digital twin \cite{Ref_wu2021digital}, full immersiveness \cite{Ref_yang2019wireless}, multi-sense experience, and blockchain \cite{Ref_xiong2018when}, \ac{6G}  requires significantly stringent performance, e.g., hyper rates on the order of terabits-per-second (\si{\tera\bps}) \cite{Ref_david20186g}, ultra-reliability, near-zero latency, and massive connectivity density, far beyond what its predecessors can offer \cite{Ref_Zong20196g}.  

Starting from 1985, when 3G (known in ITU jargon as IMT-2000) was under development, the \ac{ITU-R} has been actively engaged in the standardization of each generation of cellular systems. Like the visions for 4G (a.k.a IMT-Advanced) and 5G (a.k.a IMT-2020), in its recommendations M.1645 \cite{Ref_itu20034Gvision} and M.2083 \cite{Ref_non2015imt}, respectively, the ITU-R started the development process for 6G by defining the IMT-2030 vision as the first step. The first recommendation for IMT-2030  was completed on 22nd June 2023 during the 44th ITU-R WP5D meeting in Geneva. As shown in \figurename \ref{Figure_6Gscenarios}, the three IMT-2020 usage scenarios, i.e., enhanced Mobile Broadband (eMBB), Ultra-Reliable Low-Latency Communications (URLLC), and massive Machine-Type Communications (mMTC) are upgraded to \textit{Immersive Communication}, \textit{Hyper Reliable and Low-Latency Communication}, and \textit{Massive Communication}, respectively. In addition, three new usage scenarios -- \textit{Integrated Sensing and Communication},  \textit{Ubiquitous Connectivity}, and \textit{Integrated AI and Communication} -- are introduced.  

\begin{figure}[!tbph]
\centering
\includegraphics[width=0.43\textwidth]{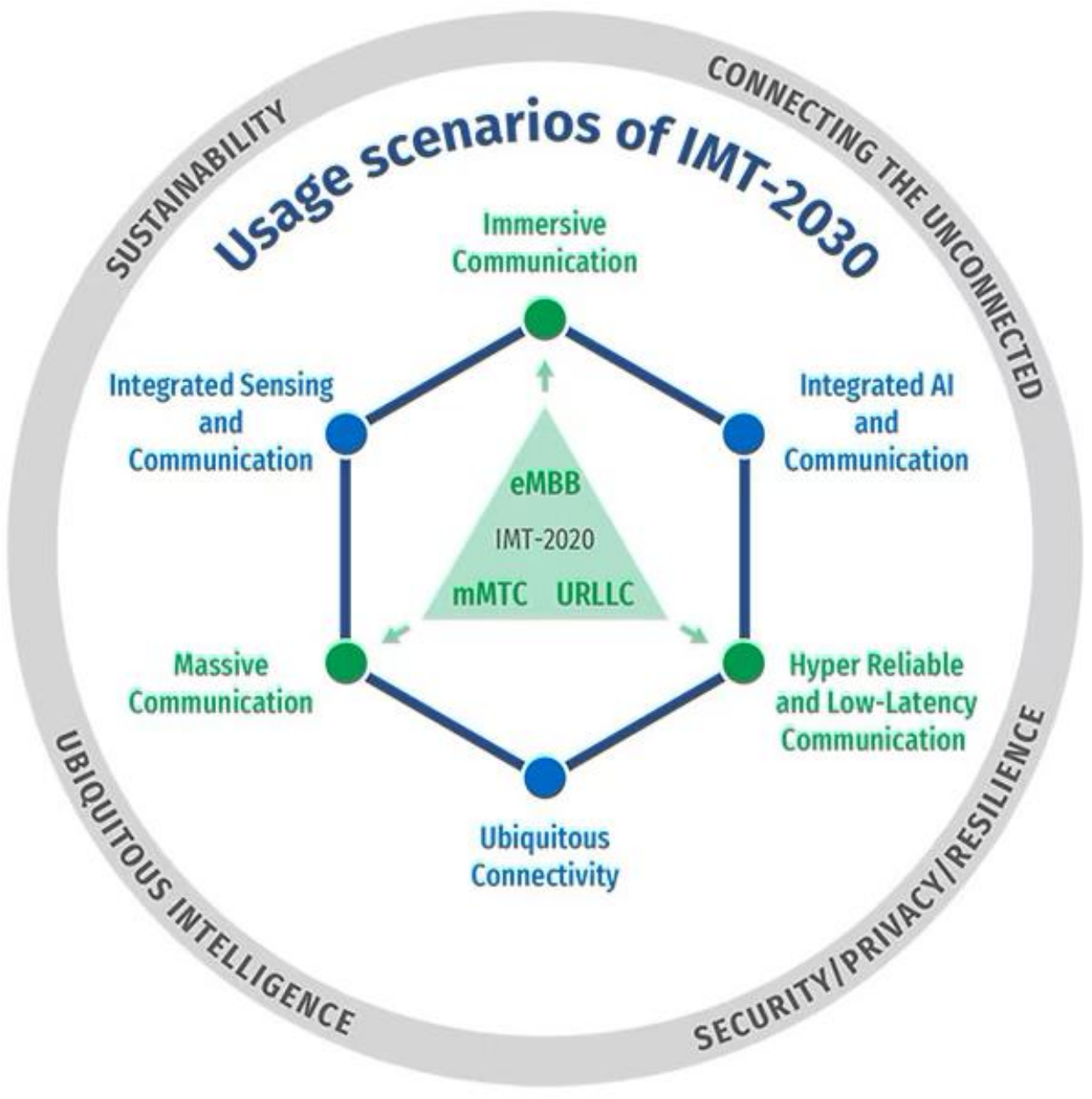}
\caption{Six usage scenarios for IMT-2030 and four overarching aspects, specified by the ITU-R in June 2023 \cite{Ref_IMT2030requirements}. }
\label{Figure_6Gscenarios}
\end{figure}

In what follows, this paper focuses on the usage scenario of \textit{Integrated Sensing and Communication}. Driven by the continuous progress in frequency assignment, antennas, devices, and signal processing, 6G will be a dual-functional system that is able to not only communicate but also sense as well \cite{Wymeersch2021}. This usage scenario enables 6G and beyond to \textit{see} the physical world through \ac{EM} waves \cite{zhang2021overview}. It offers high-resolution sensing, localization, imaging, and environment reconstruction capabilities to improve communications performance. Also, it supports a wide range of novel applications beyond communications such as object tracking, security screening, remote sensing, process monitoring, \ac{SLAM},  gesture recognition, and activity detection \cite{wang2022integrated}, distinguishing 6G from the traditional communication-oriented cellular systems from the first generation (1G) to 5G.

\subsection{\textit{Why do \ac{6G} and beyond need the \ac{THz} band?}}
The \ac{THz} band has attracted a lot of interest in recent years and is recognized as a promising enabler for \ac{6G} \cite{Ref_rappaport2019wireless}. Prior to stepping into technical details, we would like to first clarify a fundamental question that might cause some confusion or disputes in some prior literature. That is, \textit{why do we need to exploit the \ac{THz} band in \ac{6G} and beyond}? We try to address this aspect from the perspectives of both \ac{THz} communications and \ac{THz} sensing, as well as their synergy.

\subsubsection{\ac{THz} Communications}
At the \ac{WRC} held in 2019, a.k.a WRC-19,  the \ac{ITU-R} assigned a total of \SI{13.5}{\giga\hertz} spectrum, consisting of a group of high-frequency bands, for the deployment of \ac{5G} \ac{mmWave} communications \cite{Ref_dahlman20215gNR}, as we will elaborate in Sec. \ref{subsecSOTAspectrm}.
Despite the spectral abundance of \ac{mmWave} bands, it might not be sufficient to meet the growing need for bandwidth over the next decade. There are enormous spectral resources at higher frequencies that were already used for a wide variety of non-cellular applications, such as remote sensing, radio astronomy, and radar \cite{demirhan2022radar}, to name a few. With the advancement in antenna technology and \ac{RF} components, these frequencies previously considered unsuitable for mobile communications due to their unfavorable propagation characteristics become technologically usable \cite{Ref_nagatsuma2016advances}. 

Fig. \ref{Figure_EW} illustrates the whole \ac{EM} spectrum, consisting of radio, microwave, \ac{IR}, visible light, ultraviolet, X-rays, and Gamma rays, from the lower to higher frequencies \cite{Ref_jiang2023full}. It is noted that the definition of the \ac{EM} spectrum in the general case differs from the naming of frequency bands from the perspective of wireless communications, as shown in the figure. 
Based on these considerations, it is argued that the THz band is a suitable candidate to realize Tbps communications under the current level of hardware and signal-processing technologies. The reasons are explained as follows: 
\begin{itemize}
    \item \textbf{Spectrum scarcity of the sub-\SI{6}{\giga\hertz} band}: The favourable propagation characteristics of sub-\SI{6}{\giga\hertz} frequencies facilitate the use of sophisticated transmission technologies such as \ac{MMIMO} \cite{Ref_jiang2021cellfree}, \ac{NOMA} \cite{Ref_liu2022developing}, and high-order modulation like \ac{1024QAM} to achieve high spectral efficiency. However, spectrum scarcity and non-continuity pose a significant challenge to achieving higher rates.  Even if the sub-\SI{6}{\giga\hertz} band ultimately determines a bandwidth of \SI{1}{\giga\hertz}  for \ac{IMT} services, a Tbps link can only be realized under the extreme spectral efficiency of \SI{1000}{\bps/\hertz}, as suggested by the Shannon capacity $R=B\log(1+S/N)$. However, such high performance is impractical in the foreseeable future. In comparison,  as specified in ITU-R M.2410 \cite{Ref_non2017minimum},  the peak spectral efficiency for IMT-2020 is \SI{30}{\bps/\hertz} (in ideal conditions). 
    \item \textbf{Insufficient mmWave bandwidth below \SI{100}{\giga\hertz}}: At WRC-19, the \ac{mmWave} spectrum in 24.25-27.5 \si{\giga\hertz}, 37-43.5 \si{\giga\hertz}, 45.5-47 \si{\giga\hertz}, 47.2-48.2 \si{\giga\hertz} and 66–71 \si{\giga\hertz}, was assigned to IMT services \cite{Ref_marcus2019ITU}.  There is a challenge from the perspective of \ac{RF} implementation, which is typically constrained by the limit of around $10\%$ relative bandwidth. That is to say, \ac{mmWave} technologies below \SI{100}{\giga\hertz} can support a single \ac{RF} transceiver with a maximal bandwidth of \SI{10}{\giga\hertz} due to the nonlinearity of \ac{RF} components. Thus, \si{\tera\bps} can only be reached with spectral efficiency of \SI{100}{\bps/\hertz}. This is currently infeasible for high-frequency signal transmission, which is prone to the use of low-order modulation and single-carrier techniques due to the constraints of mmWave components \cite{Ref_lopez2019opportunities}. Therefore, it is argued that the potential for realizing Tbps communications relies on massively abundant frequencies above \SI{100}{\giga\hertz}.
    \item \textbf{Constraint of optical sensing}: Optical bands at \ac{IR} \cite{Ref_zou2022highcapacity}, visible-light \cite{Ref_amiad2021towards}, and ultraviolet frequencies \cite{Ref_vavoulas2019survey} offer enormous spectral resources. Like \ac{THz} frequencies,  the application of \ac{OWC} faces some challenges, such as eye-safety constraints, atmospheric (fog, rain, dust, or pollution) absorption, high diffusion loss, low optical-emitter output power, photonic phase noise, \ac{LoS} reliance, and beam misalignment \cite{Ref_song2022terahertz}. Many recent research progresses have been made in \ac{OWC} \cite{Ref_haas2020optical}, making it a valuable solution for communications. Nevertheless, lightwave does not exhibit comparable \textit{sensing}, \textit{imaging}, and \textit{positioning}  capabilities as \ac{THz} signals \cite{Ref_sarieddeen2020next}. From the perspective of integrated sensing and communications, the \ac{THz} band is considered a suitable option to efficiently realize a dual-functional 6G system. 
    \item \textbf{Adverse health effects at extreme high bands}: 
    Ionizing radiation, including ultraviolet, X-rays, and Gamma rays, poses a significant risk to human health, as it has strong energy power to dislodge electrons and generate free radicals that can lead to cancer \cite{Ref_jiang20226GbookChapter6}. The adverse health effects of ionizing radiation are controllable if used with care, so extreme high-frequency signals have been used in specific fields such as radiotherapy, photography, semiconductor manufacturing, and nuclear medicine, among others.  However, it is still too dangerous for personal communications \cite{Ref_rappaport2019wireless}. 
\end{itemize}

\begin{figure}[!tbph]
\centering
\includegraphics[width=0.49\textwidth]{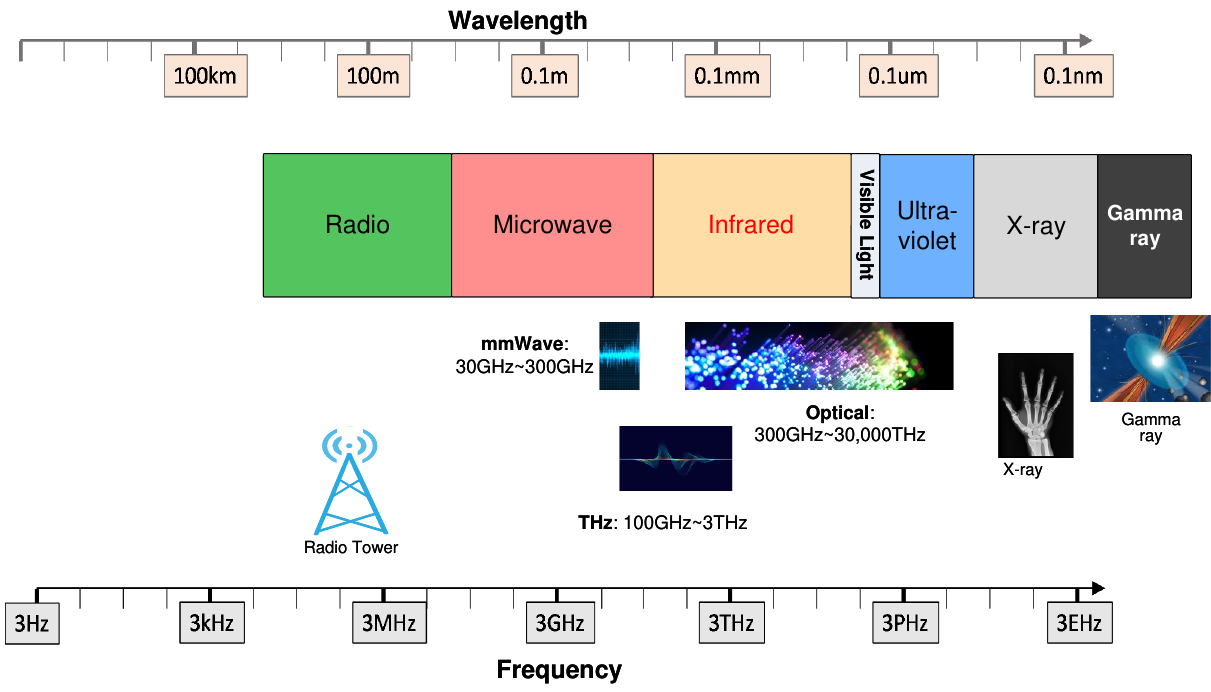}
\caption{The electromagnetic spectrum and the positions of \ac{mmWave}, \ac{THz}, and optical bands. }
\label{Figure_EW}
\end{figure}

Unlike ionizing radiation, \ac{THz} frequencies are non-ionizing because their photon energy is not sufficient (0.1 to 12.4meV, which is over three orders of magnitude weaker than ionizing photon energy levels) to release an electron from an atom or a molecule, where typically 12eV is required for ionization. 
The \ac{THz} band offers abundant spectral resources, ranging from tens of gigahertz to several terahertz, depending on the transmission distance. This makes the available bandwidth more than ten times greater than that of \ac{mmWave} bands, while the operating frequency is at least one order of magnitude below the optical bands. In addition, the technologies required to make \si{\tera\bps}-level transmission over the \ac{THz} band a reality is rapidly advancing \cite{Ref_boulogeorgos2018terahertz}. For example, novel antennas and components are under development by exploiting cutting-edge materials like graphene, to overcome \textit{the \ac{THz} gap}, where the operating frequency is too high for conventional \textit{electronic} transmitters and too low for \textit{photonic} emitters \cite{Ref_song2022terahertz}, as we will elaborate in Sec. \ref{Sec:THz_Transceiver}. Moreover,  \ac{UMMIMO} and lens antenna arrays to generate high-gain beams, with the aid of appropriate beam alignment techniques, compensating for the large free-space loss, atmospheric absorption, and weather effects, are other promising technologies, as introduced in Sec. \ref{Sec_Channel_Effect} and Sec. \ref{Sec_BF}.  

\subsubsection{\ac{THz} Sensing}

The spatial resolution of a propagated signal becomes much finer at higher frequencies, thereby enabling high-definition spatial differentiation  \cite{Ref_sarieddeen2020next}. In addition to \ac{THz} communications, \ac{THz} sensing (including positioning, imaging, and spectroscopy) exploits the tiny wavelength on the order of micrometers and the frequency-selective resonances of various materials over the measured environment to gain unique information based on the observed signal signature \cite{Ref_chaccour2022seven}. Compared with wireless sensing over other bands, \ac{THz} sensing offers the following advantages:
\begin{itemize}
    \item \textbf{High resolution and penetration capabilities}: Although low-frequency signals are able to sense, detect, and localize objects, as the radar \cite{demirhan2022radar} and \ac{GNSS},  \ac{THz} sensing/positioning can improve the resolution due to the small wavelength, even for objects hidden from direct view. THz waves are able to penetrate a variety of non-conductive materials, e.g., plastics, fabrics, paper, ceramics, and dielectric substances. This allows THz sensing to detect hidden objects, structural defects, and layers beneath surfaces, making it useful in security screening, quality control, process monitoring, and material characterization  \cite{Ref_zandonella2003terahertz}.
    \item \textbf{Non-ionizing radiation}: Compared to X-rays and Gamma rays, \ac{THz} waves have much lower photon energy, making them non-ionizing \cite{Ref_mamrashev2018detection}. This implies that \ac{THz} sensing is generally considered safe for biological samples and humans, allowing for non-destructive and non-invasive imaging and diagnosing.
    \item \textbf{Low environmental interference}: In contrast to visible or \ac{IR} radiation \cite{Ref_kahn1997wireless}, \ac{THz} waves are less vulnerable to environmental factors such as ambient light, fog, or smoke. It allows \ac{THz} sensing for outdoor environments or adverse conditions, expanding its usability in fields such as remote sensing, atmospheric monitoring, and outdoor stand-off security screening of dangerous items like firearms, bombs, and explosive belts hidden beneath clothing   \cite{Ref_sarieddeen2020next}.
    \item \textbf{Spectroscopic analysis}: \ac{THz} waves interact with molecules in a characteristic manner, leading to unique spectral fingerprints. THz spectroscopy provides valuable information about molecular vibrations and rotational transitions, enabling the identification and analysis of chemical substances, including explosives, drugs, and biomolecules \cite{Ref_abbasi2016nano}. It is particularly effective for identifying substances with distinct \ac{THz} absorption or reflection properties.
\end{itemize}

\renewcommand{\arraystretch}{1.2}
\begin{table*}[t]
	\centering
	\caption{A comparison of this survey with the existing works.}
	\begin{tabular}{|l|l|c|c|c|c|c|c|c|c|}
		\hline  \hline
		\multirow{3}{*}{\textbf{Reference}} & 
		\multirow{3}{*}{\textbf{Year}} & 
		\multicolumn{8}{c|}{\textbf{Content Coverage}} \\ \cline{3-10}
		{\ } & {\ } & \textbf{ \ac{6G}}    & \textbf{ Sensing}&  \textbf{ISAC} & \textbf{\begin{tabular}[c]{@{}c@{}} \ac{THz} \\Channel \end{tabular}} & \textbf{\begin{tabular}[c]{@{}c@{}} \ac{THz} Ant. \& \\Beamforming \end{tabular}} & \textbf{\begin{tabular}[c]{@{}c@{}} \ac{THz} Device \&\\Transceiver \end{tabular}} &  \textbf{\begin{tabular}[c]{@{}c@{}} Synergy w. \\6G Key Tech. \end{tabular}} & \textbf{\begin{tabular}[c]{@{}c@{}} \ac{THz} Trial \&\\Experiment \end{tabular}} \\ \hline \hline
  Mukherjee and Gupta \cite{Ref_mukherjee2008terahertz} & 2008 &  & &  &  & \checkmark & \checkmark &  & \\ \hline
  Ostmann and Nagatsuma \cite{Ref_ostmann2011reivew} & 2011 &  & &  & \checkmark & \checkmark & \checkmark &  & \checkmark\\ \hline
Nagatsuma \emph{et al.} \cite{Ref_nagatsuma2016advances} & 2016 &  & &  &  &  & \checkmark &  & \\ \hline
  K. M. S. Huq \emph{et al.} \cite{Ref_huq2019terahertz} & 2019 & \checkmark & &  & $\circ$ &  & $\circ$ & $\circ$ & \\ \hline
  K. Tekbiyik \emph{et al.} \cite{Bio_tekbiyik2019terahertz} & 2019 &  & &  & \checkmark & \checkmark & \checkmark &  & \checkmark\\ \hline
  Z. Chen \emph{et al.} \cite{Ref_chen2019survey} & 2019 &  & &  &  & \checkmark & \checkmark &  & \\ \hline 
  M. Naftaly \emph{et al.} \cite{Ref_naftaly2019industrial} & 2019 &  & \checkmark &  &  &  &  &  & \\ \hline 
  Y. He \emph{et al.} \cite{Ref_he2020overview}& 2020 &  & &  &  & \checkmark &  &  & \\ \hline 
    S. Ghafoor \emph{et al.} \cite{Ref_ghafoor2020mac}& 2020 & $\circ$ & &  & \checkmark & $\circ$ &  &  & \\ \hline 
  B. Ning \emph{et al.} \cite{Ref_ning2022prospective} & 2021 &  & &  &  & \checkmark &  &  & \\ \hline
  F. Lemic \emph{et al.} \cite{Ref_lemic2021survey} & 2021 & $\circ$ & &  & \checkmark & \checkmark &  &  & \\ \hline
H. Sarieddeen \emph{et al.} \cite{Ref_sarieddeen2021overview}& 2021 & $\circ$ & \checkmark&  &  & \checkmark &  & \checkmark & \\ \hline
E. Castro‑Camus1 \emph{et al.} \cite{Ref_castrocamus2021recent}& 2021 &  & \checkmark&  &  &  &  &  & \\ \hline
C.-X. Wang \emph{et al.} \cite{Ref_Wang2021Key}& 2021 & \checkmark & &  & \checkmark & \checkmark & $\circ$  &  & \\ \hline
			D. Moltchanov \emph{et al.} \cite{Ref_moltchanov2022tutorial} & 2022 & \checkmark & $\ $ & $\ $ & \checkmark & $\circ$ & $\ $ &  & \\ \hline	
   C. Chaccour \emph{et al.} \cite{Ref_chaccour2022seven} & 2022 & \checkmark & \checkmark & \checkmark &  &  &  & \checkmark & \\ \hline
   C. Han \emph{et al.} \cite{Ref_han2022terahertz} & 2022 &  &  &  & \checkmark &  &  &  & $\circ$\\ \hline
   D. Serghiou \emph{et al.} \cite{Ref_serghiou2022terahertz} & 2022 & \checkmark & &  & \checkmark &  &  &  & \\ \hline
   I. F. Akyildiz \emph{et al.} \cite{Akyildiz2022} & 2022 & \checkmark & &  & \checkmark &  & \checkmark &  & \checkmark\\ \hline
   A. Shafie \emph{et al.} \cite{Ref_shafie2022terahertz} & 2022 & $\circ$ & &  &  & $\circ$ & $\circ$ & \checkmark & \\ \hline 
		\textbf{This survey paper} & 2023 & \checkmark & \checkmark & \checkmark & \checkmark & \checkmark & \checkmark & \checkmark & \checkmark \\ \hline
		\multicolumn{10}{|l|}{\begin{tabular}[l]{@{}l@{}} Note:  \end{tabular}} \\  
		\multicolumn{10}{|l|}{\begin{tabular}[l]{@{}l@{}} For each column, the \checkmark symbol means that this aspect is discussed in detail in the reference, the $\circ$ symbol means that this aspect is only mentioned\\ briefly, and the blank indicates that this aspect is not considered.  \end{tabular}} \\ \hline \hline
	\end{tabular}
	\label{Table_paper_compwj} 
\end{table*}

\subsubsection{Synergy between \ac{THz} communications and \ac{THz} sensing} Based on these considerations, the \ac{THz} band offers not only massive spectral resources for wireless communications but also unique advantages for sensing, positioning, imaging, and spectroscopy \cite{Ref_sarieddeen2020next}. Hence, it attracted a lot of interest recently as a key enabler for implementing \ac{ISAC} for \ac{6G} and beyond \cite{Ref_li2021integrated}. On top of implementing \ac{THz} communications and \ac{THz} sensing in a unified system, these dual-functional wireless networks offer a great synergy through \textbf{sensing-aided communication} \cite{Anzhong2022, Jiguang2021} and \textbf{communication-aided sensing} \cite{demirhan2022integrated}, as we will elaborate in Sec. \ref{SEC:ISAC}.

Using sensing information in communications may be one of the significant benefits of \ac{ISAC}, which enables a more deterministic and predictable propagation channel. It facilitates the design of efficient communication algorithms and protocols, such as sensing-aided channel estimation \cite{Ref_chen2023sensing},  predictive beamforming served by sensing \cite{Ref_liu2020radar, Ref_mu2021integrated}, fast beam alignment and tracking \cite{Ref_chen2022enhancing},  and link blockage mitigation \cite{Ref_wang2018internet}. On the other hand, mobile communication networks also provide significant opportunities and benefits for network sensing or sensing as a service \cite{Ref_wang2017multimedia}. Nodes share sensing results through the mobile network, where multiple network nodes (base stations, user equipment, etc.) can act as a collaborative sensing system \cite{Ref_he2022collaborative}. This collaboration, achieved through sensing data fusion, reduces measurement uncertainties and provides larger coverage areas, as well as higher sensing accuracy and resolutions.

\subsection{Motivations and Contributions} 
Recently, the wireless community has published several research articles and surveys on \ac{THz} communications. As listed in Table \ref{Table_paper_compwj}, the existing surveys tend to focus on specific aspects of \ac{THz} communications, such as antenna fabrication \cite{Ref_he2020overview}, propagation characterization \cite{Ref_lemic2021survey}, measurement \cite{Ref_han2022terahertz}, channel modeling \cite{Ref_serghiou2022terahertz}, beamforming \cite{Ref_ning2022prospective}, and hardware \cite{Ref_chen2019survey}. 
These surveys are more beneficial for researchers who are focusing on a particular aspect of \ac{THz} communications and need an exhaustive collection of the existing research outcomes in this aspect. On the other hand, some magazine articles such as \cite{Ref_huq2019terahertz, Ref_shafie2022terahertz} offer overviews that cover relatively wide ranges of aspects but are rather concise on many topics. Therefore, a comprehensive survey, which can provide researchers with a holistic view of all the necessary ingredients to build a THz system, is still missing. In addition, prior literature primarily concentrated on \ac{THz} communications from the perspective of conventional wireless systems and applications, while \ac{THz} sensing has received much less attention. Last but not least, most of the current literature does not take into account the particular demands, applications, requirements, and scenarios of \ac{6G}.  

\begin{figure*}[!tbph]
\centering
\includegraphics[width=0.92\textwidth]{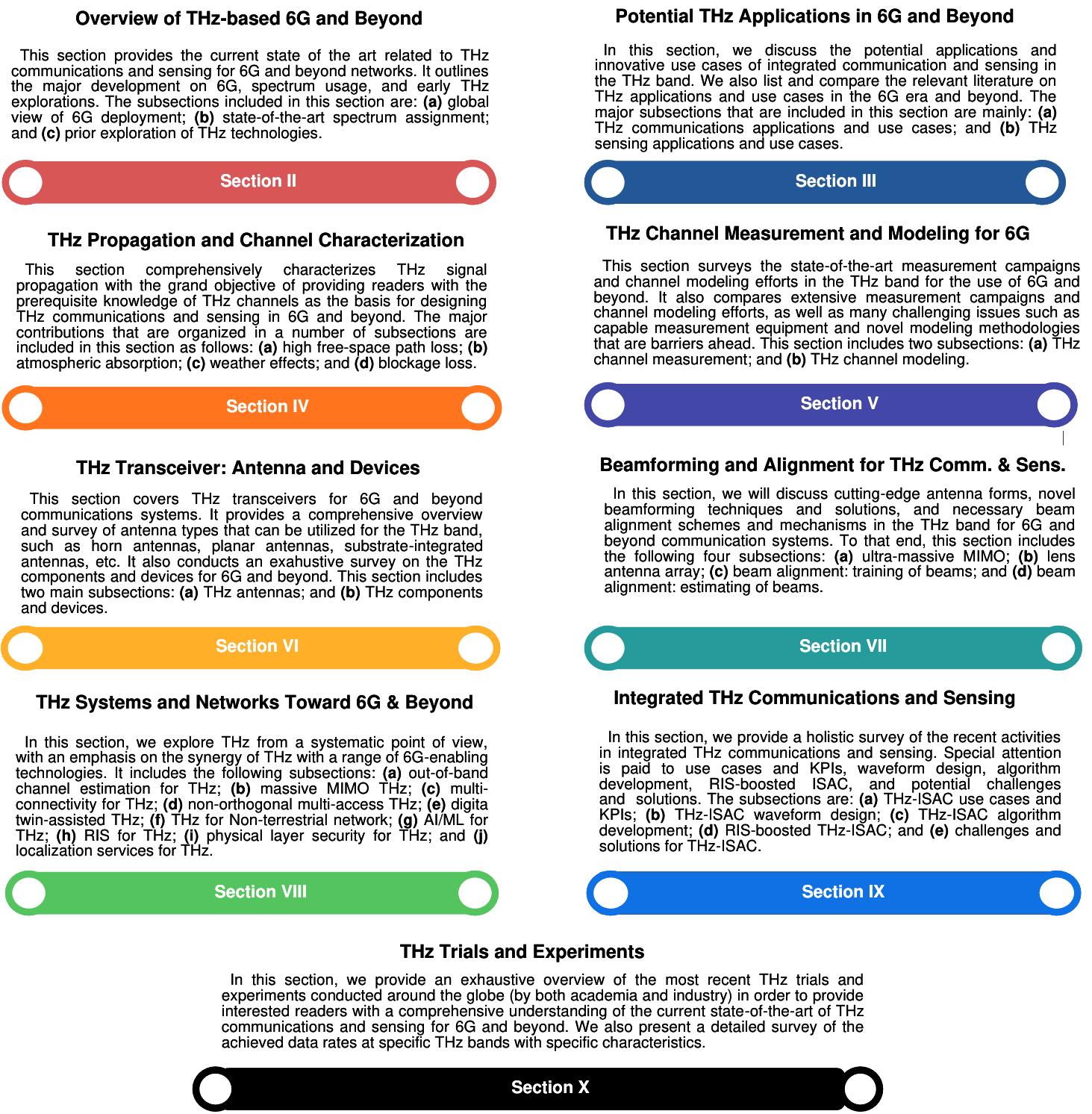}
\caption{Outline of the structure of this survey.} 
\label{Diagram_outline}
\end{figure*}
 
Therefore, this article presents a comprehensive treatment and technology survey about \ac{THz} communications and sensing. We clarify the advantages of \ac{THz} over other bands for \ac{6G} and beyond, potential \ac{THz}-based 6G applications,  \ac{THz} signal propagation characterization, \ac{THz} channel modeling, \ac{THz}
measurement, \ac{THz} antennas,  photonic-electronic devices for \ac{THz} transceiver, beamforming, beam alignment, \ac{THz} networking, the synergy with other potential \ac{6G} technologies, \ac{THz}-based integrated sensing and communications, and \ac{THz} experimental test-beds.  Table \ref{Table_paper_compwj} compares this work with the published works in terms of the topics covered. From an application and implementation perspectives, this work can provides researchers in \ac{THz} communications, \ac{THz} sensing, and \ac{6G} with a holistic view of the current state of the art and highlight the issues and challenges \cite{Ref_jiang2023terahertz} that are open for further research.

The major contributions of this survey include:
\begin{itemize}
    \item First, this work aims to answer a fundamental question: \textit{Why do \ac{6G} and beyond need to exploit the \ac{THz} band}? It clarifies the comparative advantages of \ac{THz} over other frequency bands for communications and sensing in the scenarios envisioned for \ac{6G} and beyond.
    \item  This paper provides a state-of-the-art overview of related fields by summarizing the global \ac{6G} development, latest spectrum assignment for \ac{IMT}, and early exploration efforts in \ac{THz} technologies. 
    \item This article envisions potential \ac{THz}-based communications and sensing applications for \ac{6G} and beyond. 
    \item This survey comprehensively characterizes \ac{THz} signal propagation, including the \textit{path loss, atmospheric absorption, weather effects, and blockage}. 
    \item This paper reports up-to-date \ac{THz} measurement campaigns by means of three measurement methods, i.e., \textit{frequency-domain \ac{VNA}, time-domain sliding, and \ac{TDS}}. 
    \item This article overviews both deterministic and statistical \ac{THz} channel models.
    \item This survey provides readers with the necessary knowledge required to design and build transceivers for \ac{THz} communications and sensing, including the recent advances in \ac{THz} antennas, \ac{THz} electronic devices, and \ac{THz} photonic devices. 
    \item This work elaborates on how to compensate for the large propagation loss through beamforming over large-scale antenna arrays. The fundamentals of \ac{UMMIMO}, lens antenna array, beam tracking, beam estimation, and beam alignment are introduced as well. 
     \item From a systematic perspective, this survey explores the paradigms for \ac{THz} networking, with an emphasis on the synergy of \ac{THz} communications and sensing with other \ac{6G}-enabling technologies, covering \ac{MMIMO}, \ac{UMMIMO}, \ac{NOMA}, \ac{RIS}, non-terrestrial networks, digital twins, \ac{AI} and \ac{ML}. Moreover, security, localization, integrated communications and sensing,  multi-connectivity, and channel awareness for \ac{THz} networks, are discussed.
     \item This article discusses the building blocks, opportunities, challenges, and potential solutions for \ac{ISAC} over the \ac{THz} band, elaborating its unique advantages, use cases, \acp{KPI}, joint waveform design, and efficient algorithm design.
    \item Last but not least,  the latest advances in \ac{THz} trials and experiments are reported to provide readers with an insightful view of practical aspects of \ac{THz} communications and sensing.
\end{itemize}

\subsection{The Structure of this Survey}
Overall, this survey aims to provide researchers with a holistic view of the current state of the art about all aspects required to design and build \ac{THz}-based wireless communications and sensing systems for \ac{6G} and beyond. Also, this work highlights the challenges that are open for future research. To improve the readability, an outline of the survey is illustrated in \figurename \ref{Diagram_outline}.

\section{Overview of THz-based 6G Systems}
Intending to facilitate an insightful view and knowledge, this section summarizes the current state of the art in the related fields. First, Sec. \ref{sec:potential} offers a global view of \ac{6G} development, followed by the status of spectrum usage for \ac{IMT} services worldwide in Sec. \ref{subsecSOTAspectrm}. Then, the early \ac{THz} exploration efforts are listed in Sec. \ref{Sec:THZexplore}. 

\subsection{Global View of \ac{6G} Development} \label{sec:potential}
At the beginning of 2019, South Korea's three network operators and U.S. Verizon were in a dispute with each other, vying for the title of being the world's first provider of the \ac{5G} communication services. This event marked the arrival of the \ac{5G} era \cite{Ref_dahlman20215gNR}. In the past few years, the term ‘5G’ has remained one of the most prominent buzzwords in the media, drawing unprecedented attention from the whole society. Apart from continuously enhancing network capacity and improving system performance as previous generations had done, \ac{5G} expands mobile communications services from human-centric to human-and-things, as well as from the consumer market to vertical industries \cite{Ref_ortiz2020survey}.  This has substantially increased the potential scale of mobile subscriptions from billions (i.e. equivalent to the world's population) to almost countless interconnectivity among humans, machines, and things.

In 2020, the outbreak of the COVID-19 pandemic led to a significant loss of human lives worldwide and imposed unprecedented challenges on societal and economic activities.  However, this public health crisis has underscored the unique role of telecommunication networks and the digital infrastructure in keeping society operational and families connected. This is particularly relevant for the values of \ac{5G} applications, such as remote health care, online education, mobile working, autonomous vehicles, unmanned delivery, and smart manufacturing \cite{Ref_siriwardhana2020fight}.
In July 2018, the \ac{ITU-T} standardization sector had established a focus group called \textit{Technologies for Network 2030}  with the aim of studying the capabilities of networks for 2030 and beyond \cite{Ref_li2019blueprint}.

The \ac{EC} initiated the beyond \ac{5G} program in 2020, under its Horizon 2020 calls --- \textit{ICT-20 \ac{5G} Long Term Evolution} and \textit{ICT-52 Smart Connectivity beyond \ac{5G}} --- where a batch of pioneer research projects was sponsored. 
At the beginning of 2021, the \ac{EC} launched its \ac{6G} flagship research project \textit{Hexa-X} \cite{Ref_hexax}, followed by the second phase of European level \ac{6G} research \textit{Hexa-X-II}  in early 2023 \cite{Ref_hexaxii}. The \ac{EC} has also announced its strategy to accelerate investments in ‘\textit{Gigabit Connectivity}’ including \ac{5G} and \ac{6G} to shape Europe's digital future \cite{Ref_EU2020shaping}. In October 2020, the \ac{NGMN} alliance announced its new ‘\textit{6G Vision and Drivers}’ project, intending to provide early and timely guidelines for global \ac{6G} activities. The first report for this project was published in April 2021 \cite{Ref_ngmn20216Gdrivers}. At its meeting in February 2020, the \ac{ITU-R} decided to start studying future technology trends for the evolution towards IMT-2030 \cite{Ref_itu2020future}. 

Motivated by the revolutionary force of \ac{5G}, the governments of many countries recognized the significance of mobile communications technologies for driving economic prosperity and sustainable growth. In the past years, many countries have set up research initiatives officially or announced ambitious plans for the development of \ac{6G}. The world's first \ac{6G} effort, ‘\textit{6G-Enabled Wireless Smart Society and Ecosystem (6Genesis) Flagship Program}’, was carried out by the University of Oulu in April 2018, as part of the Academy of Finland's flagship program \cite{Ref_6Genesis}. This project focuses on groundbreaking \ac{6G} research, with four interrelated strategic areas including wireless connectivity, distributed computing, devices and circuit technology, and services and applications. In September 2019, the world's first \ac{6G} white paper ‘\textit{key drivers and research challenges for 6G ubiquitous wireless intelligence}’  was published as an outcome of the first \ac{6G} Wireless Summit \cite{Ref_Aazhang2019key}. Subsequently, a series of white papers have been published, covering twelve specific areas of interest, such as \ac{ML}, edge intelligence, localization, sensing, and security. 

In October 2020, the Alliance for Telecommunications Industry Solutions (ATIS) established the ‘\textit{Next G Alliance}’, an industry-led initiative aimed at advancing North American mobile technology leadership in \ac{6G} over the next decade \cite{Ref_nextG2022market}. Founding members of the initiative include leading companies such as AT\&T, T-Mobile, Verizon, Qualcomm, Ericsson, Nokia, Apple, Google, Facebook, and Microsoft. The Next G Alliance places a strong emphasis on technology commercialization and seeks to encompass the full lifecycle of \ac{6G} research, development, manufacturing, standardization, and market readiness. In addition to this, SpaceX, a U.S. company known for its revolutionary reusable rockets, announced the Starlink project in 2015 \cite{Ref_starlink}. This project aims to deploy a very large-scale \ac{LEO} communications satellite constellation to offer ubiquitous internet access services across the whole planet. The \ac{FCC} approved its initial plan of launching 12,000 satellites, and an application for 30,000 additional satellites is currently under consideration. As of 31 October 2023, the number of Starlink satellites in orbit has reached \num{5011}, and the service is commercially available in many countries and regions. Although Starlink may not replace \ac{5G} or be considered \ac{6G}, the impact of such a very large-scale \ac{LEO} satellite constellation on \ac{6G} and beyond should be taken into account by the mobile industry.

In November 2019, the Chinese Ministry of Science and Technology kicked off the research and development efforts for \ac{6G} technology, in collaboration with five other ministries or national institutions. The event also marked the establishment of a working group, named \textit{IMT-2030(6G) Promotion Group}, responsible for managing and coordinating the program, and an expert group comprising 37 top researchers from academia, research institutes, and industry. In June 2021, the IMT-2030(6G) Promotion Group released its white paper titled ‘\textit{6G Vision and Candidate Technologies}’, outlining the state-of-the-art research findings of the group \cite{Ref_imt20302021}. It covers the \ac{6G} vision, the driving forces behind its development, potential use cases, ten candidate technologies, and additional insights.

In late 2017, the Japanese Ministry of Internal Affairs and Communications formed a working group to investigate next-generation wireless technologies. Their research findings indicated that \ac{6G} should offer transmission rates at least ten times faster than \ac{5G}, near-instant connectivity, and massive connection of up to ten million devices per square kilometer. In December 2020, Japan established the \textit{Beyond \ac{5G} Promotion Consortium (B5GPC)} with the objective of expediting the development of \ac{6G} while enhancing the country's international competitiveness through industry-academia-government collaboration. B5GPC published its inaugural white paper ‘\textit{Beyond 5G white paper: Message to the 2030s}’ in March 2022 \cite{Ref_b5gc2022}, summarizing the requirements and expectations of each industry for \ac{6G}, the necessary capabilities, and technological trends. South Korea announced its ambition to set up the world's first \ac{6G} trial in 2026. In addition, South Korea has unveiled the \textit{K-Network 2030} initiative, which aims to sponsor the development of key \ac{6G} technologies, e.g., developing cloud-native networks on South Korean-made \ac{AI} chips, launching a low-orbit communications satellite by 2027, and creating an \ac{ORAN} ecosystem for domestic firms. 

The German Federal Ministry of Education and Research (BMBF) announced in February 2021 a new funding program called ‘6G Vision’ as part of Germany's broader initiative to establish the country as a leader in \ac{6G} technology. In August 2021, under the umbrella organization and networking of a leading project named \textit{the \ac{6G} platform},  four \ac{6G} research hubs, i.e., \ac{6G}-life, \ac{6G}EM, \ac{6G} RIC, and Open6GHub,  were built \cite{Ref_fitzek20226g}. A total budget of approximately 250 million euros was assigned,  covering 160 research groups in 21 universities and 15 research institutes, as well as more than 40 small and medium enterprises. In the subsequent year, eighteen \ac{6G} industry projects, such as \ac{6G}-ANNA, \ac{6G}-TakeOff, \ac{6G}-Terafactory, and \ac{6G}-Next, and seven projects on resilience, e.g., HealthNet, AKITA, and ConnRAD, were established \cite{Ref_schotten2023}. 

In 2021, the National Agency for Research (ANR) in France launched the \textit{France 2030 plan} \cite{Ref_f2030plan}, a comprehensive national acceleration strategy focused on the evolution of communication technologies. This forward-looking initiative prioritizes digital transition, telecommunications, and global innovation. In May 2023, the France 2030 plan introduced the pivotal program PEPR-NF - Networks of the Future. This program has a shared goal of advancing technologies for beyond 5G and future network infrastructures while considering their environmental and societal impacts, as well as data security. The PEPR-NF program comprises ten interconnected projects, with one of them being NF-SYSTERA (Devices and SYStems for high-speed links in the sub-TERAhertz range). NF-SYSTERA aims to explore frequency bands beyond \SI{90}{\giga\hertz} for future wireless communication systems in the sub-THz and THz range.

\begin{table*} 
\centering
\caption{Evolution of Cellular Systems. \label{tab:evolution}}{%
\begin{tabular}{||c|c|c|c|c|c||} \hline \hline 
\multirow{2}{*}{} & \multicolumn{5}{c|}{\textbf{Mobile Generation}}\\ \cline{2-6}
&\textbf{1G} & \textbf{2G}& \textbf{3G} & \textbf{4G}& \textbf{5G} \\ \hline \hline
Main Standard & AMPS & GSM & WCDMA & LTE-Advanced & NR \\ \hline  
First Deployment Year&1979 & 1991& 2000 & 2009& 2019 \\ \hline
Peak Data Rate& \SI{10}{\kilo\bps} (signalling) & \SI{384}{\kilo\bps}  & \SI{2}{\mega\bps} & \SI{1}{\giga\bps}& \SI{20}{\giga\bps} \\ \hline
Signal Bandwidth&\SI{30}{\kilo\hertz} & \SI{200}{\kilo\hertz} & \SI{5}{\mega\hertz}& \SI{100}{\mega\hertz}& \SI{1}{\giga\hertz} \\ \hline
Frequency&\multirow{2}{*}{\SI{800}{\mega\hertz}} & \multirow{2}{*}{\SI{900}{\mega\hertz} \& \SI{1800}{\mega\hertz}} & \multirow{2}{*}{below \SI{2.1}{\giga\hertz}}& \multirow{2}{*}{Sub-\SI{6}{\giga\hertz}}& Sub-\SI{6}{\giga\hertz}/ \\ 
Bands& &  & & & millimeter wave (mmWave) \\ \hline \hline 
\end{tabular}}{
}
\end{table*}

\subsection{Up-to-date Spectrum Usage for IMT} \label{subsecSOTAspectrm}

Over the past few decades, the evolution of mobile communications has considered the following key criteria:
\begin{itemize}
    \item \textit{the signal bandwidth becomes increasingly wide};
    \item \textit{the operating frequency band is increasingly high}; and
    \item \textit{the spectral demand is increasingly large}.
\end{itemize}
Each new generation of cellular systems demanded more spectral resources and utilized a larger channel bandwidth to support more system capacity and realize a higher data rate than its predecessor. To provide an insightful view, Table \ref{tab:evolution} summarizes the evolution of mobile generations from \ac{AMPS} \cite{Ref_ehrlich1979advanced}, \ac{GSM}, \ac{WCDMA}, \ac{LTE-Advanced} \cite{Ref_jiang2013key}, to \ac{5G} \ac{NR} \cite{Ref_dahlman20215gNR} in terms of operating frequencies and signal bandwidths.

Until the \ac{4G}, cellular systems operated in low-frequency bands below \SI{6}{\giga\hertz}, which are referred to as \textit{the sub-\SI{6}{\giga\hertz} band} when high-frequency bands are considered in \ac{5G} \ac{NR} \cite{Ref_dahlman20215gNR}. 
During the \ac{ITU-R} \ac{WRC} held in 2015, also known as WRC-15, an item on the agenda was designated to identify high-frequency bands above \SI{24}{\giga\hertz} that could be used for IMT-2020 mobile services. After conducting follow-up studies after WRC-15, the \ac{ITU-R} found that ultra-low latency and high data-rate applications would require larger, contiguous spectrum blocks. At WRC-19, the ITU-R assigned several pieces of high-frequency bands for the deployment of \ac{5G} \ac{mmWave} communications worldwide. That is
\begin{itemize}
    \item 24.25-\SI{27.5}{\giga\hertz}
    \item 37-\SI{43.5}{\giga\hertz}
    \item 45.5-\SI{47}{\giga\hertz}
    \item 47.2-\SI{48.2}{\giga\hertz}
    \item 66-\SI{71}{\giga\hertz}
\end{itemize}

Meanwhile, the \ac{3GPP} specified the relevant spectrum for \ac{5G} \ac{NR}, which was divided into two frequency ranges:
\begin{itemize}
    \item FR1 - the First Frequency Range, including the sub-\SI{6}{\giga\hertz} frequency band from \SI{450}{\mega\hertz} to \SI{6}{\giga\hertz}
    \item FR2 - the Second Frequency Range, covering \SI{24.25}{\giga\hertz} to \SI{52.6}{\giga\hertz}.
\end{itemize} 
Initial \ac{mmWave} deployments are expected to operate in \SI{28}{\giga\hertz} (\ac{3GPP} \ac{NR} band n257 and n261) and  \SI{39}{\giga\hertz} (\ac{3GPP} n260) based on the \ac{TDD} mode, followed by \SI{26}{\giga\hertz} (\ac{3GPP} n258), as specified in Table \ref{tab:THz:FR2definition}.  
\begin{table}
\caption{Operating frequency bands specified by \ac{3GPP} for \ac{NR} in FR2. Source:  \cite{Ref_dahlman20215gNR}\label{tab:THz:FR2definition}}{%
\begin{tabular}{cccc} \hline \hline
\textbf{NR Band} & \textbf{Freq. Range [\si{\giga\hertz}]} & \textbf{Duplex Mode}& \textbf{Regions} \\ \hline
n257 & 26.5-29.5 & \ac{TDD} & Asia, Americas \\
n258 & 24.25-27.5 & \ac{TDD} & Asia, Europe \\
n259 & 39.5-43.5 & \ac{TDD} & Global \\
n260 & 37.0-40.0 & \ac{TDD} &Americas \\
n261 & 27.5-28.35 & \ac{TDD} & Americas \\ \hline \hline
\end{tabular}}{
}
\end{table}

\subsection{Prior Exploration of \ac{THz}}  \label{Sec:THZexplore}

The term \textit{terahertz} was initially used in the 1970s to describe the spectral line frequency coverage of a Michelson interferometer or the frequency coverage of point contact diode detectors \cite{Ref_siegel2002terahertz}. Before that, spectroscopists had coined this term for emission frequencies below the far \ac{IR} range, which is the lowest frequency part of the \ac{IR} radiation with a frequency range of about \SI{300}{\giga\hertz} to \SI{20}{\tera\hertz}.   Millimeter wave refers to the frequency band from \SIrange{30}{300}{\giga\hertz}. Hence, the border between the far \ac{IR} and \ac{THz}, and the border between \ac{mmWave} and \ac{THz}, are still rather blurry. Typically, the \ac{THz} band refers to \ac{EM} waves with a frequency band from \SIrange{0.1}{10}{\tera\hertz}. However, other definitions, e.g., \SI{300}{\giga\hertz} to \SI{3}{\tera\hertz}, are used parallelly. The difference in using frequency (\ac{THz}) and wavelength (\ac{mmWave}) for identifying the two bands leaves some ambiguity for the range from \SIrange{100}{300}{\giga\hertz}, which is also referred to as the \textit{upper-\ac{mmWave}} or \textit{sub-\ac{THz}} by some researchers. It is envisaged that \ac{5G} mainly focuses on the frequency bands below \SI{100}{\giga\hertz} while \ac{6G} and beyond will cross over this frequency point. The current trend seems to give more emphasis on using centimeter wave and low \acp{mmWave} for enhancing 6G networks, but the \ac{THz} band is certainly critical for sensing.

To avoid harmful interference to \ac{EESS} and radio astronomy operating in the spectrum between \SI{275}{\giga\hertz} and \SI{1}{\tera\hertz}, the \ac{ITU-R} WRC-15 has initiated the activity called ‘\textit{Studies towards an identification for use by administrations for land-mobile and fixed services applications operating in the frequency range 275–\SI{450}{\giga\hertz}}’. At the WRC-19 conference,  a new footnote was added to the radio regulations, allowing for the opening of the spectrum between \SI{275}{\giga\hertz} and \SI{450}{\giga\hertz} to land mobile and fixed services. Together with the already assigned spectrum below \SI{275}{\giga\hertz}, a total of \SI{160}{\giga\hertz} spectrum, containing two big contiguous spectrum bands with \SI{44}{\giga\hertz} bandwidth (i.e., from \SIrange{252}{296}{\giga\hertz}) and \SI{94}{\giga\hertz} bandwidth, respectively, is available for \ac{THz} communications without specific requirements to protect \ac{EESS} \cite{Ref_kuerner2020impact}. 

The \textit{\ac{mmWave} Coalition}, a group of innovative companies and universities united in the objective of removing regulatory barriers to technologies using frequencies ranging from \SIrange{95}{275}{\giga\hertz}, submitted comments in January 2019 to the \ac{FCC} and the National Telecommunications and Information Administration (NTIA) for developing a sustainable spectrum strategy and urged NTIA to facilitate the access to the spectrum above \SI{95}{\giga\hertz}. In March 2019, the \ac{FCC} announced that it would open up the use of frequencies between \SI{95}{\giga\hertz} and \SI{3}{\tera\hertz} in the United States, providing \SI{21.2}{\giga\hertz} of spectrum for unlicensed use and permitting experimental licensing for \ac{6G} and beyond. In 2016, the Defense Advanced Research Projects Agency (DARPA), in collaboration with prominent entities from the semiconductor and defense industries, such as Intel, Micron, and Analog Devices, established the Joint University Microelectronics Program (JUMP), comprising six research centers with the goal of addressing current and emerging challenges in the realm of microelectronic technologies. One such center, the \textit{Center for Converged TeraHertz Communications and Sensing (ComSecTer)}, is focused on the development of advanced technologies tailored to meet the requirements of the future cellular infrastructure.

The first attempt to build a wireless communications system at \ac{THz} frequencies started in 2008 with the foundation of a Terahertz Interest Group (IG\ac{THz}) under the IEEE 802.15 umbrella. In May 2014, the Task Group 3d was formed to standardize a switched point-to-point communications system operating in the frequencies from \SI{60}{\giga\hertz} to the lower \ac{THz} bands. During the meeting in March 2016, the supporting documents for IEEE 802.15.3d were approved, and the call for proposals was issued. Based on the proposal reviews and two sponsor recirculation ballots, the IEEE 802.15.3d-2017 specifications were ratified by the IEEE Standards Association (SA) Standards Board in September 2017 \cite{Ref_ieee2017standard802153d}. IEEE 802.15.3d-2017 specifies an alternative Physical (PHY) layer tailored to the lower \ac{THz} frequency band from \SIrange{252}{325}{\giga\hertz} for switched point-to-point connections. This standard aims for a maximum speed of over \SI{100}{\giga\bps} with eight bandwidth configurations from \SI{2.16}{\giga\hertz} to \SI{69.12}{\giga\hertz} and with an effective coverage from tens of centimeters to a few hundred meters \cite{Ref_petrov2020first}.

\section{Potential \ac{THz} Applications in \ac{6G} and Beyond}

The massive amount of spectrum at \ac{THz} frequencies offers opportunities for ultra-fast wireless connections \cite{Ref_huq2019terahertz}. It also introduces a new level of flexibility in mobile system design, as \ac{THz} links can be utilized for wireless backhaul among network nodes, which enables ultra-dense architectures, accelerates the deployment, and reduces the costs associated with site acquisition, installation, and maintenance. Due to the tiny wavelengths, the antenna dimension is very small, opening up possibilities for innovative applications such as nanoscale communications for nanoscale devices or nanomachines, on-chip communications, the Internet of Nano-Things, and intra-body networks \cite{Ref_lemic2021survey}. Moreover, \ac{THz} signals can be used beyond communication applications, facilitating high-definition sensing, imaging, and positioning of the surrounding physical environment \cite{Ref_sarieddeen2020next}. This offers the potential to efficiently implement integrated communications and sensing at the \ac{THz} band. Table \ref{Table_THzAPP} summarizes the available survey papers and the potential THz applications and use cases presented therein, including the present survey.

\subsection{Terahertz Communications Applications}
\paragraph{Terabit Cellular Hotspots} The proliferation of mobile and fixed users with high-throughput demands in densely populated urban areas or specific locations, such as industrial sites, necessitates the deployment of ultra-dense networks. The utilization of the \ac{THz} band can offer an abundance of spectral resources and ultra-wide bandwidth for small cells, which possess a relatively short coverage distance and high likelihood of \ac{LoS} paths, allowing for Terabit communications links. These small cells cater to both static and mobile users in both indoor and outdoor settings, providing specific applications such as ultra-high-definition video delivery, information shower, high-quality virtual reality, and holographic-type communications \cite{Ref_jiang2021road}. By incorporating conventional cellular networks operating in low-frequency bands, a heterogeneous network, consisting of a macro-base-station tier and a small-cell tier, can facilitate seamless connectivity and full transparency across a wide coverage area and global roaming, thus fulfilling the extreme performance requirements of \ac{6G} and beyond mobile networks \cite{Ref_huq2019terahertz}.

\renewcommand{\arraystretch}{1.2}
\begin{table*}[t]  \scriptsize
	\centering
	\caption{A Survey of \ac{THz} Applications and Use Cases for \ac{6G} and beyond.}
	\begin{tabular}{|l|c|c|c|c|c|c|c|c|c|c|}
		\hline \hline
		\multirow{7}{*}{\textbf{Reference}} & 		 
		\multicolumn{7}{c|}{\textbf{\ac{THz} Communications}} & \multicolumn{3}{c|}{\textbf{\ac{THz} Sensing}}\\ \cline{2-11}
		 & \rotatebox[origin=c]{90}{\parbox[c][8mm][c]{15mm}{\raggedright \textbf{ Hotspot}}} &  \rotatebox[origin=c]{90}{\parbox[c][8mm][c]{15mm}{\raggedright \textbf{ Campus}}}   & \rotatebox[origin=c]{90}{\parbox[c][8mm][c]{15mm}{\raggedright \textbf{ D2D}}} & \rotatebox[origin=c]{90}{\parbox[c][10.5mm][c]{15mm}{\raggedright \textbf{ Vehicle}}} & \rotatebox[origin=c]{90}{\parbox[c][10.5mm][c]{15mm}{\raggedright \textbf{ Security}}} & \rotatebox[origin=c]{90}{\parbox[c][10.5mm][c]{15mm}{\raggedright \textbf{ Backhaul}}} &  \rotatebox[origin=c]{90}{\parbox[c][10.5mm][c]{15mm}{\raggedright \textbf{ Nanocom.}}} & \rotatebox[origin=c]{90}{\parbox[c][10.5mm][c]{15mm}{\raggedright \textbf{ Sensing}}} & \rotatebox[origin=c]{90}{\parbox[c][10.5mm][c]{15mm}{\raggedright \textbf{ Imaging}}} & \rotatebox[origin=c]{90}{\parbox[c][10.5mm][c]{15mm}{\raggedright \textbf{ Positioning}}} \\ \hline \hline 
   H. Sarieddeen \emph{et al.} \cite{Ref_sarieddeen2020next} &   & &  &  & &  &  & \checkmark & \checkmark & \checkmark\\ \hline
   K. M.S Huq \emph{et al.} \cite{Ref_huq2019terahertz} & \checkmark  &  & $\circ$ & \checkmark & & \checkmark & $\circ$ & $\circ$ &  & \\ \hline
   K. Tekbiyik \emph{et al.} \cite{Bio_tekbiyik2019terahertz} &  &  & $\circ$ & \checkmark & \checkmark & \checkmark & \checkmark & $\circ$ & $\circ$ & $\circ$ \\ \hline
   Z. Chen \emph{et al.} \cite{Ref_chen2019survey}  & $\circ$  & &  &  & &  & \checkmark & \checkmark & \checkmark & \checkmark\\ \hline
   Y. He \emph{et al.} \cite{Ref_he2020overview} &   & &  &  & &  & \checkmark & & \checkmark & \\ \hline
   B. Ning \emph{et al.} \cite{Ref_ning2022prospective} &   & &  & \checkmark & \checkmark & $\circ$ & $\circ$ & \checkmark & $\circ$ & $\circ$ \\ \hline
   F. Lemic \emph{et al.} \cite{Ref_lemic2021survey} &   & &  &  & &  & \checkmark & & & \\ \hline
   C. Chaccour \emph{et al.} \cite{Ref_chaccour2022seven} &   & &  & \checkmark & & \checkmark & $\circ$ & \checkmark & \checkmark & $\circ$ \\ \hline
   H. Sarieddeen \emph{et al.} \cite{Ref_sarieddeen2020next} &   & &  &  & &  &  & \checkmark & \checkmark & \checkmark \\ \hline
   J. M. Jornet \emph{et al.} \cite{Bio_6804405} &   & &  &  & &  & \checkmark &  &  & \\ \hline
   J, Bo Kum \emph{et al.} \cite{Bio_8873734} &   & &  &  & & \checkmark & &  & & \\ \hline
   C. Han \emph{et al.} \cite{Bio_9794668} &   & $\circ$ &  & \checkmark & &\checkmark &\checkmark & $\circ$  & $\circ$  & $\circ$  \\ \hline
   J. M. Eckhardt \emph{et al.} \cite{Bio_9403881} &   & &  & \checkmark & &  & &  & & \\ \hline
   S. Ju \emph{et al.} \cite{Bio_9685929} &   & &  & \checkmark & &  & &  & & \checkmark \\ \hline
   G. Ke \emph{et al.} \cite{Bio_8739519} &   & & \checkmark & \checkmark & &  & &  & &  \\ \hline
   K. Rikkinen \emph{et al.} \cite{Ref_rikkinen2020THz} &   & &  &  & & \checkmark & &  & & \\ \hline
   J. M. Jornet \emph{et al.} \cite{Ref_jornet2011channel} &   & &  &  & &  & \checkmark &  &  & \\ \hline
   A. Faisal \emph{et al.} \cite{Ref_faisal2020ultramassive} &   & & $\circ$  & $\circ$  & &  & \checkmark & \checkmark & \checkmark & \checkmark \\ \hline
   J. M. Jornet \emph{et al.} \cite{Ref_jornet2013graphene} &   & &  &  & &  & \checkmark &  &  & \\ \hline
   K.O. Kenneth \emph{et al.} \cite{Ref_kenneth2019opening} &   & &  &  & &  &  & \checkmark & \checkmark & \\ \hline
   C.-X. Wang \emph{et al.} \cite{Ref_Wang2021Key} &   & &  &  & & \checkmark & \checkmark & $\circ$ & $\circ$ & $\circ$ \\ \hline
   S. Helal \emph{et al.} \cite{helal2022signal} &   & &  & \checkmark & \checkmark &  &  & \checkmark & \checkmark & \checkmark \\ \hline
   Q. H. Abbasi \emph{et al.} \cite{Ref_abbasi2016nano} &   & &  &  & &  & \checkmark &  &  & \\ \hline
   H. Park \emph{et al.}  \cite{Ref_Park2021Machine} &   & &  &  & &  &  &  & \checkmark & \\ \hline
   Akyildiz and Jornet \cite{Ref_akyildiz2010internet} &   & &  &  & &  & \checkmark &  &  & \\ \hline
   Z. Chen \emph{et al.} \cite{chen2021terahertz} &   & & &  & \checkmark & \checkmark &  & \checkmark &  & \checkmark \\ \hline
   I. B. Djordjevic \emph{et al.} \cite{Ref_Djordjevic2017OAM} &   & &  \checkmark &  &  \checkmark &  \checkmark &  &  &  & \\ \hline
   Z. Fang \emph{et al.} \cite{Ref_Fang2022Secure} &   & &   &  & \checkmark &   &  &  &  & \\ \hline
   Y. Yang \emph{et al.} \cite{Ref_yang2020terahertz} &   & &  &  & &  & \checkmark &  &  & \\ \hline
   H. Wymeersch \emph{et al.} \cite{Wymeersch2021} &   & &   &  & \checkmark &   &  & \checkmark & \checkmark & \checkmark \\ \hline 
   J. Ma \emph{et al.} \cite{Ref_ma2018security} &   & &   &  & \checkmark &   &  &  &  &  \\ \hline 
   A. A. Mamrashev \emph{et al.} \cite{Ref_mamrashev2018detection} &   & &  &  & &  & \checkmark &  &  & \\ \hline
   A. A. Boulogeorgos \emph{et al.} \cite{Ref_boulogeorgos2018terahertz} & \checkmark  &\checkmark & \checkmark  &  &  &  \checkmark &  &  &  &  \\ \hline
   M. Lotti \emph{et al.} \cite{Ref_lotti2022radio} &   & &   &  &  &   &  &  & \checkmark &  \\ \hline
   S. Helal \emph{et al.}  \cite{helal2022signal} &   & &   &  &  &   &  & \checkmark &  &  \\ \hline
   N. A. Abbasi \emph{et al.} \cite{Ref_ponce2023thz} &   & & \checkmark  &  &  &   &  &  &  &  \\ \hline
   C. Zandonella \cite{Ref_zandonella2003terahertz} &   & &   &  &  &   &  &  & \checkmark &  \\ \hline
   O. Li \emph{et al.} \cite{Ref_li2021integrated} &  \checkmark & \checkmark &   &  &  &   & \checkmark &  & \checkmark &  \\ \hline
   E. Castro‑Camus1 \emph{et al.} \cite{Ref_castrocamus2021recent} &   & &   &  &  &   &  &  & \checkmark &  \\ \hline
   \textbf{This survey paper} & \checkmark & \checkmark & \checkmark & \checkmark & \checkmark & \checkmark & \checkmark & \checkmark & \checkmark & \checkmark \\ \hline
		\multicolumn{11}{|l|}{\begin{tabular}[l]{@{}l@{}} Note:  \end{tabular}} \\  
		\multicolumn{11}{|l|}{\begin{tabular}[l]{@{}l@{}} For each column, the \checkmark symbol means that this application is discussed in detail in the reference, the $\circ$ symbol means that this application is only mentioned\\ briefly, and the blank indicates that this application is not discussed.  \end{tabular}} \\ \hline \hline
	\end{tabular}
	\label{Table_THzAPP} 
\end{table*}

\paragraph{Terabit Campus/Private Networks} \ac{THz} frequencies provide a means for implementing super-high-rate, ultra-reliable, and hyper-low-latency connectivity within a private or campus network for specific applications such as Industry 4.0 and Tactile Internet \cite{Ref_fettweis2014tactile}. This allows for seamless interconnection between ultra-high-speed optical networks and production devices with no discernible speed or delay difference between wireless and wired links. In addition, abundant bandwidths at \ac{THz} frequencies also make massive connection density a reality \cite{Ref_han2022terahertz}. These capabilities facilitate the deployment of industrial networks, linking a vast number of sensors and actuators within a factory, and campus networks providing high data-throughput, low-latency, and high-reliability connections for equipment and machines such as \ac{AGV} in a logistic center.

\paragraph{Terabit Device-To-Device and Vehicle-to-Everything}
\ac{THz} communications represent a promising tool for providing direct Tbps links between devices in close proximity \cite{Bio_9685929}. Indoor usage scenarios, such as homes or offices, can benefit from the formation of particular \ac{D2D} links \cite{Ref_jiang2017devicetodevice} among a set of personal or commercial devices \cite{Ref_ponce2023thz}. Applications such as multimedia kiosks and ultra-high-speed data transfer between personal devices can be supported with Tbps links, enabling the transfer of the equivalent content of a blue-ray disk to a high-definition large-size display in less than one second. \ac{THz} communications could also have a significant impact on Brain-Computer Interface (BCI) applications, enabling the transfer of vast amounts of collected brain-wave data to the computer that processes the data. In computer vision, \ac{THz} communications can facilitate the transfer of high-definition video data to platforms running machine learning-based analytical software. Additionally, Tbps \ac{D2D} links can be applied in outdoor settings for vehicle-to-everything scenarios \cite{Bio_8739519}, providing high-throughput, low-latency connectivity between vehicles or between vehicles and surrounding infrastructure \cite{Bio_9403881}.

\paragraph{Secure Wireless Connectivity} 
\ac{THz} channels exhibit sparsity \cite{Ref_he2021wireless} where the angular spread is remarkably smaller than that of low frequencies due to the decreased diffraction effects for small wavelengths \cite{Ref_han2022terahertz}. Additionally, to compensate for the severe signal attenuation due to large free-space path loss, atmospheric absorption, and weather effects, as detailed in Sec. \ref{Sec_Channel_Effect},  the use of large-scale antenna arrays is a good option for \ac{THz} communications and sensing. With appropriate beam steering and alignment methods, as elaborated in Sec. \ref{Sec_BF}, pencil-like beams with high gains are generated to extend the transmission range. This setting brings a unique advantage from the perspective of information security. That is, highly directional beams are able to effectively confine unauthorized users to be on the same narrow path as the intended users, therefore lowering the risks of eavesdropping \cite{Ref_Fang2022Secure}.  The equipment needed to demodulate and amplify \ac{THz} signals is large and bulky. Hence, it is difficult for an eavesdropper to intercept signals without blocking the intended recipient and therefore raising an alarm. Regardless of its inherent nature of security, the immunity from \ac{THz} eavesdropping cannot be fully guaranteed (e.g. placing a passive reflector in the beam), unless counter-measures are applied  \cite{Ref_ma2018security}. Further aspects related to secure \ac{THz} systems are discussed in Sub-Sec. \ref{Subsec:PLSforTHz}.

\paragraph{Terabit Wireless Backhaul}
The installation of fiber optical connections is typically time-consuming and costly, and it may not always be feasible to deploy public optical networks within certain buildings or areas due to property owner objections. However, the next-generation mobile network is expected to be highly heterogeneous, requiring high-throughput backhaul or fronthaul connectivity between network elements such as macro base stations, small cells, relays, and distributed antennas. Highly directive \ac{THz} links can provide ultra-high-speed wireless backhaul or fronthaul \cite{Bio_8873734}, reducing the time and cost of installation and maintenance while enabling greater flexibility in network architecture and communications mechanisms \cite{Ref_rikkinen2020THz}. 
Nowadays, in addition, mobile or fixed users in rural or remote areas suffer from worse coverage and low \ac{QoS}. If a cost-efficient and flexible solution cannot be guaranteed, the digital divide between rural areas and major cities will increase. As a wireless backhaul extension of the optical fiber \cite{Ref_boulogeorgos2018terahertz}, \ac{THz} wireless links can work well as an essential building block to guarantee a universal telecommunications service with high-quality, ubiquitous connections everywhere.

\paragraph{Terahertz Nano-Networks}
Nanotechnology enables the development of nanomachines in the size of up to a few hundred nanometers that perform simple tasks at the nanoscale in the biomedical, environmental, industrial, and military fields. Nano-communications \cite{Ref_akyildiz2010internet}, which interconnect nanomachines, can expand the potential applications and extend the range of operation.  
Existing \ac{RF} and optical transceivers suffer from several constraints such as size, complexity, and power consumption for being used in nano-communications. This motivated the use of new nano-materials to build nano-transceivers, as well as determining the frequency band to operate these nano-transceivers  
\cite{Ref_jornet2013graphene}. Hardware advances such as graphene-based plasmonic nano-antenna \cite{Ref_jornet2013graphene}, microchip emitters and detectors for \ac{THz} sensing \cite{Ref_otsuji2010plasmon},  graphene-based nano-transceiver \cite{Ref_jornet2010graphene}, and plasmonic nano-antenna array \cite{Ref_singh2020design}, further motivate the use of the \ac{THz} band although challenges like molecular absorption noise need to be tackled \cite{Ref_jornet2011channel}.

One of the early applications of nano-networks is found in the field of nanosensing \cite{Ref_akyildiz2010electromagnetic}. Nanosensors are not merely tiny sensors but nanomachines that exploit the characteristics of nano-materials to identify and measure nanoscale events. For instance, nanosensors can identify chemical compounds at concentrations as low as one part per billion, or even detect the presence of a virus or harmful bacteria. Possible use cases for \ac{THz}-based nano-networks are the following: 
\begin{itemize} 
    \item \textit{In-Vivo Body-Centric Monitoring:} Sodium, glucose, and other ions in the blood, cholesterol, cancer biomarkers, or the presence of different infectious agents can be detected utilizing nanoscale biosensors injected into the human body due to its non-invasive nature  \cite{Ref_akyildiza2014terahertz}. A set of biosensors distributed within or around the body, comprising a body sensor network, could collect relevant physical or biochemical data related to human health for in-vivo body monitoring, as well as long-term health treatment \cite{Ref_abbasi2016nano}. In addition to nano-communications, \ac{THz} waves exhibit unique advantages for imaging purposes. For example, their non-ionization nature and ability to penetrate dielectrics to a suitable depth are useful for non-destructive disease diagnosis and biomedical imaging  \cite{Ref_roh2022terahertz}.   
    \item \textit{Nuclear, Biological, and Chemical Defense:} Chemical and biological nanosensors are able to detect harmful chemicals and biological threats in a distributed manner. One of the main benefits of using nanosensors rather than classical macroscale or microscale sensors is that a chemical composite can be detected in a concentration as low as one molecule and much more timely than classical sensors \cite{Ref_mamrashev2018detection}. 
    \item \textit{Internet-of-Nano-Things:} Using \ac{THz} nano-communications to interconnect nanoscale machines, devices, and sensors with existing wireless networks \cite{Ref_jornet2011channel} and the Internet allows for the realization of a truly cyber-physical system that can be named as the \ac{IoNT} \cite{Ref_akyildiz2010internet}. The \ac{IoNT} enables disruptive applications that may change how humans work or live. 
    \item \textit{On-Chip Communication:} \ac{THz} communications can provide an efficient and scalable approach to inter-core connections in on-chip wireless networks using planar nano-antenna arrays to create ultra-high-speed links \cite{Ref_yang2020terahertz}. This novel approach may expectedly fulfill the stringent requirements of the area-constraint and communication-intensive on-chip scenario by its high bandwidth, low latency, and low overhead. 
\end{itemize}

\renewcommand{\arraystretch}{1.2}
\begin{table*}[t]  \scriptsize 
	\centering
	\caption{Potential supports of \ac{THz} sensing and communications for IMT-2030 usage scenarios.}
	\begin{tabular}{|l|c|c|c|c|c|c|c|c|c|c|}
		\hline \hline
		\multirow{15}{*}{\textbf{6G Usage Scenarios}} & 		 
		\multicolumn{7}{c|}{\textbf{\ac{THz} Communications}} & \multicolumn{3}{c|}{\textbf{\ac{THz} Sensing}}\\ \cline{2-11}
		 & \rotatebox[origin=c]{90}{\parbox[c][7mm][c]{30mm}{\raggedright \textbf{ Tbps Cellular Hotspot}}} &  \rotatebox[origin=c]{90}{\parbox[c][7mm][c]{30mm}{\raggedright \textbf{ Tbps Campus Networks}}}   & \rotatebox[origin=c]{90}{\parbox[c][7mm][c]{30mm}{\raggedright \textbf{ Tbps D2D Commun.}}} & \rotatebox[origin=c]{90}{\parbox[c][7mm][c]{30mm}{\raggedright \textbf{ Tbps Vehicle Networks}}} & \rotatebox[origin=c]{90}{\parbox[c][7mm][c]{30mm}{\raggedright \textbf{ Secure THz Commun.}}} & \rotatebox[origin=c]{90}{\parbox[c][7mm][c]{30mm}{\raggedright \textbf{ Tbps Wireless Backhaul}}} &  \rotatebox[origin=c]{90}{\parbox[c][7mm][c]{30mm}{\raggedright \textbf{ THz Nano-Networks}}} & \rotatebox[origin=c]{90}{\parbox[c][7mm][c]{30mm}{\raggedright \textbf{ THz Sensing}}} & \rotatebox[origin=c]{90}{\parbox[c][7mm][c]{30mm}{\raggedright \textbf{ THz Imaging}}} & \rotatebox[origin=c]{90}{\parbox[c][7mm][c]{30mm}{\raggedright \textbf{ THz Positioning}}} \\ \hline  \hline 
   \textbf{Immersive Communication} & \checkmark  & $\circ$& \checkmark & \checkmark &  & \checkmark &  & $\circ$ &  & \checkmark\\ \hline
   \textbf{Hyper Reliable and Low-Latency Communication} &  $\circ$ & \checkmark& \checkmark & \checkmark & \checkmark& \checkmark &  & $\circ$ &  & $\circ$\\ \hline
   \textbf{Massive Communication} &   & $\circ$ &  &  &  &  & \checkmark &  &  & \checkmark\\ \hline
   \textbf{Integrated AI and Communication} &  \checkmark &\checkmark &\checkmark  &\checkmark  & \checkmark& \checkmark &  & \checkmark & \checkmark & \checkmark\\ \hline
   \textbf{Integrated Sensing and Communication} &   & &  &  & $\circ$& $\circ$ & \checkmark & \checkmark & \checkmark & \checkmark\\ \hline
   \textbf{Ubiquitous Connectivity} &   & & \checkmark &  & & \checkmark &  &  &  & \checkmark\\ \hline
		\multicolumn{11}{|l|}{\begin{tabular}[l]{@{}l@{}} Note:  \end{tabular}} \\  
		\multicolumn{11}{|l|}{\begin{tabular}[l]{@{}l@{}} For each entry, the \checkmark symbol means that this \ac{THz} use can explicitly contribute to the designated 6G usage scenario, the $\circ$ symbol means a potential contribution,\\ and the blank means no contribution at all.  \end{tabular}} \\ \hline \hline
	\end{tabular} 
	\label{Table_THzfor6GScenarios} 
\end{table*}

\subsection{Terahertz Sensing Applications}
\paragraph{Terahertz Sensing} At \ac{THz} frequencies, the spatial resolution of a signal becomes much finer due to the tiny wavelengths, allowing for high-definition spatial differentiation \cite{Wymeersch2021}. \ac{THz} sensing techniques take advantage of the frequency-selective resonances of various materials in the measured environment, as well as the small wavelengths, typically on the order of micrometers \cite{Ref_li2021integrated}. This enables the extraction of unique information based on the observed signal signature. \ac{THz} signals can penetrate non-conducting materials like plastics, fabrics, paper, wood, and ceramics, but they face challenges when penetrating metallic materials or when water heavily attenuates their radiation power. The specific strength and phase variations of \ac{THz} signals caused by different thicknesses, densities, or chemical compositions of materials enable the accurate identification of physical objects \cite{helal2022signal}. 
\paragraph{Terahertz Imaging} 
Using \ac{THz} radiation to form images has many technical advantages over microwaves and visible light. \ac{THz} imaging \cite{Ref_Park2021Machine} exhibits high spatial resolution due to smaller wavelengths and ultra-wide bandwidths with moderately sized hardware than imaging using low frequencies. Compared with infrared and visible light, \ac{THz} waves have better penetration performance, making common materials relatively transparent to \ac{THz} signals.  There are many security screening applications, such as checking postal packages for concealed objects, allowing \ac{THz} imaging through envelopes, packages, parcels, and small bags to identify potential hazardous items \cite{Ref_zandonella2003terahertz}. \ac{THz} radiation is non-ionizing, and therefore it has no known health risks to biological cells except for heating, which has motivated its application for imaging of the human body, where ionizing radiations, i.e., ultraviolet, X-ray, and Gamma-ray, cannot be utilized due to high health risks. Therefore, \ac{THz} imaging is suitable for the stand-off detection of items such as firearms, bombs, and explosive belts hidden beneath clothing in airports, train stations, and border crossings \cite{Ref_castrocamus2021recent}.
\paragraph{Terahertz Positioning} 
It is envisioned that \ac{6G} and beyond systems are required to offer high-accurate positioning and localization in both indoor and outdoor environments, in addition to communications services, which \ac{GNSS} and conventional multi-cell-based localization techniques using low-frequency bands fail to provide. Devices incorporating \ac{THz} sensing and \ac{THz} imaging will likely provide centimeter-level localization anywhere \cite{Ref_sarieddeen2019terahertz}. On the other hand, leveraging \ac{THz} imaging for localization has unique benefits compared to other methods.  \ac{THz} imaging can localize users in \ac{NLoS} areas, even if their travel paths to the base station experience more than one reflection (e.g., multiple bounces).   
High-frequency localization techniques are based on the concept of \ac{SLAM} \cite{Ref_lotti2022radio}, in which the accuracy is improved by collecting high-resolution images of the environment, where \ac{THz} imaging can provide such high-resolution images. \ac{SLAM}-based techniques consist of three main steps: imaging the surrounding environment, estimation of ranges to the user, and fusion of images with the estimated ranges.  Since \ac{SLAM} deals with relatively slow-moving objects, there is sufficient time to process high-resolution \ac{THz} measurement. Such measurement can hold sensing information, resulting in complex state models comprising the fine-grained location, size, and orientation of target objects, as well as their electromagnetic properties and material types \cite{Ref_sarieddeen2020next}.

To summarize how \ac{THz} communications and \ac{THz} sensing can support \ac{6G} and beyond systems, Table \ref{Table_THzfor6GScenarios} shows the inter-connection between \ac{THz} applications and IMT-2030 usage scenarios (see \figurename \ref{Figure_6Gscenarios}).

\section{\ac{THz} Propagation and Channel Characterization} \label{Sec_Channel_Effect}
Compared with low-frequency channels, \ac{THz} channels are much less understood, as they possess several distinct characteristics  \cite{Ref_Jeon2022, Ref_han2022terahertz}.
Like microwave and \ac{mmWave}, \ac{THz} signals suffer from \ac{FSPL}, as inherent attenuation when an electromagnetic wave is radiated from an isotropic antenna. Unfortunately, \ac{THz} antennas have a weak ability to capture the radiation power due to their small aperture. This leads to a \ac{FSPL} that proportionally grows with the increase of the carrier frequency. 

Since the wavelength of a \ac{THz} wave falls into the same order of magnitude as the dimensions of molecules in the atmosphere and human tissue, strong molecular absorption and particle scattering, which are negligible in low-frequency bands, become significant \cite{Ref_serghiou2022terahertz}. To be specific, water vapor and oxygen molecules suspended in the atmosphere impose an incredible loss of thousands of \si{\decibel} per kilometer in the worst case. In addition to this gaseous absorption from water molecules, liquid water droplets, in the form of suspended particles into clouds, rain-falling hydrometeors, snowflakes, and fogs, can attenuate the signal strength since their dimensions are comparable to the \ac{THz} wavelength. Furthermore, surrounding physical objects become sufficiently large in size for scattering, and ordinary surfaces are also too rough to make specular reflections. As a result, a \ac{THz} wave is susceptible to being blocked by buildings, furniture, vehicles, foliage, and even the human body. 

Characterizing the propagation of THz signals is mandatory for designing transmission algorithms,  developing network protocols, evaluating system performance, and deploying commercial networks.  Therefore, this section comprehensively characterizes \ac{THz} signal propagation, including path loss, atmospheric absorption, weather effects, and blockage, aiming to provide the readers with the prerequisite knowledge of \ac{THz} channels for designing \ac{THz} systems for communications and sensing.

  \subsection{High Free-Space Path Loss}

When an isotropic radiator feeds an \ac{EM} wave into free space, the energy evenly spreads over the surface of an ever-increasing sphere. The metric \textit{\ac{EIRP}}  indicates the maximal energy in a particular direction relative to a unity-gain isotropic antenna. Hence, it equals the product of the transmit power $P_t$ and the transmit antenna gain $G_t$ in the direction of a receive antenna. The \textit{law of conservation of energy} states that the total energy contained on the surface of a sphere of any radius $d$ remains constant \cite{Ref_tse2005fundamentals}. Power flux density, namely the power flow per unit area of the incident field at the antenna, is equivalent to the \ac{EIRP} divided by the surface area of a sphere with radius $d$, i.e., $\frac{P_tG_t}{4\pi d^2}$. The received power captured by a receive antenna is proportional to its aperture $A_{r}$, and it is equal to
\begin{equation} \label{Eqn:THz:RxPower}
P_r = \left(\frac{P_tG_t}{4\pi d^2}\right)A_{r}.
\end{equation}
Meanwhile, the gain of a receive antenna $G_r$ depends on its effective aperture area according to the following relationship 
\begin{equation} \label{Eqn:THz:EffArea}
A_{r} =    G_r \left( \frac{\lambda^2}{4\pi}\right),
\end{equation}
where $\lambda$ stands for the wavelength of the transmitted signal.
Substituting \eqref{Eqn:THz:EffArea} into \eqref{Eqn:THz:RxPower} yields the well-known Friis transmission equation introduced by Harald T. Friis in 1946 \cite{Ref_friis1946note}, i.e.,
\begin{equation} \label{Eqn:THz:PathLoss}
P_r=  P_tG_tG_r \left ( \frac{\lambda}{4\pi d} \right)^2.
\end{equation}
  
Due to the large dynamic range across several orders of magnitude,  we usually express the strength of signals and noise in decibels (dB). Free-space path loss is defined as the ratio between the transmit and receive power on a logarithmic scale: 
\begin{equation} \label{Eqn:THz:PLinDecibel}
\mathrm{PL}=10\lg \frac{P_t}{P_r} =  20\lg\left(\frac{4\pi d}{\lambda}\right)-10\lg \left( G_t G_r\right),
\end{equation}
which implies that the \ac{FSPL} increases \SI{20}{\decibel} per decade (ten times) as a function of the carrier frequency. For example, the loss increases by \SI{20}{\decibel} from \SI{30}{\giga\hertz} to \SI{300}{\giga\hertz} under the assumption that $d$, $G_t$, and $G_r$ are kept fixed. Note that the high path loss at THz frequencies is due to the small aperture area of the receive antenna, which is proportional to the square of wavelength, based on \eqref{Eqn:THz:EffArea} and assuming that $G_r$ is kept fixed.

\ac{FSPL} does not completely account for the realistic characteristics of wireless propagation because the physical environment of a terrestrial wireless communication system is distinct from free space. In addition to the \ac{LoS} path, an electromagnetic wave may be reflected, diffracted, and scattered by the surrounding objects in an urban or indoor scenario, creating \ac{NLoS} paths between a pair of transmit and receive antennas. Due to the differences in energy attenuation, propagation delays, and phase rotations, the additional copies of an electromagnetic wave, which are referred to as multi-path components, cause an extra drop in the received signal power. The novel use cases for the \ac{THz} band in \ac{6G} and beyond networks, such as kiosk downloading, nano-scale networks, and wireless backhaul, introduce many peculiarities, which need to be well investigated. Hence, extensive measurement and accurate modeling of the path loss in terms of frequency, distance, and propagation environment have to be carried out. In this context, existing research works are summarized in Table \ref{tableChannelProp}.

\begin{table*}[!t]
    \centering 
    \caption{Summary of research works on the propagation characteristics}
    \label{tableChannelProp}
    \scriptsize 
    \begin{tabular}{|c|c|c|p{13cm}|}
        \hline \hline
        \textbf{Effects} & \textbf{Year} & \textbf{Reference} & \textbf{Major Contributions} \\ \hline \hline
         \multirow{17}{*}{\begin{tabular}[c]{@{}c@{}}Free-Space\\Path Loss\end{tabular}} & 2020 & \cite{Bio_9135643} & Employed a channel sounder that covers \SI{140}{\giga\hertz} - \SI{220}{\giga\hertz} and a frequency extender for THz channel measurement and built a platform for the exploration of THz communications within a short distance up to \SI{5}{\meter}. \\ \cline{2-4}
        & 2020 & \cite{Bio_9148631} & Presented the first set of double-directional outdoor measurement over a 100 m distance in urban scenarios based on RF-over-Fiber (RFoF) extensions in the \SI{141}{\giga\hertz} - \SI{148.5}{\giga\hertz} range. \\ \cline{2-4}
        & 2021 & \cite{Bio_9473566} & Presented the results in an urban environment of the above-listed reference on a linear moving route for a distance up to 15 m in the \SI{140}{\giga\hertz} - \SI{141}{\giga\hertz} range. Analyzed how key channel parameters change as they move from short to longer distances. \\ \cline{2-4}
        & 2021 & \cite{Bio_9500510} & Unveiled that metallic-covered surfaces lead to a considerable enhancement of multi-path, indicating a critical impact of building materials.  Relying on pronounced sparsity for \ac{THz} system design might not always be valid in this type of environment. \\ \cline{2-4}
        & 2021 & \cite{Bio_9473756} & Wideband channel measurement campaigns at \SI{140}{\giga\hertz} and \SI{220}{\giga\hertz} are conducted in indoor scenarios including a meeting room and an office room.  Developed single-band close-in path loss models to investigate the large-scale fading characteristics.  \\ \cline{2-4}
        & 2021 & \cite{Bio_9466322} & A wideband channel measurement campaign between \SI{130}{\giga\hertz} and \SI{143}{\giga\hertz} is investigated in a typical meeting room. \\ \cline{2-4}
        & 2021 & \cite{Bio_9500677} & Introduced outdoor UMi propagation measurement at \SI{142}{\giga\hertz} along a 39 m × 12 m rectangular route, where each consecutive and adjacent receiver location is 3 m apart from each other. \\ \cline{2-4}
        & 2022 & \cite{Bio_9839013} & Investigated the \ac{LoS} and \ac{NLoS} path loss by carrying out measurement in indoor scenarios at frequencies ranging from \SI{130}{\giga\hertz} to \SI{320}{\giga\hertz} with a frequency-domain \ac{VNA}-based sounder. \\ \cline{2-4}
        & 2022 & \cite{Bio_9838910} & Path-loss analyses in a factory building based on sub-\ac{THz} channel measurement with a maximal distance of \SI{40}{\meter} using directional horn antennas at \SI{142}{\giga\hertz}. Facilitated emerging sub-THz applications for future smart factories.  \\ \cline{2-4}
        & 2022 & \cite{Bio_9685929} & Investigated the \ac{UMi} large-scale path loss at 28, 38, 73, and \SI{142}{\giga\hertz}. Introduced a detailed spatial statistical \ac{MIMO} channel generation procedure based on the derived empirical channel statistics.  \\ \hline
        \multirow{13}{*}{\begin{tabular}[c]{@{}c@{}}Atmospheric\\Absorption\end{tabular}}
         & 1982 & \cite{Bio_7768108} & Calculated the absorption values due to atmospheric oxygen and water vapor with frequencies spanning from \SIrange{1}{340}{\giga\hertz}. \\ \cline{2-4}
        & 1987 & \cite{Bio_Liebe1987Mill} & Performed laboratory measurement of water vapor attenuation at \SI{138}{\giga\hertz} and formed an empirical propagation model that utilizes a local line base to address frequencies up to \SI{1}{\tera\hertz}. \\ \cline{2-4}
        & 2011 & \cite{Ref_jornet2011channel} & Revealed that oxygen molecules are the main cause of atmospheric absorption. Investigated channel capacity of the \ac{THz} band for different power allocation schemes, including a scheme based on the transmission of femtosecond-long pulses. \\ \cline{2-4}
        & 2016 & \cite{Ref_kokkoniemi2016discussion} & Studied the molecular absorption noise from different perspectives and gave their derivations and the general ideas behind the noise modeling in the higher frequency bands, such as \ac{THz} band (\SIrange{0.1}{10}{\tera\hertz}). \\ \cline{2-4}
        & 2019 & \cite{Bio_tekbiyik2019terahertz} &  Discussed open issues and the state-of-the-art solutions to these issues for \ac{THz} communication system design. Highlighted that propagation losses at \ac{THz} frequencies are more heavily affected by atmospheric absorption compared to that of the \ac{mmWave}. \\ \cline{2-4}
        & 2020 & \cite{Bio_9269931} & Showed that high atmospheric absorption can be alleviated by a proper choice of the carrier frequency, e.g., the band from \SIrange{275}{325}{\giga\hertz} while having drastically more bandwidth than \ac{5G}. \\ \cline{2-4}
        & 2021 & \cite{Bio_9391508} & A wireless backhaul network for a given ultra-dense mobile network is automatically planned and analyzed in the ThoR project. \\ \cline{2-4}
        & 2021 & \cite{Bio_9411010} & Developed an automatic planning algorithm for backhaul links operating at \SI{300}{\giga\hertz}, which is tested and evaluated in a realistic scenario of an ultra-dense network, taking into account various atmospheric effects. \\ \hline
        \multirow{18}{*}{\begin{tabular}[c]{@{}c@{}}Weather\\Effects\end{tabular}} & 2015 & \cite{Bio_7169630} & Measured \ac{THz} pulse propagation through a \SI{186}{\meter} distance under different weather conditions such as rain falling at \SI{3.5}{\milli\meter\per\hour^{}} and snow falling at \SI{2}{\centi\meter\per\hour^{}}, demonstrating the potential of \ac{LoS} THz communications, sensing, and imaging through fog and smoke. \\ \cline{2-4}
        & 2015 & \cite{Bio_THzcontrolledrain} & Measured the effects of rain attenuation on \SIrange{0.1}{1}{\tera\hertz} frequencies through THz time-domain spectroscopy and a rain chamber, which was designed to generate controllable and reproducible rain conditions. \\ \cline{2-4}
        & 2015 & \cite{Bio_6971163} & Experimentally demonstrated the propagation of \ac{THz} signals through \SI{137}{\meter} of dense fog with approximate visibility of \SI{7}{\meter}, and reported the observed \ac{THz} attenuation. \\ \cline{2-4}
        & 2017 & \cite{Bio_7944215} & Quantified the attenuation of \SI{150}{\giga\hertz} and \SI{300}{\giga\hertz} \ac{THz} waves in the sand for the outdoor scenario of low-THz sensing. \\ \cline{2-4}
        & 2018 & \cite{Bio_8557275} & Assessed the attenuation through various intensities of snowfall experimentally at \SI{300}{\giga\hertz}, which is characterized by measuring the ratio of the received power from the target through the snow precipitation and through the same path with no precipitation.\\ \cline{2-4}
        & 2019 & \cite{Bio_norouzian2019experimental} & Assessed the attenuation through various intensities of snowfall experimentally at low-\ac{THz} frequencies (\SIrange{100}{300}{\giga\hertz}), showing that snow attenuation at \SI{300}{\giga\hertz} is less than \SI{20}{\decibel\per\kilo\meter^{}} for snowfall rates below \SI{20}{\milli\meter\per\hour^{}}. \\ \cline{2-4}
        & 2019 & \cite{Bio_golovachev2019propagation} & Theoretically and experimentally studied the effects of water droplets suspended in the atmosphere on the propagation of \ac{mmWave} and \ac{THz} waves, using a frequency-modulated continuous-wave high-resolution radar operating at \SI{330}{\giga\hertz}.  \\ \cline{2-4}
        & 2019 & \cite{Bio_9017627} & The characteristics of rain attenuation have been investigated using raindrop size distribution collected in Indonesia. Reports indicating that the regional variation of rain attenuation should be considered. \\ \cline{2-4}
        & 2019 & \cite{Ref_juttula2019rain} & Evaluated the rain induced interference at \SI{300}{\giga\hertz}. Results are calculated also at \SI{60}{\giga\hertz} for comparison. Analyzed possible co-channel interference due to rain droplets, showing that the typical interference levels remain modest. \\ \cline{2-4}
        & 2020 & \cite{Bio_8826596} & Studied rain attenuation with different rainfall rates at \ac{mmWave} (\SI{77}{\giga\hertz}) and low-\ac{THz} (\SI{300}{\giga\hertz}) frequencies. Revealed that the measured results at \SI{77}{\giga\hertz} best agree with the ITU-R P838 model whereas the calculation based on Mie scattering and the Weibull distribution are best fit to the measured data at \SI{300}{\giga\hertz}. \\ \hline
        \multirow{16}{*}{\begin{tabular}[c]{@{}c@{}}Blockage\\Loss\end{tabular}} & 2014 & \cite{Ref_rappaport2014millimeter} & Developed models with variations in the \ac{LoS} probability expressions across different channel measurement campaigns. \\ \cline{2-4}
        & 2018 & \cite{Bio_8570041} & Presented a novel spatial dynamic channel sounding system based on phased array transmitters and receivers operating at \SI{60}{\giga\hertz}. Verified that the blockage duration is dependent on the density and speed of dynamic blockers that can last longer than \SI{100}{\milli\second}. \\ \cline{2-4}
        & 2018 & \cite{Bio_smallTHzpetrov} & Studied the impact of micro-mobility such as shakes and rotations of user equipment even when the user is in a stationary position. \\ \cline{2-4}
        & 2019 & \cite{Bio_8721173} & Proposed a novel spatially-consistent human body blockage state generation procedure, which extends the standardized 3D channel model by 3GPP to capture the correlation between the \ac{LoS} links and the reflected cluster states affected by human body blockage. Showed that a dynamic blockage causes an extra loss of around \SIrange{15}{40}{\decibel}. \\ \cline{2-4}
        & 2020 & \cite{Bio_9135643} & Employed a channel sounder that covers \SI{140}{\giga\hertz} - \SI{220}{\giga\hertz} and a frequency extender for THz channel measurement and built a platform for the exploration of THz communications within a short distance up to \SI{5}{\meter}. Pointed out that the reflection from rough surfaces like concrete or brick walls attenuates THz signals with a power drop ranging from \SIrange{40}{80}{\decibel}. \\ \cline{2-4}
        & 2021 & \cite{Bio_9403881} & Reported a comprehensive measurement campaign with the aim of analyzing the wave propagation at \SI{300}{\giga\hertz} in typical vehicular deployments. Showed that the blockage caused by vehicles leads to a loss of \SIrange{25}{60}{\decibel} over the frequency band of \SI{300}{\giga\hertz}. \\ \cline{2-4}
        & 2021 & \cite{Bio_stepanov2021accuracy} & Proposed several models characterized by various degrees of details to capture micromobility patterns of different applications. Assumed that the user behavior is more regulated as a micro-mobility. It shows that drift to the origin is a critical property that has to be captured by the model. \\ \cline{2-4}
        & 2022 & \cite{Bio_9601228} & Measured and statistically investigated the micromobility process of various applications including video viewing, phone calls, virtual reality viewing, and racing games. Suggested that micro-mobility follows a Markov pattern where user behavior is not controlled. \\ \hline \hline
    \end{tabular} 
 \end{table*}

\subsection{Atmospheric Absorption}
\begin{figure*}[!tbph]
\centering
\includegraphics[width=0.86\textwidth]{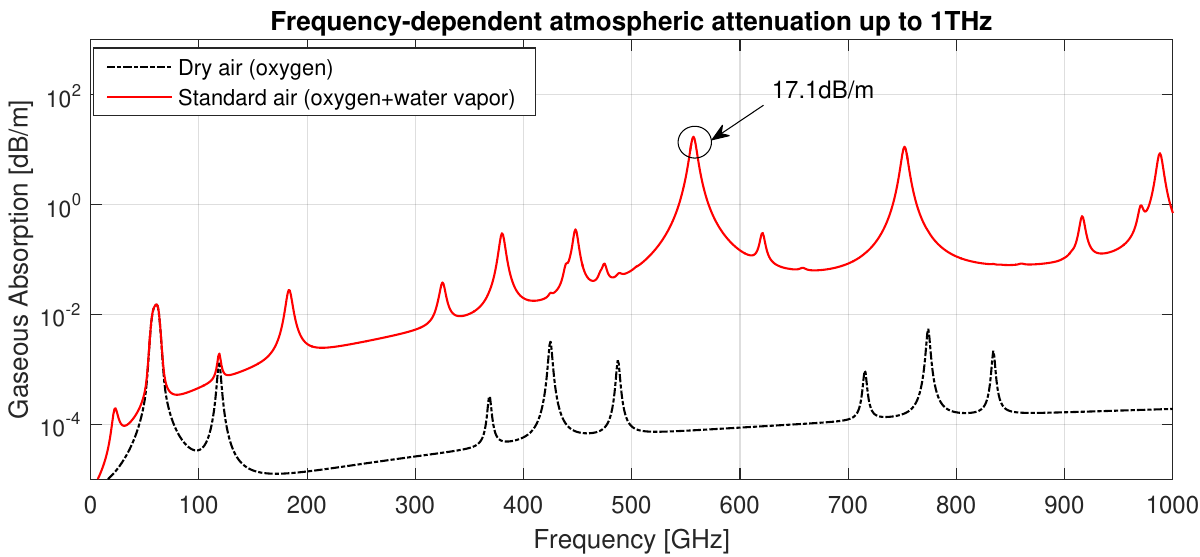}
\caption{Illustration of gaseous absorption from \SIrange{1}{1000}{\giga\hertz}, where the legend \textit{Standard Air} denotes a normal atmosphere condition with air pressure 1013.25hPa, temperature \SI{15}{\degreeCelsius}, and water-vapor density \SI{7.5}{\gram\per\meter^3}, according to \cite{Ref_WJ_itu2019attenuation}, while \textit{Dry Air} considers the effect of oxygen absorption only with a water-vapor density of \SI{0}{\gram\per\meter^3}. Except for frequencies centered at \SI{60}{\giga\hertz} and \SI{118.7}{\giga\hertz}, the effect of water vapor dominates.} 
\label{Diagram_atmosphericAbsorption}
\end{figure*}

Although gaseous molecules absorb some energy of an \ac{EM} wave, atmospheric absorption is negligible over the sub-6G band such that traditional cellular systems do not take it into account when calculating the link budget. However, this effect substantially magnifies for the \ac{THz} wave, and the absorption loss becomes extremely large at certain frequencies. 
Such attenuation arises from the interaction of an \ac{EM} wave with a gaseous molecule \cite{Ref_weide2000gasabsorption}. When the \ac{THz} wavelength approaches the size of molecules in the atmosphere, the incident wave causes rotational and vibrational transitions in polar molecules. These processes have a quantum nature where resonances take place at particular frequencies depending on the internal molecular structure, leading to large absorption peaks at certain frequencies \cite{Ref_slocum2013atmospheric}. 

As a main gaseous component of the atmosphere, oxygen plays a major role in atmospheric absorption under clear air conditions. In addition, water vapor suspended in the air strongly affects the propagation of an electromagnetic wave. The attenuation caused by water vapor dominates the \ac{THz} band, with the exception of a few specific spectral regions where the effect of oxygen is more evident. A more extensive study of atmospheric absorption is often carried out in radio astronomy and remote sensing. However, from the perspective of wireless communications, the absorption of some additional molecular species, e.g., oxygen isotopic species, oxygen vibrationally excited species, stratospheric ozone, ozone isotopic species, ozone vibrationally excited species, a variety of nitrogen, carbon, and sulfur oxides, is usually negligible compared with that of water vapor and oxygen  \cite{Ref_siles2015atmospheric}.

Responding to the need to accurately estimate the gaseous absorption at any air pressure, temperature, and humidity, the ITU-R conducted a study item and recommended a mathematical procedure to model these attenuation characteristics. As a combination of the individual spectral lines from oxygen and water vapor, along with small additional factors for the non-resonant Debye spectrum of oxygen below \SI{10}{\giga\hertz}, pressure-induced nitrogen absorption over \SI{100}{\giga\hertz}, and a wet continuum to account for the excess absorption from water vapor, the ITU-R P676 model \cite{Ref_WJ_itu2019attenuation} has been built to generate the values of atmospheric attenuation at any frequency from \SIrange{1}{1000}{\giga\hertz}. Alternatively, the high-resolution transmission molecular absorption (HITRAN) database \cite{Ref_gordon2021HITRAN}, which is a compilation of spectroscopic parameters, can be used to predict and analyze the transmission in the atmosphere. To compute the atmospheric absorption at \ac{THz}, it needs to extract spectroscopic data from the HITRAN database and then apply radiative transfer theory.

The atmospheric attenuation from \SIrange{1}{1000}{\giga\hertz} is illustrated in Fig.~\ref{Diagram_atmosphericAbsorption}. Assume the air is perfectly dry with a water vapor density of \SI{0}{\gram\per\meter^3}, then only the effect of oxygen molecules exists, as indicated by the \textit{Dry Air} curve in the figure. 
On the other hand, the \textit{Standard Air} line shows the usual atmospheric condition at the sea level (with an air pressure of 1013.25~hPa, the temperature of \SI{15}{\degreeCelsius}, and a water-vapor density of \SI{7.5}{\gram\per\meter^3}). Except for two frequency windows centered around \SI{60}{\giga\hertz} and \SI{118.7}{\giga\hertz}, where many oxygen absorption lines merged, the attenuation due to water vapor dominates the \ac{THz} band. As we can see, this absorption can bring a peak loss of \SI{17.1}{\decibel\per\meter^{}} at \SI{560}{\giga\hertz}, accounting for an extreme level of approximately \SI{17000}{\decibel\per\kilo\meter^{}}, which is prohibitive for wireless communications. In contrast, the attenuation at the sub-\SI{6}{\giga\hertz} band is less than  \SI{0.0001}{\decibel\per\meter^{}}, seven orders of magnitude smaller. 

Early in the 1980s, some researchers reported their studies for various atmospheric conditions and elevation angles. Ernest K. Smith calculated the absorption values due to atmospheric oxygen and water vapor for frequencies spanning from \SIrange{1}{340}{\giga\hertz} \cite{Bio_7768108}. Hans J. Liebe and Donald H. Layton \cite{Bio_Liebe1987Mill} performed laboratory measurement of water vapor attenuation at \SI{138}{\giga\hertz} and formed an empirical propagation model that utilizes a local line base to address frequencies up to \SI{1}{\tera\hertz}. For some studies focusing on very short-range indoor coverage or nano-communications \cite{Ref_Kokkoniemi2015frequency}, this effect is typically not a major factor. However, for macro-scale \ac{THz} communications and sensing, especially in outdoor environments, atmospheric absorption should be taken into consideration. We provide a summary of state-of-the-art studies in Table \ref{tableChannelProp}.
 
\subsection{Weather Effects}

Besides the gaseous absorption, an additional atmospheric impact in an outdoor environment is the weather \cite{Bio_reviewthz}. Extensive studies focused on satellite communications channels since the 1970s provided many insights into the propagation characteristics of \ac{mmWave} and \ac{THz} signals under various weather conditions \cite{Ref_crane1980prediction}. Like water vapor in the atmosphere, the outcomes revealed that liquid water droplets, in the form of suspended particles in clouds, fogs, snowflakes, or rain falling hydrometeors, absorb or scatter the incident signals since their physical dimensions are in the same order as the \ac{THz} wavelength. Such attenuation is not as strong as the path loss and atmospheric absorption but still needs to be taken into account for proper channel characterization \cite{Ref_weng2021millimeter}.

A cloud is an aggregate of tiny water particles (with a dimension as small as \SI{1}{\micro\meter}) or ice crystals (from \SIrange{0.1}{1}{\milli\meter}). Water droplets, in the form of raindrops, fogs, hailstones, and snowflakes, are oblate spheroids with radii up to a few tens of millimeters or generally perfect spheres with radii below \SI{1}{\milli\meter}. The size of water droplets is comparable to the \ac{THz} wavelength (\SIrange{0.1}{1}{\milli\meter}). As a result, water droplets attenuate the power of \ac{THz} waves through absorption and scattering. 
The ITR-R provided a power-low equation to model the rain attenuation as a function of distance, rainfall rate in millimeters per hour (\si{\milli\meter\per\hour^{}}), and the mean dimension of raindrops \cite{Ref_itu2005specific}. \figurename~\ref{Diagram_RainAttenu} shows the rain attenuation described by the ITU-R P838 model from \SIrange{1}{1000}{\giga\hertz} and rain rate from light rain (\SI{1}{\milli\meter\per\hour^{}}) to heavy rain (\SI{200}{\milli\meter\per\hour^{}}). 

Such attenuation can be treated as an additional loss that is simply added on top of the path loss and gaseous absorption.  Besides the ITU-R model, there are other models, such as a simplified one given in \cite{Ref_smulders1997characterisation} to describe the rain attenuation. The measurement at \SI{28}{\giga\hertz} demonstrated that heavy rainfall with a rain rate of more than \SI{25}{\milli\meter\per\hour^{}} brings an attenuation of about \SI{7}{\decibel\per\kilo\meter^{}}. Extreme attenuation of up to \SI{50}{\decibel\per\kilo\meter^{}} occurs at a particular frequency of \SI{120}{\giga\hertz} and an extreme rain rate of \SIrange{100}{150}{\milli\meter\per\hour^{}}. As a rule of thumb,  rain provides an excess attenuation of approximately \SIrange{10}{20}{\decibel} over a distance of \SI{1}{\kilo\meter} at the \ac{THz} band. Furthermore, the attenuation of cloud and fog can be calculated by the ITU-R P840 model \cite{Ref_itu2019attenuation} under the assumption that the signals go through a uniform fog or cloud environment. 

\begin{figure}
\centering
\includegraphics[width=0.48\textwidth]{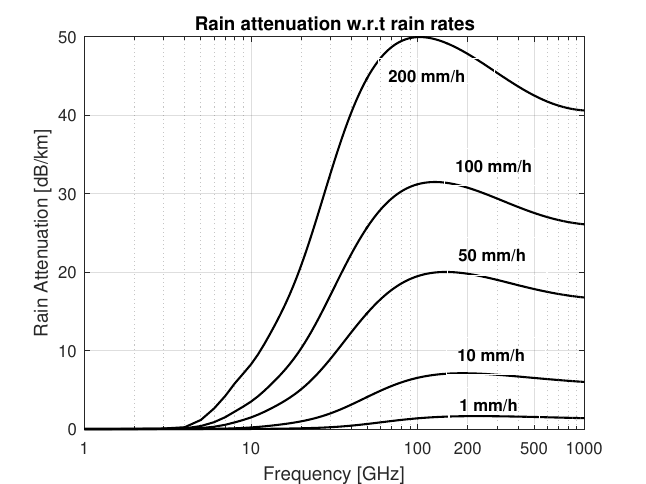}
\caption{Rain attenuation measured in \si{\decibel\per\kilo\meter^{}} for terrestrial communications links as a function of the rain rate and frequency, covering the range from \SIrange{1}{1000}{\giga\hertz}. The rain rate is measured in millimeters per hour (\si{\milli\meter\per\hour^{}}), averaging over a period of time such as one hour. The values from light rain (\SI{1}{\milli\meter\per\hour^{}}) to heavy rain (\SI{200}{\milli\meter\per\hour^{}}) are illustrated. The peak of attenuation occurs on the frequency band from \SIrange{100}{300}{\giga\hertz} since the wavelength in this band matches the size of raindrops.} 
\label{Diagram_RainAttenu}
\end{figure}

In addition to \ac{mmWave} signals \cite{Ref_weng2021millimeter}, many recent efforts of investigating \ac{THz} waves in the presence of rains and fogs have been conducted, as listed in Table \ref{tableChannelProp}. Besides, the propagation loss during snowfall has been studied through measurement campaigns. F. Norouzian \emph{et al.} assessed the attenuation through various intensities of snowfall experimentally at \SI{300}{\giga\hertz} in their work \cite{Bio_8557275}, as well as sub-\ac{THz} frequencies (\SIrange{100}{300}{\giga\hertz})  in \cite{Bio_norouzian2019experimental}. Unlike other weather conditions, there is no theoretical basis for snow because of the challenging nature of defining snowflake shape and size distributions. An equivalent ITU-R model to calculate the snow attenuation does not exist and the snow particles are merely considered as raindrops. As a rule of thumb, measurement results indicate that snow attenuation at \SI{300}{\giga\hertz} is less than \SI{20}{\decibel\per\kilo\meter^{}} for snowfall rates below \SI{20}{\milli\meter\per\hour^{}} \cite{Bio_norouzian2019experimental}. Last but not least, much work has been done to investigate one of the most common types of contaminant in outdoor environments, i.e., dust or sand. The authors of \cite{Bio_7944215} quantified the attenuation at \SI{150}{\giga\hertz} and \SI{300}{\giga\hertz} in the presence of sand for outdoor THz sensing.

\subsection{Blockage Loss}
Due to the small wavelength of \ac{THz} waves, the dimensions of surrounding physical objects are sufficiently large for scattering, while specular reflections on ordinary surfaces become difficult. On the other hand, \ac{THz} systems rely heavily on pencil beams to extend the effective propagation distance. As a result, a direct path between the transmitter and receiver is desired. However, the \ac{LoS} \ac{THz} link is highly susceptible to being blocked by macro objects, such as buildings, furniture, vehicles, and foliage, and micro-objects, e.g., humans, in comparison with traditional sub-6G signals \cite{Ref_rappaport2014millimeter}. 

A single blockage might cause a signal loss of a few tens of dB. The extent of foliage loss is related to the depth of vegetation \cite{Ref_smulders1997characterisation}, where \SI{17}{\decibel}, \SI{22}{\decibel}, and \SI{25}{\decibel} are observed at \SI{28}{\giga\hertz}, \SI{60}{\giga\hertz}, and \SI{90}{\giga\hertz}, respectively. The blockage loss due to vehicles \cite{Bio_smallTHzpetrov} is determined by the vehicle type and geometry, from \SI{20}{\decibel} in the front-shield glass to \SI{50}{\decibel} in the engine area.  The human body blockage imposes a more profound influence because of the dynamic movement of humans and the close interaction of \ac{THz} devices with humans. The loss attributed to the self-body blockage \cite{Bio_8570041} is expected to reach approximately \SI{40}{\decibel} in the \ac{THz} band. Such blockage losses can dramatically decay the signal power and may even lead to a thorough outage. Hence, it is necessary to clarify the traits of blockage and find effective solutions to avoid being blocked or quickly recover the connection when a link gets blocked. 

Statistical models can be applied to estimate the impact of blockage loss. For instance, the self-body blockage loss may be approximated as a Boolean model where a human is treated as a three-dimensional cylinder with centers forming a \ac{2D} \ac{PPP}. A \ac{LoS} blockage probability model assumes that a link of distance $d$ will be \ac{LoS} with probability $p_{L}(d)$ and \ac{NLoS} otherwise. The expressions of $p_{L}(d)$ are usually obtained empirically for different settings. The blockage probability for a \ac{LoS} link with a self-body blockage can be estimated by the method given in \cite{Ref_tripathi2021millimeter}. In \cite{Ref_3gpp2018study}, 3GPP specified an urban macro-cell scenario, where a calculation method for a \ac{LoS} blockage probability is given. The same model applies to the urban micro-cell scenario, with a smaller distance range. There are some variations in the expressions of the \ac{LoS} probability  across different channel measurement campaigns and environments, e.g., the model developed by NYU \cite{Ref_rappaport2014millimeter}. Some other prior works also revealed quantitative results to estimate the blockage loss at \ac{THz} frequencies, as given in Table \ref{tableChannelProp}.

\section{\ac{THz} Channel Measurement and Modeling for \ac{6G} and Beyond}

Novel \ac{6G} usage scenarios such as kiosk downloading, nano-communications, wireless backhauling, and integrated communications and sensing pose many peculiarities in transmission distances,  hardware capabilities, and propagation environments \cite{Ref_han2022terahertz}. 
These unique propagation characteristics and particular requirements motivate the research community to perform further research on wireless channels. Extensive measurement campaigns and channel modeling efforts are expected for the success of deploying \ac{6G} and beyond. On the other hand, many challenging issues such as appropriate measurement equipment and novel modeling methodologies are barriers ahead \cite{Ref_jornet2011channel}. This section summarizes the state of the art in channel measurement campaigns and presents up-to-date channel models of the \ac{THz} band for \ac{6G} and beyond use cases.  

\subsection{\ac{THz} Channel measurement}
Profound knowledge of propagation characteristics and proper channel models are prerequisites for designing transmission algorithms,  developing network protocols, and evaluating the performance of \ac{THz} communications and sensing. Channel measurement are the most appropriate methods for a full understanding of \ac{THz} signal propagation and the subsequent modeling. Due to the unique propagation characteristics, innovative measurement equipment is required. There are two major kinds of measurement devices for \ac{THz} channels, i.e. the \acf{VNA} \cite{Ref_peng2020channel} and the \ac{CS} \cite{Ref_rey2017channel}. 

Both devices acquire channel state information by transmitting a reference signal and processing the corresponding received signal at the receiver. However, their implementations are distinct. The \ac{VNA} operates in the \textit{frequency domain} \cite{Ref_peng2020channel}, which measures the \ac{CTF} for a specific narrowband channel at each time and sequentially scans all frequency points within the band of interest. The \ac{CTF} of each narrowband channel is modeled by a scalar, and the wideband \ac{CTF} is obtained by aggregating a large number of narrowband CTFs. The channel impulse response (CIR) is the inverse discrete Fourier transform (IDFT) of the wideband \ac{CTF}. This method inherits the advantages of narrowband channel measurement, such as high precision due to individual calibrations for each frequency point, and low measurement noise. But it takes a long time and cannot capture dynamic channel effects. 

The channel-sounding approach in \cite{Ref_rey2017channel} operates in the \textit{time domain}, taking advantage of the classical technique called \ac{DSSS}. The transmitter of a \ac{CS}  sends a maximal-length sequence (m-sequence) as a stimulus signal so as to approximate a Dirac-impulse-shaped auto-correlation function. Since a received signal is the convolution of a transmitted signal and a time-varying channel,  cross-correlating the received signal with a delayed version of the m-sequence yields the CIR of a measured wideband channel. Time-domain \ac{CS} works much faster than a \ac{VNA} with frequency scanning and can capture channel dynamic variations. However, its precision is disturbed by strong thermal noise, which is proportional to the signal bandwidth.

When deciding on a measurement technique for \ac{THz} channels, several factors such as the signal bandwidth, speed, distance, power consumption, cost, and complexity of the measurement system need to be taken into consideration.  According to the aforementioned two basic methods, three techniques} are employed for measuring THz channels \cite{Ref_han2022terahertz}, i.e., \textit{frequency-domain measurement using \ac{VNA}, time-domain measurement using sliding correlation, and time-domain spectroscopy}. We comparatively introduce these techniques in the following, while an overview of recent measurement campaigns is offered in Table \ref{table_THzmeasure}. 

\begin{table*}[!t]
    \centering 
    \caption{Summary of THz channel measurement campaigns}
    \label{table_THzmeasure}
    \scriptsize 
    \begin{tabular}{|c|c|c|p{12cm}|}
        \hline \hline
        \textbf{Techniques} & \textbf{Year} & \textbf{Reference} & \textbf{Main Features and Contributions} \\ \hline \hline
        \multirow{30}{*}{\begin{tabular}[c]{@{}c@{}}Frequency-Domain\\\ac{VNA} Measurement\end{tabular}} & 2015 & \cite{Bio_6898846} & Developed a measurement system for the frequency band of \SI{280}{}-\SI{320}{\giga\hertz} using an N5224A PNA \ac{VNA} and VDI transceivers (Tx210/Rx148), which was utilized to measure short-range scenarios in a desktop environment up to \SI{0.7}{\meter}. \\ \cline{2-4}
        & 2016 & \cite{Bio_7636947} & Utilized the system in \cite{Bio_6898846} to measure short-range scenarios in computer motherboards at frequencies of \SI{300}{}-\SI{312}{\giga\hertz}. Found that a few centimeters of vertical misalignment between the transmitter and the receiver result in a path loss greater than \SI{5}{\decibel}. \\ \cline{2-4}
        & 2016 & \cite{Bio_7511280} & Measured THz \ac{LoS} scenarios with the Tx-Rx distance ranging from \SI{0.01}{\meter} to \SI{0.95}{\meter} over the frequency band of \SI{260}{}-\SI{400}{\giga\hertz}, which was conducted using a VNA in conjunction with a sub-harmonic mixer. \\ \cline{2-4}
        & 2019 & \cite{Bio_khalid2019statistical} & Performed measurement campaigns for \ac{NLoS} scenarios with the presence of a reflective surface, in addition to \ac{LoS}. \\ \cline{2-4}
        & 2020 & \cite{Bio_8889517} & Utilized their above-developed system to measure short-range scenarios in a data center with a propagation distance of \SI{0.4}{}-\SI{2.1}{\meter} at \SI{300}{}-\SI{320}{\giga\hertz}. \\ \cline{2-4}
        & 2020 & \cite{Bio_9039668} & Evaluated the \ac{THz} wave propagation in a realistic data-center environment. The measurement were taken at a Tx-Rx distance of \SI{1.75}{\meter} and \SI{2.28}{\meter}, within the frequency range of \SI{300}{}-\SI{312}{\giga\hertz}. \\ \cline{2-4}
        & 2020 & \cite{Bio_9148631} & Extended the separation between the transmitter and receiver to a distance of \SI{100}{\meter} with the use of \ac{RFoF} technique. \\ \cline{2-4}
        & 2020 & \cite{Bio_9205476} & Established a channel measurement system covering the frequency range of \SI{500}{}-\SI{750}{\giga\hertz}, which utilized the Keysight PNA-X \ac{VNA}, which was configured with VDI extender heads. Investigated the effect of linear and angular displacement between the transmitter and receiver. \\ \cline{2-4}
        & 2020 & \cite{Bio_9145314} & Investigated a wideband channel measurement between\SI{130}{\giga\hertz} and \SI{143}{\giga\hertz} in a typical meeting room for future \ac{THz} wireless communication access networks. Utilized directional antennas for resolving the multipath components in the angular domain.  Analyzed the line-of-sight path loss, power-delay-angular profiles, temporal and spatial features, and correlations among \ac{THz} multipath characteristics. \\ \cline{2-4}
        & 2020 & \cite{Bio_9135643} & Made advancements to the existing \SI{140}{\giga\hertz} VNA-based measurement equipment, enabling it to support frequencies ranging from \SI{140}{\giga\hertz} to \SI{220}{\giga\hertz}. Conducted indoor \ac{LoS} measurement in an office setting, with the measurement distance ranging from \SIrange{0.5}{5.5}{\meter}.  \\ \cline{2-4}
        & 2021 & \cite{Bio_9473756} & Analyzed the large-scale fading characteristics of \ac{THz} signals in indoor scenarios and proposed two multi-band path loss models. Investigated coherent and non-coherent beam combination methods for reducing path loss.  \\ \cline{2-4}
        & 2021 & \cite{Bio_9500596} &  Discussed and compared the propagation of \ac{NLoS} multipath components in the office room scenario and its impact on the THz channel with the channel measurement in the meeting at \SI{140}{\giga\hertz}.  \\ \cline{2-4}
        & 2021 & \cite{Bio_9466322} & Investigated a wideband channel measurement campaign between\SI{130}{\giga\hertz} and \SI{143}{\giga\hertz} in a typical meeting room, proposed a combined MPC clustering and matching procedure with ray-tracing techniques to investigate the cluster behavior and wave propagation of \ac{THz} signals. \\ \cline{2-4}
        & 2022 & \cite{Bio_9838312} & Built a VNA-based measurement system covering the \ac{THz} band from \SIrange{260}{400}{\giga\hertz}. Indoor channel measurement in the frequency range of \SI{306}{}-\SI{321}{\giga\hertz} were performed in an L-shaped hallway and a long corridor at the campus, with distances ranging from \SI{7.7}{}-\SI{25}{\meter} and \SI{5}{}-\SI{31}{\meter}, respectively. \\ \hline
        \multirow{23}{*}{\begin{tabular}[c]{@{}c@{}}Time-Domain\\Sliding Correlation\\Measurement\end{tabular}} & 2007 & \cite{Bio_4455844} & Considered channel measurement, simulation, and antenna design for \ac{THz} frequencies up to \SI{300}{\giga\hertz}. \\ \cline{2-4}
        & 2017 & \cite{Bio_7887768} & Developed a measurement system that can switch between two modes: sliding correlation and real-time spread spectrum. \\ \cline{2-4}
        & 2018 & \cite{Bio_8647921} & Summarized wireless communication research and activities above \SI{100}{\giga\hertz}, provided the design of a \SI{140}{\giga\hertz} wideband channel sounder system, and proposed indoor wideband propagation measurement and penetration measurement for common materials at \SI{140}{\giga\hertz}. \\ \cline{2-4}
        & 2019 & \cite{Bio_8761205} & Provided an analysis of radio wave scattering for frequencies ranging from the microwave to the Terahertz band (e.g., \SI{1}{\giga\hertz}–\SI{1}{\tera\hertz}). Focused on the study of how the \ac{THz} waves at \SI{140}{\giga\hertz} reflect and scatter. \\ \cline{2-4}
        & 2019 & \cite{Bio_8739519} & Conducted channel measurement and modeling for a variety of specific scenarios such as train-to-train (T2T) and infrastructure-to-infrastructure (I2I). \\ \cline{2-4}
        & 2019 & \cite{Bio_8684885} &  Measured, simulated, and characterized the train-to-infrastructure (T2I) inside-station channel at the \ac{THz} band for the first time. Provided the foundation for future work that aims to add the T2I inside-station scenario into the standard channel model families. \\ \cline{2-4}
        & 2019 & \cite{Bio_8740079} & Carried out measurement in an actual data center at \SI{300}{\giga\hertz} with a channel sounder, dividing the environment into inter-rack and intra-rack components. \\ \cline{2-4}
        & 2020 & \cite{Bio_9295992} & Conducted channel measurement and modeling for an intra-wagon scenario, covering the frequency range from \SI{60}{\giga\hertz} to \SI{300}{\giga\hertz}. \\ \cline{2-4}
        & 2020 & \cite{Bio_9135389} &  Investigated reflection and penetration losses of \ac{THz} frequencies at \SI{300}{\giga\hertz} band in vehicular communications. Revealed that the vehicle body is extremely heterogeneous in terms of the propagation losses. \\ \cline{2-4}
        & 2021 & \cite{Bio_9403881} &  A comprehensive measurement campaign is reported with the aim of analyzing the wave propagation at \SI{300}{\giga\hertz} in typical vehicular deployments.\\ \cline{2-4}
        & 2021 & \cite{Bio_9397335} & Utilized a channel sounder at \SI{300}{\giga\hertz} that uses the sliding correlation method with m-sequences of order 12, to observe the propagation of \ac{mmWave} and \ac{THz} waves in railway scenarios. \\ \cline{2-4}
        & 2021 & \cite{Bio_9411894} & Developed a corresponding indoor channel simulator, which can recreate 3-D omnidirectional, directional, and m\ac{MIMO} channels for arbitrary carrier frequency up to \SI{150}{\giga\hertz}. \\ \cline{2-4}
        & 2022 & \cite{Bio_9685929} &  Investigated the \ac{UMi} large-scale path loss at 28, 38, 73, and \SI{142}{\giga\hertz}. Introduced a detailed spatial statistical \ac{MIMO} channel generation procedure based on the derived empirical channel statistics. \\ \hline
        \multirow{12}{*}{\begin{tabular}[c]{@{}c@{}}Time-Domain\\Spectroscopy\\ Measurement\end{tabular}} & 2007 & \cite{Bio_4380579} &  Proposed a new approach for modeling the reflective properties of building materials in \ac{THz} communication systems, which uses modified Fresnel equations to account for losses due to diffuse scattering in materials with rough surfaces. \\ \cline{2-4}
        & 2008 & \cite{Bio_4512142} & Presented reflection \ac{THz}-TDS measurement and matching transfer matrix simulations of the frequency-dependent reflection coefficient of multi-layer building materials in the frequency range from \SI{100}{\giga\hertz} to \SI{500}{\giga\hertz} for a set of angles, both in TE- and TM-polarization. \\ \cline{2-4}
        & 2011 & \cite{Bio_5873179} & Investigated the influence of diffuse scattering at \SI{300}{\giga\hertz} on the characteristics of the communication channel and its implications on the \ac{NLoS} propagation path.  \\ \cline{2-4}
        & 2015 & \cite{Bio_THzcontrolledrain} & Measured the attenuation of \ac{THz} signals due to weather conditions with the \ac{THz}-TDS equipment, conducted theoretical analysis, and summarized the impact of various weather factors on \ac{THz} communications links. \\ \cline{2-4}
        & 2017 & \cite{Bio_7829286} &  Reviewed VNA and TDS measurement techniques and discussed the different issues involved in making measurement using these systems. Briefly discussed the operating principles of electro-optic sampling (EOS). \\ \cline{2-4}
        & 2019 & \cite{Bio_8736035} & Used the \ac{THz}-TDS to assess the interference between \ac{THz} devices in the \SI{300}{\giga\hertz} frequency band and applied stochastic geometric techniques to model and analyze the interference. \\ \hline \hline
    \end{tabular} 
\end{table*}

\subsubsection{Frequency-Domain VNA Measurement}
A \ac{VNA} is a measuring device utilized to assess the response of a component or network at one port, in response to an incoming wave at another port. The frequency-domain channel measurement performed using \ac{VNA} are based on the principle of linear systems. It is important to note that commercial VNAs typically have a limitation on the frequency range that is less than \SI{67}{\giga\hertz}, requiring the use of up-conversion modules for measurement in the \ac{THz} band. Most of the VNA-based measurement campaigns were focused on frequencies ranging from \SI{140}{\giga\hertz} to \SI{750}{\giga\hertz} and utilized directional horn antennas with antenna gains between \SIrange{15}{26}{\dB i}. The separation between the transmitter and receiver (i.e. the Tx-Rx distance) varied from a minimum of \SI{0.1}{\meter} to a maximum of \SI{14}{\meter}, and in some cases, was extended to a distance of \SI{100}{\meter} with the use of the \ac{RFoF} technique \cite{Bio_9148631}. 

\subsubsection{Time-Domain Sliding Correlation Measurement}
Some studies have been conducted to measure the characteristics of \ac{THz} waves using the \ac{SC} method over the frequency range up to \SI{300}{\giga\hertz}. A team from NYU developed a measurement system that can switch between two modes: sliding correlation and real-time spread spectrum \cite{Bio_7887768}. Using the \ac{SC} mode, they focused on the study of how the \ac{THz} waves at \SI{140}{\giga\hertz} reflect and scatter  \cite{Bio_8761205}. They measured \ac{THz} channels in indoor scenarios including ofﬁces, conference rooms, classrooms, long hallways, open-plan cubicles, elevators, and factory buildings, some results are reported in the literature such as \cite{Bio_8647921, Bio_9013236, Bio_9411894, Bio_9838910}. Besides, this team also conducted an outdoor wideband measurement campaign in an urban microcell environment \cite{Bio_9500482,Bio_9500677,Bio_9685929}. Further examples based on this measurement method can be found in Table \ref{table_THzmeasure}.

\subsubsection{Time-Domain Spectroscopy Measurement}
\Ac{TDS} is the most straightforward method for measuring impulse responses. It involves transmitting a train of extremely narrow pulses, where the period of the pulse train is greater than the maximum excess delay of the channel. The amplitude of a sampling instance can be considered as the amplitude of the CIR at the time of the exciting pulse, at a delay equal to the difference between the pulse transmission and the sampling time \cite{Ref_han2022terahertz}. By sampling the received signal at a high speed in the time domain, the impulse response can be directly calculated. \ac{THz}-TDS makes use of an extensive and scalable bandwidth in the \ac{THz} frequency band. However, the large setup size and low power output limit its application scenarios \cite{Bio_reviewthz}. To mitigate these aspects, lens antennas are often used at both the transmitter and receiver to enhance the intensity of the pulse signal. The lens beam is highly concentrated, making it well-suited for measuring material properties such as reflection, scattering, and diffraction in the \ac{THz} frequency range.  TDS for \ac{THz} is primarily utilized for channel measurement over short distances, typically less than a few meters.

\subsection{\ac{THz} Channel Modeling}

Developing a wireless communication system requires an accurate channel model that fully captures the major propagation characteristics of the operating carrier frequency. It allows wireless researchers and engineers to assess the performance of different transmission algorithms and medium-control protocols without having to conduct expensive and time-consuming real-world field measurements on their own. A large number of channel models, focusing on the sub-\SI{6}{\giga\hertz} frequency band for traditional cellular systems \cite{Ref_3gpp2018study}, have been built through curve fitting or analytical analysis based on field measurement data. These models reflect all propagation effects, both known and unknown, and therefore work well. Given the peculiarities of \ac{THz} signal propagation, it is necessary to develop appropriate \ac{THz} channel models for research, development, performance evaluation, and standardization \cite{Ref_yong2007IEEE80215} of \ac{THz} communications and sensing in \ac{6G} and beyond. 

Two widely used techniques for developing appropriate channel models are deterministic \cite{Ref_han2018propagation} and stochastic \cite{Ref_saleh1987statistical} methods. The former methods utilize the electromagnetic laws of wave propagation to determine the received signal strength at a particular location. The most popular deterministic modeling approach is known as ray tracing \cite{Ref_glassner1989raytracing}. The parameters of each ray, such as the attenuation, angle of departure, angle of arrival, propagation delay, and Doppler shift, can be computed taking into account the geometric optic rules of propagation including the computation of path losses via the Friis transmission equation, the Fresnel equation for reflections, Khirchoff's scattering theory, and the universal theory of diffraction. Ray tracing is highly applicable for various static \ac{6G} applications at the \ac{THz} band, e.g., indoor hot spots, wireless backhaul, and nano-networks. 

However, the ray-tracing approach suffers from high computational complexity and long simulation time. Also, accurate information about the geometric environment, the exact knowledge of the boundary conditions, and the properties of different objects are required \cite{Ref_glassner1989raytracing}.  To alleviate the complexity of channel modeling, stochastic methods are applied to provide a statistical description of the propagation channel. These models are derived from empirical data and need much less computational complexity in comparison with the deterministic ones \cite{Bio_7913686}. By using stochastic models, channel data can be generated easily without profound channel knowledge, allowing researchers and engineers to focus on their design and simulation works.

\begin{table*}[!t]
\renewcommand{\arraystretch}{1.3}
\caption{Summary of \ac{THz} Channel Modelling for \ac{6G} and Beyond}
\label{table_THzChannelModel}
\centering
\scriptsize
\begin{tabular}{|c|c|c|c|p{10.5cm}|}
\hline \hline
\textbf{}               & \textbf{Methods}                                                                             & \textbf{Year} & \textbf{References} & \textbf{Contributions} \\ \hline \hline
\multirow{18}{*}{ \rotatebox[origin=c]{90}{\parbox[c][][c]{20mm}{\raggedright \textbf{Deterministic }}}}  & \multirow{11}{*}{Ray Tracing}                                                                 & 2012              & Saadane \textit{et al.}~\cite{Bio_SaadaneVisiTree}                    & Presents and validates a ray-tracing method for \ac{UWB} indoor propagation channels.                       \\ \cline{3-5} 
                                &                                                                                              & 2018              & Sheikh \textit{et al.}~\cite{Bio_8454694}                    & Presents a novel \ac{3D} ray-tracing algorithm based on the Beckmann-Kirchhoff model to model diffuse scattering mechanisms in non-specular directions at terahertz frequencies                       \\ \cline{3-5}
                                &                                                                                              & 2018              & Virk \textit{et al.}~\cite{Bio_8345613}                   & Presents a new method of on-site permittivity estimation for accurate site-specific radio propagation simulations, which is important for cellular coverage analysis.                       \\ \cline{3-5} 
                                &                                                                                              & 2019              & Guan \textit{et al.}~\cite{Bio_8684885}                    & Presents measurement, simulation, and characterization of the train-to-infrastructure inside-station channel at the Terahertz band for the first time.                        \\ \cline{3-5}
                                &                                                                                              & 2019              & Gougeon \textit{et al.}~\cite{Bio_8880819}                   & Describes the BRAVE project, which aims to explore the use of higher frequencies in the sub-THz domain for future wireless communications systems. Discusses the challenges of channel modeling at these frequencies and presents extensions made to a ray-based deterministic tool to address these challenges.                       \\ \cline{3-5} 
                                &                                                                                              & 2020              & Sheikh \textit{et al.}~\cite{Bio_8859609}                   & Explores the use of massive MIMO systems at terahertz frequencies. Develops a 3D ray-tracing modeling approach to investigate the impact of surface roughness on THz channel capacity, and calculate channel capacities for both line-of-sight and non-line-of-sight scenarios with different surface roughnesses.                         \\ \cline{2-5} 
                                & \multirow{5}{*}{\begin{tabular}[c]{@{}c@{}}FDTD\end{tabular}}      & 2000              & Wang \textit{et al.}~\cite{Bio_855493}                   &  Develops a technique that combines the ray-tracing and FDTD methods for site-specific models of indoor radio wave propagation.                        \\ \cline{3-5}
                                &                                                                                              & 2007              & Zhao \textit{et al.}~\cite{Bio_4057499}                   &  Describes an FDTD-based method for modeling the radio channel in an UWB indoor environment.                        \\ \cline{3-5}
                                &                                                                                              & 2014              & Fricke \textit{et al.}~\cite{Bio_6902134}                   & Analyzes the propagation effects in electromagnetic wave propagation modeling for intra-device communications. Presents a new approach based on FDTD calculations.                        \\ \cline{2-5} 
                                & \multirow{3}{*}{\begin{tabular}[c]{@{}c@{}}Channel\\Measurement\end{tabular}}                        & 2022              &  Rappaport \textit{et al.}~\cite{Bio_rappaport2022radio}                  & Offers comprehensive, practical guidance on RF propagation channel characterization at mmWave and sub-terahertz frequencies, with an overview of both measurement systems and current and future channel models.                       \\ \hline
\multirow{20}{*}{\rotatebox[origin=c]{90}{\parbox[c][][c]{20mm}{\raggedright \textbf{Statistical}}} }   & \multirow{15}{*}{Physical}  & 1999              & Kunisch \textit{et al.}~\cite{Bio_797367}                   & Describes a study where radio waves at \SI{60}{\giga\hertz} carrier frequency with a bandwidth of \SI{960}{\mega\hertz} were measured in an indoor environment. Using the data collected, the authors determine parameters for a multipath model based on a well-known statistical indoor channel model by Saleh \textit{et al.} (1987).                       \\ \cline{3-5}
                                &                                                                                              & 2002              & Zwick \textit{et al.}~\cite{Bio_1021910}                    &  Introduces a new stochastic channel model for indoor propagation that is designed specifically for future communications systems with multiple antennas such as spatial-division multiple access, spatial filtering for interference reduction, or MIMO. The model is designed to provide an accurate representation of the indoor propagation channel, taking into account various factors that affect communications performance in such systems.                       \\ \cline{3-5}
                                &                                                                                              & 2010              & Azzaoui \textit{et al.}~\cite{Bio_5464245}                    & Introduces a statistical model of the \ac{UWB} channel impulse response at \SI{60}{\giga\hertz}. Considers the transfer function as an $\alpha$-stable random process.                        \\ \cline{3-5}
                                &                                                                                              & 2013              & Priebe \textit{et al.}~\cite{Bio_6574880}                   & Introduces a new stochastic \SI{300}{\giga\hertz} indoor channel model to facilitate fast system simulations and adequate design of upcoming terahertz communications systems. A complete scenario-specific parameter set is provided for the considered environment.                       \\ \cline{3-5}
                                &                                                                                              & 2014              & Gustafson \textit{et al.}~\cite{Bio_6691924}                   & Presents clustering results for a double-directional 60 GHz MIMO channel model and derives a model that is validated with measured data. Suggests that when creating these models, it's important to consider small details in the environment.                      \\ \cline{3-5}
                                &                                                                                              & 2015              & Samimi \textit{et al.}~\cite{Bio_7248689}                    & Presents an omnidirectional spatial and temporal 3-dimensional statistical channel model for \SI{28}{\giga\hertz} dense urban non-line of sight environments.                        \\ \cline{2-5}   
                                & \multirow{8}{*}{Analytical}                                                                  & 2003              & Svantesson \textit{et al.}~\cite{Bio_1202728}                   & Reports on a study that uses measurement taken at Brigham Young University to investigate the statistical properties of indoor MIMO channels. Investigates whether the covariance matrix can be modeled as a Kronecker product of correlations at the transmitter and receiver, using a likelihood ratio test.                        \\ \cline{3-5}
                                &                                                                                              & 2015              & Sun \textit{et al.}~\cite{Bio_7093168}                    &   Investigates MU-MIMO systems with a large number of antennas at the base station. Proposes a \ac{BDMA} transmission scheme that serves multiple users via different beams, which performs near-optimally, and that the proposed pilot sequences have advantages over other sequences.   \\ \cline{3-5}                              
                                &                                                                                              & 2017              & You \textit{et al.}~\cite{Bio_7913686}                    &  Proposes a new method for wideband massive MIMO transmission over mmWave/Terahertz bands called \ac{BDMA} with per-beam synchronization (PBS) in time and frequency. Investigates beam scheduling to maximize the ergodic achievable rates for both uplink and downlink BDMA and develop a greedy beam scheduling algorithm.                          \\ \hline
\multirow{15}{*}{\rotatebox[origin=c]{90}{\parbox[c][][c]{20mm}{\raggedright \textbf{Hybrid}}}}        & \multirow{12}{*}{\begin{tabular}[c]{@{}c@{}}Deterministic\\Hybrid\end{tabular}}           & 2000              & Wang \textit{et al.}~\cite{Bio_855493}                   & Presents a hybrid technique based on combining ray-tracing and FDTD methods for site-specific modeling of indoor radio wave propagation.                       \\ \cline{3-5}        
                                &                                                                                              & 2006              & Reynaud \textit{et al.}~\cite{Bio_4109307}                   & Presents an original approach, combining the advantages of UTD and FDTD methods.                       \\ \cline{3-5}
                                &                                                                                              & 2008              & Thiel \textit{et al.}~\cite{Bio_4589145}                    & Introduces a novel method for analyzing wave propagation in a building consisting of a heterogeneous mixture of homogeneous and periodic walls is presented. Walls are discretized into finite-size building blocks. FDTD approach is used to compute their electromagnetic response in a periodic arrangement as well as in corner and terminal locations.                         \\ \cline{3-5} 
                                &                                                                                              & 2020              & Lecci \textit{et al.}~\cite{Bio_9322374}                   & Introduces a detailed mathematical formulation for quasi-deterministic models/Q-D at \ac{mmWave} frequencies, that can be used as a reference for their implementation and development. Moreover, it compares channel instances obtained with an open-source National Institute of Standards and Technology Q-D model implementation against real measurement at \SI{60}{\giga\hertz}, substantiating the accuracy of the model.                       \\ \cline{3-5} 
                                &                                                                                              & 2021              & Zhu \textit{et al.}~\cite{Bio_3gpp2021hybrid}                     & Reviews the main existing channel models suitable for mmWave frequency band as well as the typical channel modeling methods, illustrates the generation procedure of stochastic channel model in the 3GPP standard, he map-based hybrid channel model is demonstrated and analyzed.                      \\ \cline{2-5} 
 
                                & \multirow{6}{*}{\begin{tabular}[c]{@{}c@{}}Deterministic-\\Statistical\\ Hybrid\end{tabular}} & 2014              & Maltsev \textit{et al.}~\cite{Bio_7063558}                   & Introduces a new Q-D approach for modeling mmWave channels. The proposed channel model allows the natural description of scenario-specific geometric properties, reflection attenuation and scattering, ray blockage, and mobility effects.                       \\ \cline{3-5} 

                                &                                                                                              & 2014              &  Samimi \textit{et al.}~\cite{Bio_7037347}                   & Presents \ac{UWB} statistical spatial and omnidirectional channel models for \SI{28}{\giga\hertz} mmWave cellular dense urban \ac{NLoS} environments, developed from wideband measurement in New York City that used synthesized timing from \ac{3D} ray-tracing. \\ \cline{3-5} 
                                &                                                                                              & 2021              & Bian \textit{et al.}~\cite{Bio_9318511}                   &  Proposes a novel \ac{3D} non-stationary GBSM for 5G and beyond 5G systems.               
                                \\ \hline  \hline
\end{tabular}
\end{table*}

The state-of-the-art channel models in terms of different methodologies are surveyed next and are summarized in Table \ref{table_THzChannelModel}. The channel models are divided into three categories: \textit{deterministic, statistical, and hybrid}.
\subsubsection{Deterministic Channel Modelling}
Generally, there exist mainly three representative methods for deterministic channel modeling, including ray tracing \cite{Ref_glassner1989raytracing}, \ac{FDTD} \cite{Bio_4057499}, and channel measurement-based methods.

Let us first look at the most popular method, namely ray tracing. Visibility tree \cite{Bio_SaadaneVisiTree} and ray launching \cite{Bio_8529268} are two alternatives to achieve ray tracing. To date, ray tracing has been calibrated through field measurement, e.g., the work \cite{Bio_8684885} reports indoor and T2I inside-station scenarios at \SI{300}{\giga\hertz} \ac{THz} frequencies.  The calibration and validation for frequencies between \SI{1.0}{}-\SI{10}{\tera\hertz} remain challenging due to the lack of material parameters. For single-antenna systems, conventional ray-tracing models are suitable for analyzing point-to-point communications. By contrast, when dealing with multiple-antenna systems, performing ray tracing for each Tx-Rx link can be prohibitively complex \cite{Bio_1424633}. To reduce the computational complexity associated with multiple antennas, it is possible to perform a single ray-tracing simulation that extracts not only the amplitudes and delays but also the directions of the paths. This information can be combined with the array characteristics to generate the transfer function between each transmit and receive antenna pair, which is independent of the antenna array size \cite{Bio_7909753}. Another approach to alleviating the computational burden is to use simplified ray-tracing models. For instance, map-based models are based on ray-tracing and use a simpliﬁed \ac{3D} geometrical description of the environment \cite{Bio_7481518}, which can be more accurate if a laser is employed for scanning the environment \cite{Bio_8608834, Bio_8345613, Bio_8880819}.

\ac{FDTD} is a numerical analysis method that relies on solving Maxwell's equations directly. This technique is particularly suited for scenarios involving small and complex scatterers, where surface materials exhibit a higher degree of roughness at \ac{THz} frequencies \cite{Bio_4057499}. However, it demands many memory resources to track all the locations, as well as substantial time and computational power to update the desired estimates at successive time instants \cite{Bio_855493}. When applied to objects with large dimensions compared with the wavelength of \ac{THz} signals,  \ac{FDTD} suffers from high computational complexity. In order to apply it effectively, a database of the target environment with sufficiently high resolution is required. This database may be generated from a laser scanning for a point cloud \cite{Bio_8608834}.  In a small intra-device channel, a comparison between the ray-tracing and \ac{FDTD} method was presented in \cite{Bio_6902134}.

Last but not least, another approach, which is referred to as the channel measurement-based method relies on real-world field measurement of the target environment and large-volume data analysis \cite{Bio_rappaport2022radio}. In recent years, the trend of open-source data has motivated many researchers to make their measurement results available online. Some standardization groups, such as the NextG Channel Model Alliance \cite{Ref_NextGCM} under the National Institute of Standards and Technology (NIST), aim to make data exchange easier. The European project ARIADNE has provided initial measurement results and created channel models for D band links in LoS and NLoS office environments \cite{Ref_kokkoniemi2022initial}.
In the context of \ac{THz} channels, there are some challenges due to the volume of measured data, which is affected by both the large bandwidth and large number of elements of the antenna arrays.

\subsubsection{Statistical Channel Modelling}
Statistical approaches are used to capture the statistical behaviors of wireless channels in various scenarios \cite{Bio_6574880}. One of its main strengths is the low computational complexity, which enables fast construction of channel models based on key channel statistics and facilitates simulation-based studies of wireless communications. It is broadly classified into two categories, i.e. physical models and analytical models \cite{Bio_662642,Bio_990911,  Bio_686770}. The former models describe the statistics of double-directional channel characteristics, such as the power delay profile, arrival time, and angle distribution, which are independent of the antenna properties. In contrast, the latter methods characterize the impulse response of a channel and the characteristics of antennas in mathematical terms. 
\paragraph{Physical Models}
Early research work on statistical channel modeling for the \ac{mmWave} or \ac{THz} band focused on enhancing and adapting the well-known Saleh-Valenzuela (S-V) model through calibration \cite{Bio_6691924}, which is based on the observation that multipath components arrive in the form of clusters \cite{Bio_797367}. Some other research work maintained the model based on clustering while utilizing different distributions, instead of the Poisson process, to describe the \ac{ToA} in order to improve the accuracy compared with measurement \cite{Bio_6574880,Bio_7248689,Bio_5464245}. Another example, which is referred to as the Zwick model \cite{Bio_821698}, is based on multipath components rather than clusters and does not account for amplitude fading. In \cite{Bio_1021910}, the original Zwick model was enhanced to incorporate its applicability to \ac{MIMO} systems.

\paragraph{Analytical Models}
Analytical models take into account the channel and antenna characteristics as a whole, thereby characterizing the impulse response from the antenna elements between the transmitter (Tx) and the receiver (Rx). These individual impulse responses are organized into a matrix and the statistical properties of the matrix elements, including the correlations, are considered. The Kronecker-based model \cite{Bio_1202728} assumes that the correlation between the transmit and receive arrays is separable. However, as the number of antennas increases and single-reflection propagation is dominant in the \ac{THz} band, this assumption becomes less accurate. To overcome this, some other models account for either \ac{MIMO} or \ac{MMIMO} channels from the perspective of beams or eigenspaces. For instance, an approach called the virtual channel representation (VCR) \cite{Bio_7913686} characterizes the physical propagation by sampling rays in a beam space. These aforementioned models can be also called correlation-based stochastic models (CBSMs). Despite their limited capability in spatial determinism, CBSMs are well-suited for evaluating the performance of \ac{MMIMO} systems due to their low complexity. Unfortunately, it is challenging to properly describe \ac{MMIMO} channels, especially \ac{UMMIMO} channels over the \ac{THz} band, due to difficulties in modeling the near-field and the spatial non-stationarity. To address this issue, an enhanced method referred to as \ac{BDCM} was proposed \cite{Bio_7093168}, making the \ac{BDCM} applicable to \ac{UMMIMO} scenarios for the \ac{THz} band.

\subsubsection{Hybrid Channel Modelling}
Since different methods have specific advantages and limitations, hybrid methods are often considered for combining the benefits of deterministic and statistical models, depending on the considered scenarios \cite{Ref_Han2017Optimal}. As the deterministic methods offer high accuracy at the price of long computing times and large memory usage but the statistical methods require less computational power, most existing approaches focus on hybrid deterministic-statistical models \cite{Bio_4109307, Bio_4589145,Bio_9322374, Bio_3gpp2021hybrid}. Moreover, some methods combine two deterministic approaches, like ray tracing and FDTD, which are referred to as hybrid deterministic methods. In these methods, FDTD is typically utilized for studying regions close to complex discontinuities, while ray tracing is used to trace the rays outside the FDTD regions. The synergy between ray tracing and FDTD was presented in \cite{Bio_855493}, where the location of the receiver is restricted in the FDTD region. Later, some works such as \cite{Bio_4109307,Bio_4589145} extended it for improving the time efficiency and modeling multiple interactions between ray tracing and FDTD.

Although statistical channel models are highly efficient, they struggle to accurately capture spatial consistency and the temporal evolution of cluster correlations. These aspects motivated the development of some hybrid models that incorporate both statistical and geometrical approaches. This hybrid approach enables the inclusion of some channel features that are impossible to characterize through a stochastic model. In 2002, a quasi-deterministic (Q-D) channel model \cite{Bio_1267850} was initially proposed for \ac{mmWave} channels, which was adopted by the IEEE 801.11ad standardization group for indoor scenarios at 60 GHz  \cite{Bio_5504964}. Moreover, the Q-D channel model has been successfully applied to other wireless standards, such as the \ac{mmWave} Evolution for Backhaul and Access (MiWEBA) \cite{Bio_7063558} and the IEEE 802.11ay \cite{Bio_7499315}, which is an evolution of the IEEE 802.11ad standard. 

Another hybrid method called geometry-based stochastic channel model (GSCM) incorporates a geometrical component during the stochastic modeling process \cite{Bio_990911, Bio_686770, Bio_6574880}. Although the placement of scatterers is stochastic, the simplified ray-tracing method employed in GSCM is deterministic. A new non-stationary GSCM, called the beyond \ac{5G} channel model (B5GCM), was recently introduced in \cite{Bio_9318511}, where the correlation functions were derived based on a general \ac{3D} space-time-frequency model. This model can be categorized into two types: regular-shaped GSCM, which is primarily used for theoretical analysis, such as correlation functions, and irregular-shaped GSCM that can better replicate measured results. Notably, the COST259, COST273, and COST2100 models \cite{Bio_4027578} take advantage of hybrid models. The well-known 3GPP Spatial Channel Model (SCM) \cite{Bio_3gpp2018hybrid} and WINNER II model \cite{Bio_8045088} also considered this approach. Furthermore,  Samimi et al. \cite{Bio_7037347} utilized temporal clusters and spatial lobes to handle the temporal and spatial components.

\section{\ac{THz} Transceiver: Antennas and Devices} \label{Sec:THz_Transceiver}

This section aims to provide readers with the necessary knowledge required to design and build \ac{THz} transceivers, which is recognized as one of the most challenging issues to block the practical use of THz communications and sensing. We discuss the cutting-edge antenna technologies that are appropriate for transmitting and receiving \ac{THz} signals, along with the fundamental and innovative aspects of photonic-electronic devices and components used in constructing \ac{THz} transceivers.

\subsection{\ac{THz} Antennas}
The conventional concepts of EM antennas can also be applied in the \ac{THz} regime, as mentioned in the references like \cite{Ref_xu2020areview}. Nevertheless, the exceptionally high \ac{THz} frequencies raise specific limitations and issues that must be carefully considered. The tiny wavelength necessitates the use of extremely small structures, imposing disruptive challenges in the manufacturing processes. Fortunately, this reduction in structure size opens up opportunities for the implementation of innovative manufacturing techniques like \ac{LTCC}, antenna-on-chip design, \ac{SIW}, and others. Another major issue is the skin effect in conductive materials \cite{
Ref_he2020overview}. This phenomenon occurs as the skin depth, indicating the depth to which current penetrates in a conductive material, decreases significantly. As a result, the conductivity of metallic materials decreases, leading to extra losses in the antenna system. To address this challenge, researchers have been investigating the utilization of novel materials, such as graphene. Graphene, with its outstanding electrical and thermal properties, shows great potential in alleviating the skin effect and enhancing the performance of THz antennas. Its high electron mobility and low resistivity render it an appealing choice for overcoming the constraints associated with conventional conductive materials.

To provide readers with an insightful view, we present the fundamentals of different THz antenna techniques and a summary of  current research advances in the following.

\subsubsection{Horn Antennas}
Horn antennas are widely used in wireless applications due to their favorable characteristics. They belong to the category of high directivity antennas, capable of achieving gains of up to \SI{25}{\dB i}. Similar to hollow waveguides, these antennas have low power loss,  making them appropriate for low-noise and high-power applications. Furthermore, they can operate across a wide frequency range, which is advantageous for broadband signals.
Various manufacturing technologies have been considered to address the challenges posed by the short wavelength of THz signals. Among these, the authors of \cite{Ref_zhang2016metallic} introduced elective laser-melting 3D printing technology to produce conical horn antennas, which are capable of operating in the frequency bands of E (\SIrange{60}{90}{\GHz}), D (\SIrange{110}{170}{\GHz}), and H (\SIrange{220}{325}{\GHz}). Another appropriate manufacturing technology for THz horn antennas is \ac{LTCC}, as reported in \cite{Ref_tajima2014_300ghz}. By utilizing substrate-integrated waveguide technology, a horn structure is created within the multi-layer LTCC substrate. It achieves a peak gain of \SI{25}{\dB i} and supports a bandwidth of \SI{100}{\giga\hertz}. Remarkably, the dimensions of the \ac{LTCC} horn antenna are compact, measuring only $5\times5\times2.8\si{\milli\meter}$.  

\subsubsection{Planar Antennas}
Planar transmission line technology, commonly used for \ac{THz} frequencies, offers flexible antenna designs. Among these designs, patch antennas based on microstrip line technology are particularly well-known. However, planar antennas generally exhibit inferior performance compared to horn antennas. This is primarily due to the \ac{EM} wave partially propagating within a dielectric substrate, resulting in increased overall loss. Additionally, the small thickness in \ac{THz} planar structures leads to substrate mode issues \cite{Ref_rebeiz1992millimeter}, where some of the wave energy is trapped in the substrate and cannot be effectively utilized. Despite these drawbacks, planar antennas remain popular, primarily due to the flexibility in producing the planar structures. The planar shapes can be easily designed and realized \cite{Ref_sharma2009rectangular, Ref_dhillon2017thz, Ref_rubani2019design, Ref_llatser2012graphene}. Moreover, patch antennas can be efficiently packaged to build an antenna array, allowing flexible control on the antenna pattern \cite{Ref_nissanov2020high, Ref_naghdehforushha2018high, Ref_jha2012microstrip}. Another advantage of THz planar antennas is the feasibility of designing high-gain on-chip antennas \cite{Ref_alibakhshikenari2021study, Ref_seok2008a410ghz}. Sandwiched between the silicon and polycarbonate substrates, a 15-element array comprising circular and rectangular patches on the top surface of the polycarbonate layer achieved a gain of \SI{11.71}{\dB i} \cite{Ref_alibakhshikenari2020highgain}.

\subsubsection{Substrate-Integrated Antennas}
The small wavelength of THz signals causes some challenges in the design of planar antennas. However, it turns into an advantage by leveraging the capability of low-cost manufacturing and the flexibility of planar structures to build \ac{SIW}. The key idea behind \ac{SIW} is to connect the upper and the lower conductive layers of a substrate by the via walls, as discussed in a comprehensive review of this technology in \cite{Ref_wu2012substrate}. This process builds a rectangular waveguide inside the substrate, with the distance between vias defining the upper cutoff frequency. The \ac{SIW} technology can be integrated with a variety of antennas and array designs, such as horn antennas \cite{Ref_cai2017alow, Ref_jin2018eband}, slot antennas \cite{Ref_hu2012asige, Ref_xu2012_140ghz}, patch antennas \cite{Ref_alibakhshikenari2021high, Ref_fan2017alow}, and others \cite{Ref_nissanov2021high, Ref_althuwayb2020chip, Ref_camblor2012sub, Ref_alibakhshikenari2022acomprehensive}.

\subsubsection{Carbon-Based Antennas}
Graphene and carbon nanotubes, which are rolled sheets of graphene, are two promising materials for use over the \si{\THz} frequency band. Due to the skin effect, metallic materials, especially copper, suffer low conductivity and therefore inferior performance. In contrast, graphene features high conductivity due to the propagation of plasmon mode in a high-frequency range. The conductivity can be tuned not only by chemical doping, but also by applying electric and magnetic fields, enabling the production of tunable antennas \cite{Ref_hartmann2014terahertz}. These properties make graphene suitable for fabricating \si{\THz} antennas. The most discussed use is to replace the metallic conductors with graphene or carbon nanotubes in, for example, planar antenna structures \cite{Ref_lu2021carbon, Ref_correasserrano2017graphene}. Besides, the approach of building a dipole antenna based on carbon nanotubes was reported \cite{Ref_hao2006infrared, Ref_mahmoud2012characteristics}.

\begin{figure*}
    \centering
    \subfloat[]{%
        \scalebox{.7}{%
            \begin{circuitikz}
                \ctikzset{blocks/fill=white}
                \large
                \draw
                    (0,0)
                        node[oscillator,anchor=right](lo1){}
                        node[right]{$f_\mathrm{IF}$}
                    (lo1.up)
                        -- ++ (0,3.25)
                        node[mixer](m1){}
                        -- ++ (2,0)
                        node[adder](add){}
                        -- ++ (0,-1.25)
                        -- ++ (-1.25,0)
                        node[mixer](m2){}
                    (m2.down)
                        -- ++ (0,-.75)
                        node[twoportshape, t={\makebox[1.1em][l]{\SI{90}{\degree}}}]{}
                        to[short, -*] ++ (-.75,0)
            
                    (m1.left)
                        to[short, -o] ++ (-0.75,0)
                        node[above]{$I(t)$}
                    (m2.left)
                        to[short, -o] ++ (-1.5,0)
                        node[above]{$Q(t)$}
                ;
                \draw
                    (add)
                        to[bandpass] ++ (3,0)
                        node[mixer](m3){}
                        to[bandpass] ++ (3,0)
                        to[amp, boxed] ++ (0.0001,0)
                        -- ++ (1.25,0)
                        node[dinantenna]{}
                        
                    (m3.down)
                        -- ++ (0,-1.5)
                        node[oscillator,anchor=up](lo2){}
                    (lo2.right)
                        node[right]{$f_\mathrm{LO, \si{\THz}}$}
                ;
            \end{circuitikz}%
        }%
    }%
    \hfill%
    \subfloat[]{%
        \scalebox{.7}{%
            \begin{circuitikz}
                \ctikzset{blocks/fill=white}
                \large
                \draw
                    (0,0)
                        node[oscillator,anchor=right](lo1){}
                        node[right]{$f_\mathrm{IF}$}
                    (lo1.up)
                        -- ++ (0,3.25)
                        node[mixer](m1){}
                        -- ++ (-2,0)
                        coordinate (div)
                        to[short, *-] ++ (0,-1.25)
                        -- ++ (2.75,0)
                        node[mixer](m2){}
                    (m2.down)
                        -- ++ (0,-.75)
                        node[twoportshape, t={\makebox[1.1em][l]{\SI{90}{\degree}}}]{}
                        to[short, -*] ++ (-.75,0)
                    (m1.right)
                        to[short, -o] ++ (1.5,0)
                        node[above]{$I(t)$}
                    (m2.right)
                        to[short, -o] ++ (0.75,0)
                        node[above]{$Q(t)$}
                ;
                \draw
                    (div)
                        to[bandpass] ++ (-3,0)
                        node[mixer](m3){}
                        to[bandpass] ++ (-3,0)
                        to[amp, boxed] ++ (0.0001,0)
                        -- ++ (-1.25,0)
                        node[dinantenna]{} 
                    (m3.down)
                        -- ++ (0,-1.5)
                        node[oscillator,anchor=up](lo2){}
                    (lo2.right)
                        node[right]{$f_\mathrm{LO, \si{\THz}}$}
                ;
            \end{circuitikz}%
        }%
    }%
    \caption{Architecture of superheterodine (a) transmitter and (b) receiver design including IQ-modulator.}
    \label{fig:transceiver}
\end{figure*}
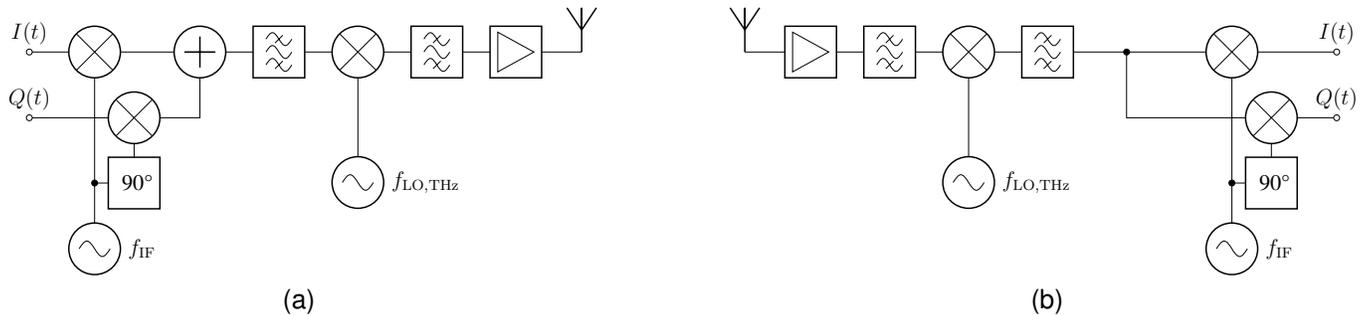

\subsection{\ac{THz} Components and Transceiver Design}
For a long time, there has existed a terahertz gap \cite{Bio_tekbiyik2019terahertz} between the microwave and infrared regions. The gap arises because traditional electronics are not efficient at generating or detecting THz signals, while optical techniques like lasers and photonic detectors are generally not applicable due to limitations related to the wavelengths involved. On the one hand, these frequencies are too low for \textit{photonic} devices. On the other hand, they are too high for \textit{electronic} devices. In this regard, the \si{\THz} band used to be unreachable for either \textit{photonic} or \textit{electronic} technologies. 
Prior \ac{THz} research was concentrated on applications such as imaging or spectrometry \cite{Ref_zandonella2003terahertz}. This fact is attributed to two major reasons:
\begin{itemize}
    \item These applications require comparably high output power but are not demanding on the receiver side, since the amplitude rather than the phase of a signal acts as the information carrier. 
    \item The equipment or operational conditions are not so challenging, compared with small, lightweight, cheap, and battery-powered mobile terminals. Some technologies, especially photonic generation and detection are applicable. 
\end{itemize}

In contrast, \ac{THz} communications and sensing require the capability of accurate phase recovery, especially the case for digital modulation when both \ac{IQ} branches are exploited. Furthermore, compactness and low energy consumption play critical roles. Thus, the utilization of semiconductor-based integrated circuits is advantageous for \ac{THz} communications and sensing applications in \ac{6G} and beyond systems. We present the fundamentals and state-of-the-art advances in photonic and electronic components as follows.

\subsubsection{Electronic Devices}

There are two classes of electronic devices capable of generating \si{\THz} radiation. The first class is high-power devices primarily based on the principle of electron tubes, capable of producing signal power ranging from \SI{10}{\W} to \SI{1}{\MW}. These high-power devices are typically necessary for specialized applications, such as satellite communications. Another class consists of semiconductor devices, constructed from various semiconductor materials, which provide a more compact and cost-effective solution. While they may have limitations in terms of transmission power, semiconductor devices are suitable for \ac{THz} communications and sensing \cite{Ref_rieh2011an}.  III-V-semiconductors such as \ac{GaAs}, \ac{GaN}, and \ac{InP}, are commonly used to fabricate THz components. In addition, \ac{SiGe} based devices show promising performance. Compared to the aforementioned technologies, the \ac{CMOS} technology has limitations such as low transmit power and high noise figure. However, recent research argues that the \ac{CMOS} technology may also be suitable for THz communications and sensing applications \cite{Ref_kenneth2019opening}.

Most semiconductor-based \si{\THz} transceivers follow the conventional design for lower frequencies, such as the heterodyne transceiver approach, as shown in \figurename~\ref{fig:transceiver}. This entails passing a modulated baseband signal to the front-end circuitry. Then, it is up-converted to the \si{\THz} carrier frequency by mixing with the \ac{LO} signal. The signal is amplified and sent over an antenna. At the receiver, the same steps are applied in a reverse order, and the down-converted signal is passed to the baseband circuitry for further processing. The most critical aspect lies in \ac{LO} generation, as it involves synthesizing a THz signal. To achieve this, a multiplexer is employed to upconvert a high-frequency signal into the \si{\THz} range. However, it's important to note that each multiplexer stage introduces additional inter-modulation products, which in turn raise the noise figure and distort the generated signal. Consequently, the number of multiplexer stages becomes a limiting factor for THz signal generation.

\begin{table*}
    \centering
    \caption{overview of Semiconductor transmitter and receiver Technologies}
    \label{tab:transceiver}
    \begin{tabular}{c|c|c|c|c|c|c}  \hline \hline
      \textbf{Semiconductor Techno.}                  & \textbf{Freq. [\si{\GHz}] }  & \textbf{Transmit Power [\si{\dBm}]}  & \textbf{Noise Figure [\si{\dB}]} & \textbf{Throughput [\si{\giga\bps}]} & \textbf{Reference} &  \textbf{Pub. Year} \\ \hline
        \hline
        
        \SI{70}{\nm} \ac{InP}       & \num{300}         & \num{10}      & \num{9.8}     & \num{20}  & \cite{Ref_song2016demonstration}  & 2016\\ 
        \SI{80}{\nm} \ac{InP}       & \num{300}         & \num{-3}      & N/A           & \num{120} & \cite{Ref_hamada2020_300ghz}      & 2020\\ 
        \SI{130}{\nm} \ac{InP}      & \num{590}         & \num{-2}      & ---           & ---       & \cite{Ref_urteaga2016a130nm}      & 2016\\ 
        \SI{25}{\nm} \ac{InP}       & \num{850}         & \num{-0.3}    & \num{12.7}    & ---       & \cite{Ref_leong2017_850ghz}       & 2017\\ 
        \SI{130}{\nm} \ac{SiGe}     & D-band            & \num{9}       & \num{9}       & \num{48}  & \cite{Ref_carpenter2016adband}    & 2016\\
        \SI{130}{\nm} \ac{SiGe}     & D-band            & \num{10}      & \num{10}      & ---       & \cite{Ref_elkhouly2020dband}      & 2020\\
        \SI{130}{\nm} \ac{SiGe}     & \num{240}         & \num{8.5}     & ---           & \num{50}  & \cite{Ref_grzyb2017a240}          & 2017\\
        \SI{130}{\nm} \ac{SiGe}     & \num{300}         & \num{-4.1}    & ---           & ---       & \cite{Ref_yu2021a300ghz}          & 2021\\
        \SI{130}{\nm} \ac{SiGe}     & \num{400}         & \num{-20}     & ---           & ---       & \cite{Ref_hu2012asige}            & 2012 \\
        \SI{28}{\nm} \ac{CMOS}      & \num{240}         & \num{0.7}     & ---           & ---       & \cite{Ref_londhe2022a232}         & 2022 \\
        \SI{40}{\nm} \ac{CMOS}      & \num{300}         & \num{-5.5}    & ---           & \num{105} & \cite{Ref_takano2017a105gbps,Ref_fujishima2017_300ghz}     & 2017\\
        \SI{40}{\nm} \ac{CMOS}      & \num{300}         & ---           & \num{27}      & \num{32}  & \cite{Ref_hara2017a32gbps,Ref_fujishima2017_300ghz}        & 2017\\
        \SI{28}{\nm} \ac{CMOS}      & \num{390}         & \num{-5.4}    & ---           & \num{28}  & \cite{Ref_dheer2020ahighspeed}    & 2020\\        
        \hline \hline
    \end{tabular}
\end{table*}

Table~\ref{tab:transceiver} offers a summary of the literature on the transmitter and receiver front end in terms of different semiconductor technologies. Among III-V semiconductor materials, \ac{InP} components demonstrate superior results in terms of output power and noise figure. Furthermore, \ac{InP} transistors can reach a maximum oscillating frequency $f_\mathrm{max}$ of up to \SI{1.5}{\THz} \cite{Ref_mei2015first}. A transceiver design based on a high-electron-mobility transistor (HEMT) was reported. HEMT supports \ac{THz} frequencies up to \SI{850}{\GHz}, with particular attention given to the frequency band around \SI{300}{\GHz} \cite{Ref_kim2015_300ghz, Ref_song2016demonstration, Ref_hamada2018_300ghz}. 

\ac{SiGe} and \ac{CMOS} are both silicon-based technologies suitable for the \ac{THz} frequency band. Due to the intrinsic characteristics of silicon, the maximum oscillation frequency is limited below \SI{1}{\THz}. There are studies that show \acp{HBT} based on \SI{130}{\nm} \ac{SiGe} process reaching $f_\mathrm{max}$ of \SI{620}{\GHz} \cite{Ref_ruecker2019device} as well as \SI{720}{\GHz} \cite{Ref_heinemann2016anexperimental}. The \ac{SiGe} components fabricated for \acp{HBT} exhibit advantages such as good linearity, high gain, and low noise. However, the power gain is limited, which prevents the operation above \SI{500}{\GHz}. Recent research focused particularly on the components and devices for THz communications and sensing over the D-band, which refers to the range of frequencies around \SIrange{110}{170}{\giga\hertz}. As a result, novel transceiver design was introduced such as those presented in \cite{Ref_elkhouly2020dband,Ref_karakuzulu2021full,Ref_zhao2012a160ghz,Ref_carpenter2016adband}. The authors of \cite{Ref_karakuzulu2023afout} demonstrated a complete communication link that realizes a high throughput of \SI{200}{\giga\bps}. Some researchers present their transceiver design operating around \SI{300}{\GHz} \cite{Ref_londhe2022a232,Ref_fujishima2017_300ghz,Ref_han2015asige,Ref_eissa2018wideband,Ref_eissa2020frequency}. A detailed survey on \ac{SiGe} transceiver for different purposes is provided in \cite{Ref_kissinger2021millimeter}.

THz devices built from CMOS exhibit lower performance compared to other semiconductor technologies. \ac{CMOS} \acp{FET} can only reach a maximum oscillating frequency of around \SI{450}{\GHz}, with limited power gain. Nevertheless, CMOS technology is known for its low production costs, which is its major advantage. The transceivers operating at \SI{240}{\GHz} \cite{Ref_londhe2022a232,Ref_thyagarajan2015a240ghz,Ref_kang2015a240ghz} or \SI{300}{\GHz} \cite{Ref_fujishima2019ultrahigh, Ref_katayama2016a300ghz, Ref_abdo2020a300ghz, Ref_zhong2018a300ghz} were implemented. The authors of \cite{Ref_takano2017a105gbps,Ref_hara2017a32gbps,Ref_fujishima2017_300ghz} showcased some CMOS transcivers operating in \SI{300}{\GHz}, which achieve high data rates such as \SI{105}{\giga\bps} or \SI{32}{\giga\bps}. The highest reported operating frequency is \SI{390}{\GHz}, reaching a data throughput of \SI{28}{\giga\bps}. Further achievements of \ac{CMOS} transceiver are summarized in \cite{Ref_kenneth2019opening, Ref_zhong2018cmos, Ref_fujishima2021overview}.

\subsubsection{Photonic Devices}

\begin{figure*}
    \centering
    \subfloat[]{%
        \scalebox{.7}{%
            \begin{circuitikz}[
                ld/.style={circuitikz/bipoles/twoport/width=1.7},
                mod/.style={circuitikz/bipoles/twoport/width=1.3},
                pd/.style={circuitikz/bipoles/twoport/width=1.3}
            ]
            \ctikzset{blocks/fill=white}
            \large
                \draw
                    (0,0)
                        node[twoportshape, anchor=right, ld, t=\parbox{5em}{\small\centering Wavelength-\\ tunable laser}](ld1){}
                        -- ++ (1.5,0)
                        -- ++ (0,-.75)
                        node[circle,fill=black](comb){a}
                        -- ++ (0,-.75)
                        -- ++ (-1.5,0)
                        node[twoportshape, anchor=right, ld, t=\parbox{5em}{\small\centering Wavelength-\\ tunable laser}](ld2){}
                    (ld1.right) node[below right]{$f_1$}
                    (ld2.right) node[below right]{$f_2$}
                ;
                \draw
                    (comb)
                        -- ++ (2,0)
                        node[twoportshape, mod, t=\parbox{4em}{\small\centering Optical\\modulator}](mod){}
                        -- ++ (2.5,0)
                        node[twoportshape, pd, t=\parbox{4em}{\small\centering UTC-PD}](pd){}
                        -- ++ (1.75,0)
                        node[dinantenna]{}
                    (mod.n)
                        to[short, -o] ++ (0,.75)
                        node[right]{$s(t)$}
                    (pd.right) node[below right]{$f_{\si{THz}} = f_1 - f_2$}
                ;
            \end{circuitikz}%
        }%
    }%
    \hfill%
    \subfloat[]{%
        \scalebox{.7}{%
            \begin{circuitikz}[
                ld/.style={circuitikz/bipoles/twoport/width=1.7},
                mod/.style={circuitikz/bipoles/twoport/width=1.3},
                pd/.style={circuitikz/bipoles/twoport/width=1.3}
            ]
                \ctikzset{blocks/fill=white}
                \large
                \draw
                    (0,0)
                        node[twoportshape, anchor = right, ld, t=\parbox{5em}{\small\centering Wavelength-\\ tunable laser}](ld1){}
                        (ld1.right)
                        -- ++ (2,0)
                        node[twoportshape, mod, t=\parbox{4em}{\small\centering Optical\\modulator}](mod){}
                        -- ++ (2,0)
                        -- ++ (0,-.75)
                        node[circle,fill=black](comb){a}
                        -- ++ (0,-.75)
                        -- ++ (-4,0)
                        node[twoportshape, anchor=right, ld, t=\parbox{5em}{\small\centering Wavelength-\\ tunable laser}](ld2){}
                    (mod.n)
                        to[short, -o] ++ (0,.75)
                        node[right]{$s(t)$}
                    (ld1.right) node[below right]{$f_1$}
                    (ld2.right) node[below right]{$f_2$}
                ;
                \draw
                    (comb)
                        -- ++ (2,0)
            
                        node[twoportshape, pd, t=\parbox{4em}{\small\centering UTC-PD}](pd){}
                        -- ++ (1.75,0)
                        node[dinantenna]{}
                    (pd.right) node[below right]{$f_{\si{THz}} = f_1 - f_2$}
                ;                
            \end{circuitikz}%
        }%
    }%
    \caption{Configuration of photo-mixing transmitters with (a) Double sub-carrier modulation and (b) Single sub-carrier modulation \cite{Ref_nagatsuma2013teratertz}. }
    \label{fig:photomixing}
\end{figure*}
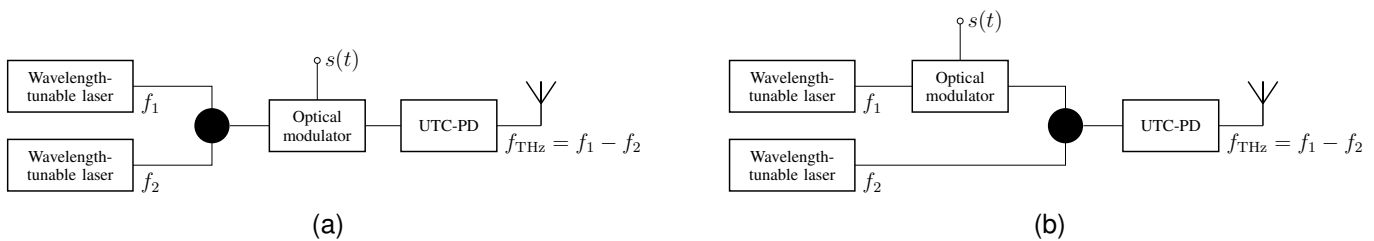
To transition from extremely high-frequency photonic radiation to a lower THz frequency, an optical frequency downconverter, commonly known as photomixing, is required. Illustrated in \figurename~\ref{fig:photomixing}, this process involves feeding two laser signals with frequencies $f_1$ and $f_2$ towards a photomixing diode, such as the \ac{UTC-PD} \cite{Ref_ishibashi2020uni}. Subsequently, the photodiode, akin to traditional high-frequency mixing, generates a \si{\THz} signal at frequency $f_{\si{\THz}} = f_1 - f_2$.
Photomixing provides high tunability and modulation bandwidth. Also, a variety of complex modulation schemes may be implemented with moderate efforts, as compared to electronic solutions. \figurename~\ref{fig:photomixing} shows two modulation approaches, modulation of both laser signals as well as the single laser signal only \cite{Ref_nagatsuma2013teratertz}.  A further advantage is the ability to deliver an optical signal over a large distance by means of an optical fiber. Thus, optical signal generation and modulation may be performed separated from the \si{\THz} signal generation, which may yield flexibility to the transmission system \cite{Ref_nagatsuma2016advances}.

Another photonic method of generating \si{\THz} signals is through \ac{QCL}. Conventional lasers emit photons through the recombination of electron-hole pairs across the band gap. Thus, the lower bound for the frequency of the emitted photons is defined by the band gap, which limits the ability of the lasers to work below the \ac{IR} band. In contrast, \ac{QCL}s are unipolar \cite{Ref_kazarinov1971possible}. The photons are emitted by quantum jumps of electrons between two different energy levels. These energy levels are defined by the structure of the quantum wells. This provides the possibility to tune the frequency of the laser. Due to the low energy gap between the energy levels, \acp{QCL} are well suited for mid-\ac{IR} applications. In recent decades, there was also a big progress on \ac{QCL} development for the far-\ac{IR} or \si{\THz} range \cite{Ref_belkin2015new}. \ac{QCL}s are able to cover the frequency band from \SIrange{1}{5}{\THz} with a peak radiation power of \SI{1}{\W} \cite{Ref_kumar2011recent,Ref_vitiello2021physics,Ref_chen2011wireless}. However, the operating frequency for \ac{QCL} is in the cryogenic range, which limits their applications.

\section{Beam-Forming and Alignment for \ac{THz} Communications and Sensing} \label{Sec_BF}
The use of the \ac{THz} band can alleviate the problem of spectrum scarcity and facilitate novel applications, such as nano-scale networks and in-device communications. However, its practical use is challenged by large propagation losses, which generally lead to  very short distances of signal transmission \cite{Ref_akyildiz2018combating}. The main causes of this problem include the high spreading loss that grows quadratically with carrier frequency, the gaseous absorption due to oxygen molecules and water vapor in the atmosphere, and the adverse effects of weather conditions, as discussed in the previous section. Such a propagation loss can reach hundreds of decibels per kilometer or even higher. Additionally, this problem is further aggravated by the following two factors \cite{Ref_peng2020channel}:
\begin{itemize}
\item \textit{Strong thermal noise}: Noise power is proportional to signal bandwidth with  the constant power density. Therefore, the unique advantage of massive bandwidth at the \ac{THz} band imposes a side effect of strong thermal noise.  
\item \textit{Hardware constraint}: The transmit power at the \ac{THz} band is quite constrained since the output power decreases with frequency and is at the level of decibel-milli-Watts in the foreseeable future. Hence,  raising power to extend the communications distance is not feasible \cite{Ref_rikkinen2020THz}.  
\end{itemize}

To extend the signal transmission distance beyond a few meters, high-gain directional antennas are necessary to compensate for such a high propagation loss in \ac{THz} communications and sensing. Thanks to tiny wavelengths,  massive numbers of elements can be tightly packed in a small area to generate high beamforming gains \cite{Ref_yang2013random, Ref_jiang2012randomBeamforming, Ref_yang2011randomVTC}. In this section, we will discuss the cutting-edge antenna forms, novel beamforming techniques, and necessary beam alignment schemes at the \ac{THz} band \cite{Ref_ning2022prospective}.

\subsection{Ultra-Massive MIMO}

Since the length of a resonant antenna is typically in the order of the wavelength at the resonance frequency, the dimension of an array with tens of elements is a few square meters and a few square centimeters at the sub-6G and \ac{mmWave} bands \cite{Ref_faisal2020ultramassive}, respectively. Moving to the \ac{THz} band, the antenna length further reduces. Hundreds of elements can be compacted in an array within a few centimeters using conventional metallic materials. However, this number is not sufficient to overcome the huge propagation loss suffered by \ac{THz} signals \cite{Ref_akyildiz2018combating}.  

Taking advantage of  \ac{SPP} waves \cite{Ref_lockyear2009microwave}, the inter-element separation of an array can be further reduced to the \ac{SPP} wavelength, which is much smaller than the \ac{EM} wavelength. Consequently,  nanomaterials that support the propagation of \ac{SPP} waves, such as graphene and metamaterials, are employed to further improve the hardware compactness.  Graphene, a one-atom-thick carbon nanomaterial with unprecedented mechanical, electrical, and optical properties, is employed to fabricate plasmonic nano-antennas with a smaller size of almost two orders of magnitude compared to metallic \ac{THz} antennas \cite{Ref_jornet2013graphene}. In particular, thousands of graphene-based nano-antennas can be embedded in a few square millimeters at \SI{1}{\tera\hertz}. The emergence of nano-antennas paves the way for building very large-scale arrays for \ac{THz} communications. In 2016, Akyildiz and Jornet \cite{Ref_akyildiz2016realizing} presented the concept of \ac{UMMIMO} communications, and demonstrated a $1024\times 1024$ system where both the transmitter and receiver are equipped with an array of $1024$ nano-antennas.

A massive number of elements impose challenges such as prohibitive power consumption and high hardware complexity \cite{Ref_jiang2021impactcellfree}. It is worth rethinking the array architecture and beamforming schemes in \ac{UMMIMO} systems at the \ac{THz} band. 
Fully digital beamforming can generate a desired beam but it leads to unaffordable energy consumption and hardware cost since each antenna in a large-scale array needs its dedicated \ac{RF} chain.  This motivates the study of analog beamforming with low complexity. By employing analog phase shifters,  only a single \ac{RF} chain is needed, substantially lowering hardware and power costs.  Nevertheless, the analog architecture supports only a single stream, limiting data rates and the number of users. As a compromise of these two forms, hybrid digital-analog architecture is the best choice for \ac{THz} from the perspective of performance and complexity trade-off. Combining an analog-shifter network with a few \ac{RF} chains, hybrid beamforming can significantly reduce hardware cost and low energy consumption, while achieving comparable performance as digital beamforming. 

Although hybrid beamforming has been extensively studied for the sub-6GHz and \ac{mmWave} bands \cite{Ref_molisch2017hybrid, Ref_zhang2019hybrid, Ref_ahmed2018survey}, the peculiarities of the \ac{THz} band, such as channel sparsity \cite{Ref_he2021wireless} and beam squint \cite{Ref_wu20223dhybrid}, impose many difficulties for designing an \ac{UMMIMO} system. Currently, many new forms of hybrid beamforming are discussed in the literature, including \ac{AoSA} to balance the power consumption and data rate, widely-spaced multi-subarray to overcome the low spatial multiplexing gain due to channel sparsity, and true-time-delay-based hybrid beamforming to address the problem of beam squint \cite{Ref_han2021hybrid}. 

\subsubsection{Array of Subarrays}
In a hybrid architecture, the connection between elements and \ac{RF} chains has two basic forms: \ac{FC} and \ac{AoSA} \cite{Ref_lin2016terahertz}. In the \ac{FC} hybrid beamforming, each element is fully connected to all \ac{RF} chains via a signal combiner, and the signal of an \ac{RF} chain radiates over all antenna elements via an individual group of phase shifters.  Any \ac{RF} chain should have the capability to drive the entire large-scale antenna array, which is power-aggressive. Particularly, the use of a large number of phase shifters and combiners will exacerbate the problems of high hardware cost and power consumption. In contrast, all elements in  \ac{AoSA} are divided into disjoint subsets called subarrays, and a subarray is only accessible to one specific \ac{RF} chain \cite{Bio_7397861}. \ac{AoSA} conducts signal processing at a subarray level with fewer phase shifters, such that hardware cost, power consumption, and signal power loss can be dramatically reduced. In addition, beamforming and spatial multiplexing can be jointly optimized by cooperating with precoding in the baseband.

Recent literature shows the strong interest to exploit an array of subarrays. For instance,   Lin and Li  published a series of works on this topic. In \cite{Bio_7036065}, they analyzed the ergodic capacity of an indoor single-user \ac{THz} communications system and obtained an upper bound, providing guidance on the design of antenna subarray size and numbers for certain long-term data rate requirements with different distances. An adaptive beamforming scheme, which considers the impact of transmission distance,  was proposed for multi-user \ac{THz} communications in \cite{Bio_7116524}. Therein the array-of-subarray structure for multi-user sub-\ac{THz} communications was considered and its spectral and energy efficiency were analyzed. They  then showcased a THz-based multi-user system for indoor usage that uses an array-of-subarrays architecture to handle hardware restrictions and channel characteristics in the \ac{THz} band, which has shown a great advantage by comparing with the \ac{FC} structure in both spectral and energy efficiency \cite{Ref_lin2016terahertz}. In \cite{Bio_9591285}, Tarboush \emph{et al.}  proposed an accurate stochastic simulator of statistical THz channels, named TeraMIMO, aiming at catalyzing the research of \ac{UMMIMO} antenna configurations. TeraMIMO adopted the \ac{AoSA} antenna structure for hybrid beamforming and accounted for spatial sparsity.

To further reduce the complexity, various alternating optimization algorithms have been proposed for the \ac{AoSA} architectures \cite{Bio_7397861}. In contrast to the \ac{FC} architecture, the \ac{AoSA} architecture has a restricted number of phase shifters that equals the number of antennas. However, since the \ac{RF} chains and antennas are connected exhaustively, the \ac{FC} architecture can achieve data rates comparable to those of the optimal fully-digital beamforming architecture. Conversely, the data rate of the \ac{AoSA} architecture is significantly lower compared to that of the \ac{FC} architecture. This is attributed to the partial interconnection between antennas and \ac{RF} chains. Hence, we need to balance the power consumption and data rate of the \ac{THz} hybrid beamforming, inspired by the challenge of designing large-scale antenna arrays in \ac{THz} \ac{UMMIMO} systems. To address this issue, some researchers introduced a new form of hybrid beamforming called \ac{DAoSA} \cite{Bio_9110897, Bio_9934006, Bio_9852650}, which features a flexible hardware connection. \ac{DAoSA} achieves a good balance between power consumption and data rates by intelligently determining the connection between sub-arrays and \ac{RF} chains.

\subsubsection{Widely-Spaced Multi-Subarray}
Due to the tiny wavelength, the \ac{THz} channel is usually sparse, consisting of a \ac{LoS} path and a few reflection paths. The transmit power concentrates on the \ac{LoS} path, and the overall angular spread of \ac{THz} signals is small.  For instance, a maximal angular spread of \SI{40}{\degree} has been observed for indoor environments in the \ac{THz} band, compared to \SI{120}{\degree} for indoor scenarios at \SI{60}{\giga\hertz} \ac{mmWave} frequencies \cite{Ref_xing2021millimeter}.  Since the number of spatial degrees of freedom is upper-bounded by the number of multipath components, the number of data streams or the potential spatial multiplexing gain is usually small, limiting the achievable bandwidth efficiency at the \ac{THz} band. A widely-spaced multi-subarray hybrid beamforming architecture is proposed \cite{Ref_yan2022joint} to overcome the low spatial multiplexing gain due to channel sparsity. Instead of critical spacing, the inter-subarray separation is over hundreds of wavelengths, reducing the correlation between the subarrays.

The \ac{WSMS} hybrid beamforming architecture is promising by exploiting intra-path multiplexing for \ac{THz} \ac{UMMIMO} systems \cite{Bio_8356240}. It was discovered in \cite{Bio_7501567} that when the distance between antennas is expanded, the planar-wave assumption becomes invalid, and it is necessary to consider the propagation of spherical waves between antennas. Previous research has examined the use of intra-path multiplexing in \ac{LoS} \ac{MIMO} architecture operating at microwave and \ac{mmWave} frequencies, which enables multiplexing gain to be achieved using just a single \ac{LoS} path \cite{Bio_9307264}. Given that the intra-path multiplexing gain is not restricted by the number of multipath, it is a highly viable and promising solution for \ac{THz} communications, which are known to exhibit significant channel sparsity \cite{Bio_9422343}. In \cite{Bio_8845161}, the authors demonstrated that the \ac{WSMS} architecture can substantially improve the spectral efficiency of \ac{THz} systems through the use of additional intra-path multiplexing gain, which sets it apart from existing hybrid beamforming that solely relies on inter-path multiplexing. As the follow-up, the authors designed an alternating optimization algorithm to maximize the sum rate \cite{Bio_10008606} under the \ac{WSMS} architecture.

\subsubsection{True-Time-Delay-Based Hybrid Beamforming}
Most of the current hybrid beamforming architectures rely on phase shifters, which are frequency-independent, inducing the same phase rotation at different frequency components of a signal. Under the ultra-wide bandwidth at the \ac{THz}, these shifters only provide correct phase shifting for a certain frequency, whereas other frequencies suffer from phase misalignment. As a result, the formed beam is squinted with a substantial power loss, e.g., \SI{5}{\decibel} reported in \cite{Ref_han2021hybrid}. To solve the problem of beam squint at the \ac{THz} band, \ac{TTD} can be applied to substitute phase shifters \cite{Ref_wu20223dhybrid}. The \ac{TTD} is frequency-dependent, and the phase rotation adjusted by \ac{TTD} is proportional to the carrier frequency and perfectly matches the ultra-wideband \ac{THz} beamforming.

According to \cite{Bio_7959180}, the \ac{TTD}-based phase shifting is aligned with the requirements of wideband \ac{THz} hybrid beamforming, given its proportional relationship with the carrier frequency. While ideal \ac{TTD}s with infinite or high resolution are capable of precise phase adjustments, they are often associated with high power consumption and hardware complexity \cite{Bio_7394105}. For the perspective of practical \ac{THz} systems, low-resolution \ac{TTD}s that strike a balance between energy efficiency and performance are more suitable, as reported in the literature such as \cite{Bio_7394105, Bio_9349090, Bio_8786920}. In \cite{Bio_9735144}, a novel hybrid precoding architecture named \ac{DPP} was introduced to mitigate the issue of beam squint in \ac{THz} communications systems. By incorporating a time delay network between digital and analog precoding, \ac{DPP} generates frequency-dependent beamforming vectors. Similarly, Gao \textit{et al.} \cite{Bio_9398864} proposed a \ac{TTD}-based hybrid beamforming that aims to address beam squint through virtual subarrays, as first presented in \cite{Bio_9500455}. The proposed algorithm achieves near-optimal performance as that of full-digital precoding. 

In order to address the limitation of \ac{TTD}, Nguyen and Kim \cite{Bio_9839132} proposed a hybrid beamforming scheme that takes into account the relationship between the number of antennas and the required delay for \ac{TTD}. They also carried out joint optimization under limited delay to create an optimal  compensation scheme for beam squint. It is noted that most research work, as mentioned above, focused on \ac{2D} hybrid beamforming, which is primarily designed for \ac{ULA}. However, \ac{ULA}s may not be suitable for \ac{UMMIMO} systems due to limited antenna aperture. In contrast, \ac{UPAs} that can accommodate a large number of elements compactly, are more potential for deploying \ac{UMMIMO} systems. 
There is a lack of research on beam squint compensation in \ac{3D} hybrid beamforming using \ac{ULA}s for \ac{THz} broadband \ac{UMMIMO} systems. Responding to this, the authors of \cite{Ref_wu20223dhybrid}  proposed a \ac{3D} beamforming architecture by leveraging two-tier \ac{TTD}, which is able to combat the beam squint effect from both the horizontal and vertical directions. The impact of the array structure on the beam squint has been analyzed in \cite{Ref_nguyen2022beam}.

\subsection{Lens Antenna Array}
Despite its high potential at the \ac{THz} band, hybrid beamforming is still confined by high hardware and power costs due to the use of many analog phase shifters. Some studies \cite{Ref_payami2016hybrid} demonstrate that the power consumed by phase shifters becomes critical. In this context, a disruptive antenna technology called \emph{lens antenna} \cite{Ref_quevedoteruel2018lens} has been studied in recent years.  

James Clerk Maxwell predicted the existence of \ac{EM} waves  in 1873 and inferred that visible light is one type of  \ac{EM} waves. To verify Maxwell's theory, early scientists who believed a radio wave is a form of invisible light, concentrated on duplicating classic optics experiments into radio. Heinrich Hertz proved the existence of  \ac{EM} waves and also first demonstrated the refraction phenomena of radio waves at \SI{450}{\mega\hertz} using a prism. These experiments revealed the possibility of focusing radio waves on a narrow beam as visible lights through an optical lens. In 1894, Oliver Lodge \cite{Ref_lodge1923origin} successfully used an optical lens to concentrate \SI{4}{\giga\hertz} radio waves. In the same year, Indian physicist Jagadish Chandra Bose \cite{Ref_mhaske2022bose} built a cylindrical sulfur lens to generate a beam in his microwave experiments over \SIrange{6}{60}{\giga\hertz}. In 1894, Augusto Righi at the University of Bologna focused radio waves at \SI{12}{\giga\hertz} with \SI{32}{\centi\meter} lenses. In World War II, the race of developing radar technology fostered the emergence of modern lens antennas. Used as a radar reflector, the famous Luneberg lens \cite{ Ref_pfeiffer2010printed} was invented in 1946, which is also attached to stealth fighters nowadays to make it detectable during training or to conceal their true  \ac{EM} signature.

As refracting visible light by an optical lens, a lens antenna uses a shaped piece of radio-transparent material to bend and concentrate  \ac{EM} waves \cite{Ref_bosiljevac2012nonuniform}.  It usually comprises an emitter radiating radio waves and a piece of dielectric or composite material in front of the emitter as a converging lens to force the radio waves into a narrow beam. Conversely, the lens directs the incoming radio waves into the feeder in a receive antenna, converting the induced electromagnetic waves into electric currents. To generate narrow beams, the lens needs to be much larger than the wavelength of the  \ac{EM} wave \cite{Ref_milne1982dipole}. Hence, a lens antenna is more suitable for \ac{mmWave} and \ac{THz} communications, with tiny wavelengths. Like an optical lens, radio waves have a different speed within the lens material than in free space so the varying lens thickness delays the waves passing through it by different amounts, changing the shape of the wavefront and the direction of the waves.

On top of lens antennas, an advanced antenna structure referred to as a lens antenna array has been developed \cite{Ref_zeng2016millimeter}. A lens antenna array usually consists of two major components: an  \ac{EM} lens and an array with antenna elements positioned in the focal region of the lens.  \ac{EM} lenses can be implemented in different ways, e.g., dielectric materials, transmission lines with variable lengths, and periodic inductive and capacitive structures. Despite its various implementation, the purpose of  \ac{EM} lenses is to provide variable phase shifting for electromagnetic waves at different angles \cite{Ref_yang2021communication}. In other words, a lens antenna array can direct the signals emitted from different transmit antennas to different beams with sufficiently separated angles of departure. Conversely, a lens antenna array at the receiver can concentrate the incident signals from sufficiently separated directions to different receive antennas \cite{Ref_zeng2017cost}.

Recent research work reported a few high-gain \ac{THz} lens antennas, such as dielectric or metallic lens antennas \cite{Bio_7859330, Bio_7987018, Bio_9131807}. Dielectric lens antennas have been demonstrated with high gain and wide operating bandwidth by integrating the dielectric lens with a standard rectangular waveguide feed \cite{Bio_7859330} or a leaky-wave feed \cite{Bio_7987018}. But their radiation efficiency needs to be improved. On the other hand, metallic lens antennas have no dielectric loss, making them suitable for \ac{THz} communications and sensing. In \cite{Bio_9131807}, a high-gain \ac{THz} antenna using a metallic lens composed of metallic waveguide delay lines was reported. For wideband signal transmission, recently, the authors of \cite{Bio_xiong2022wideband} presented a fully metallic lens antenna with a wide impedance bandwidth and high gain at the D band from \SIrange{110}{170}{\giga\hertz}. A ﬂared H-plane horn is used to achieve a large H-plane radiation aperture to further increase the radiation gain.

The deployment of \ac{MMIMO} systems entails the challenges associated with a huge amount of antenna elements \cite{Bio_6798744}.  \ac{EM} lens arrays with a reasonable number of elements can lower the required number of antennas and corresponding \ac{RF} chains while maintaining high beamforming gain. However, dielectric  \ac{EM} lenses are difﬁcult to integrate with multiple antenna techniques due to their bulky size, high insertion loss, and long focal lengths to control the beam gain. The metallic lens antennas are defined as artificial composites that obtain electrical properties from their structure rather than their constituent materials. Some studies on metallic lens antennae have been done to achieve beam gain from a single antenna, like \cite{Bio_8974617}.  Lee \emph{et al.}  proposed a large-aperture metallic lens antenna designed for multi-layer \ac{MIMO} transmission for \ac{6G}, demonstrating that a single large-aperture metallic lens antenna can achieve a beam gain of up to \SI{14}{\decibel} compared to the case without a lens. By adopting the proposed large-aperture metallic lens antenna,  system-level simulations show that the data throughput of the user equipment is effectively increased \cite{Bio_9707823}.

\subsection{Beam Alignment: Training of Beams}
The utilization of the promising \ac{THz} spectrum range is hindered by significant propagation losses imposed on its frequencies. To counteract the losses, large antenna arrays, such as \ac{UMMIMO} and lens antenna array discussed above, can be employed, but this leads to highly directional and narrow beams \cite{Ref_Yuhang2021Razor}. To ensure a satisfactory \ac{SNR} at the receiver and prevent connection loss, it is critical to maintaining degree-level alignment between the transmitter and receiver beams.
Therefore, beam alignment is a critical issue that must be addressed for establishing a reliable connection. This is accomplished by aligning the beams at the transmitter and receiver to the direction of the channel paths, where \ac{CSI} is critical for implementing the fine alignment \cite{Ref_Bile2019Power}. However, traditional channel sensing methods used at lower frequencies are not feasible at \ac{THz} frequencies due to the significant path losses that render pilot signals undetectable during the link establishment stage. 

\begin{figure}
\centering
\includegraphics[width=0.49\textwidth]{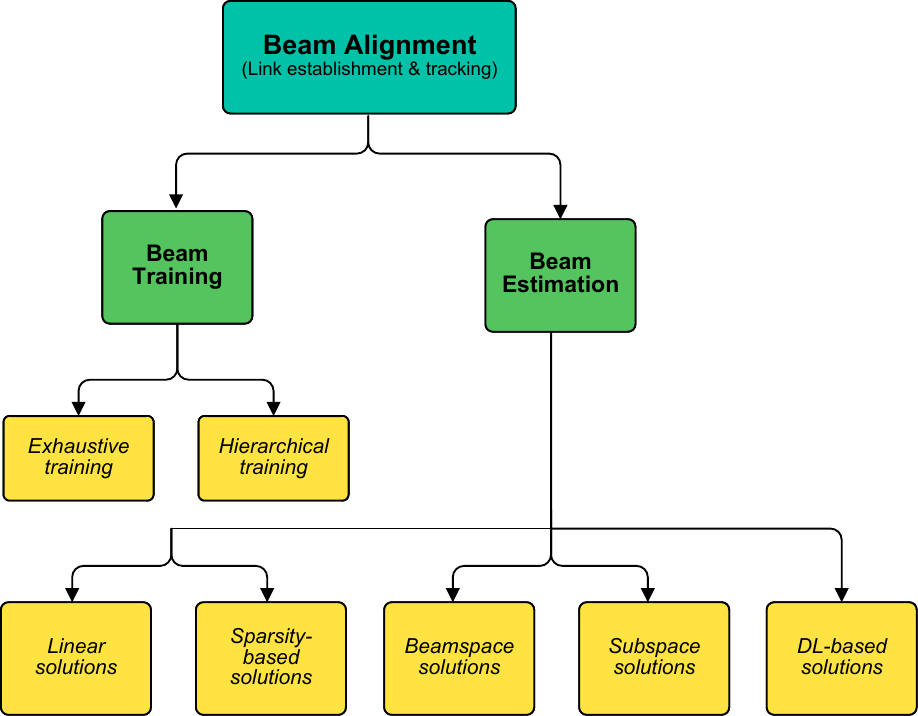}
\caption{Illustration of beam alignment techniques.} 
\label{Diagram_beamalignment}
\end{figure}

Significant research efforts have been made in recent years to understand the unique characteristics of the \ac{THz} channel and to develop appropriate beam alignment algorithms. These efforts aim to establish beam alignment during the link establishment stage \cite{Ref_jiang2022initial, Ref_jiang2022initial_ICC} and to maintain alignment while the beams are in motion (beam tracking). Two categories of beam alignment techniques have been identified: \textit{beam training} \cite{Ref_Zhenyu2016Hierarchical} and \textit{beam estimation} \cite{Ref_Yuhang2022Millidegree}, as shown in \figurename~\ref{Diagram_beamalignment}. The former involves transmitting known signals and adjusting beamforming parameters to align the beams. The latter involves estimating \ac{CSI} from received signals and using it to refine the beamforming parameters. This part surveys the state-of-the-art advances in beam tracking techniques, while beam estimation is discussed in the subsequent part.

Beam training involves scanning the channel with directional beams from a codebook to determine the beam pair at the transmitter and receiver that results in the highest \ac{SNR} of the received signal \cite{Ref_Sooyoung2013Millimeter}. Beam training can be broadly classified into two categories: exhaustive and hierarchical training, which are discussed as follows:

\subsubsection{Exhaustive Training} 
Many studies have adopted exhaustive training, which involves sequentially probing all the predefined directions in the codebook to find the optimal beam pair that maximizes the \ac{SNR} at the receiver \cite{Ref_Junyi2009Beam}. The approach is used in the IEEE 802.15.3c standard \cite{IEEE802153c}. However, the method is time-consuming and not practical at \ac{THz} frequencies, where beams from a large-scale antenna array tend to be very narrow, making it difficult to scan the entire space in a reasonable amount of time. Additionally, the accuracy of the training is limited by the codebook's resolution. 

Responding to these limitations, Tan and Dai  \cite{Ref_Jingbo2020Wideband} have improved the exhaustive search by exploiting the beam split or squint effect. They used delay-phase-precoding architecture to control the beam split effect and accelerate the tracking process \cite{Ref_Jingbo2019Delay}. By doing so, the split beams have wider angular coverage and can scan many directions in a single shot. Another approach from the \ac{RF} domain used \ac{LWA} at the transmitter and receiver of a \ac{THz} link to estimate the \ac{AoD} and \ac{AoA} \cite{Ref_Yasaman2020Single}. The angular radiation of an \ac{LWA} is frequency-dependent, and the received spectral peak determines the \ac{AoD}. The bandwidth of the received signal is proportional to the rotation angle over the \ac{AoA} of the \ac{LWA} receiver, which speeds up the channel scanning process but requires additional hardware at both the transmitter and receiver.

\subsubsection{Hierarchical Training} From a practical point of view, many studies have adopted hierarchical training to reduce training overhead \cite{Ref_Zhenyu2016Hierarchical,Ref_Song2017Multi,Ref_Junyi2009Beam,Bio_7959180,Ref_Renmin2018Coordinated}. Hierarchical algorithms are based on multi-resolution codebooks, which contain wide beams at lower levels and narrow beams at higher levels. The search begins at the lowest level and gradually moves to higher levels to find the optimal narrow beam. In \cite{Bio_7959180}, the authors proposed a subarray-based multi-resolution codebook, where beams at each level are generated by the contribution of all subarrays. In \cite{Ref_Renmin2018Coordinated}, the authors proposed an accelerated hierarchical training that concurrently scans angular space with different \ac{RF} chains. The authors of \cite{Ref_Yuh2021High} proposed a multi-modal beam pattern-based training that simultaneously radiates beams targeting at different directions using a single \ac{RF} chain. The equally spaced activation approach has been proposed to generate the steering vector for multiple beam radiation. However, the loss at \ac{THz} frequencies may render the training algorithm ineffective. The authors in \cite{Ref_Bile2019Power} and \cite{Ref_Bile2014Fast} adopted hierarchical training that utilizes the power-angular spectral correlation between sub-6GHz and \ac{THz} frequencies. 

In \cite{Ref_Giorgos2020A}, Stratidakis \emph{et al.} proposed a localization-aided hierarchical beam tracking algorithm that uses location information to reduce pilot overhead. This algorithm assumes the linear motion of a user, which may not be accurate in realistic scenarios. In \cite{Ref_Boyu2022A}, the authors proposed a unified \ac{3D} training and tracking framework based on a \ac{3D} hierarchical codebook built upon the quadruple-uniform-planar-array architecture. This proposal offers two advantages: a unified framework for training and tracking and \ac{3D} space coverage compared to \ac{2D} space coverage in most existing works. The training overhead of hierarchical algorithms is much lower compared to exhaustive ones. However, hierarchical  algorithms suffer from a high overhead of feedback messages required for coordination between the transmitter and receiver. The number of levels in multi-resolution codebooks also leads to higher training overhead, which may not enhance performance, especially in multi-hop \ac{THz} links \cite{Ref_Arian2021Reinforcement}. A new approach is proposed in \cite{Ref_Yifei2021Hierarchical}, where it is based on a multi-armed bandit algorithm and utilizes prior knowledge of channel frequency-selective fading. This algorithm is designed with a hierarchical structure to accelerate the beam alignment process. 

\subsection{Beam Alignment: Estimating of Beams}
Beam estimation is a method of acquiring channel information with the goal of reducing training overhead when compared to beam scanning techniques. The estimation process  begins with initial training, which involves collecting channel measurement. These measurement are then processed to derive the angular information of the target channel. Prior studies have proposed a variety of algorithms, based on linear estimation \cite{Ref_Suhanya2017Robust,Ref_M2015Least}, \ac{ComS}-based sparse estimation \cite{Ref_Tobias2021Angle,Ref_Mingyao2021Near,Ref_Roi2015Channel,Ref_Viktoria2018Compressive,Ref_Jie2020Fast}, beamspace-based estimation  \cite{Ref_Xinyu2017Fast,Ref_Giorgos2019A}, subspace-based estimation  \cite{Ref_Anwen20172D,Ref_Yuhang2020Millidegree,Ref_Yuhang2022Millidegree, Ref_Anwen2021Angle}, or deep learning-based estimation \cite{Ref_Yuhang2020Deep,Ref_Shuai2019Deep,Ref_Yuhang2022Millidegree}.

\subsubsection{Linear Estimation}
The authors of  \cite{Ref_Suhanya2017Robust} used an extended Kalman filter, which is a well-known example of linear estimation, to perform beam tracking for a \ac{MS}. The \ac{MS} sends training symbols over the uplink during each time slot, and the extended Kalman filter-based algorithm at the \ac{BS} iteratively estimates the channel parameters (the path gain, \ac{AoD}, and \ac{AoA}) from the observed signal. The proposed algorithm achieves milli-degree level angle estimation with moderate mobility of the \ac{MS} and antenna array size. However, the study assumes the \ac{BS} is equipped with a fully digital beamformer, which is not practical due to the high power consumption imposed by a large number of \ac{RF} chains. Additionally, the study assumes that the \ac{MS} is parallel to the \ac{BS} during tracking, such that \ac{AoD} equals \ac{AoA}, which is not realistic because the orientation of the \ac{MS} can be arbitrary in real-world scenarios. Other linear methods, such as maximum likelihood and least squares, can also be applied \cite{Ref_M2015Least}. However, these estimators require a large number of observations and do not exploit the sparsity feature of \ac{THz} channels.

\subsubsection{Compress Sensing-based Sparse Estimation}
The sparsity property of \ac{THz} channels can significantly reduce the computational complexity of beam estimation algorithms by transforming the problem into a sparse recovery problem. A technique referred to as \ac{ComS} is considered an optimal approach to solving these problems, as discussed by the authors in \cite{Ref_Viktoria2018Compressive}. They analyzed two \ac{ComS}-based algorithms, i.e., \textit{greedy compressive sampling matching} and \textit{Dantzig selector-based method}, for solving convex programs. The results show that the \ac{ComS}-based methods have higher accuracy compared to linear estimation based on least squares. The authors in \cite{Ref_Tobias2021Angle} utilized \ac{ComS}-based techniques to accelerate the training proposed in \cite{Ref_Bile2019Power}. In this approach, the estimated angles from wide beams in the first stage are refined using an L1-norm regularized least squares method to obtain accurate estimates of \ac{AoD} and \ac{AoA}, reducing the scope of the narrow beam search in the second stage. 

Yang {\it et al.} \cite{Ref_Jie2020Fast} proposed an orthogonal matching pursuit-based fast algorithm to estimate the \ac{AoA} and \ac{AoD} of a \ac{BS}-\ac{MS} link. This study considered the cost and power consumption of adopting an auxiliary fully-digital array for channel estimation and evaluated the effect of \ac{RF} imperfections and low-resolution \ac{ADC} per \ac{RF} chain. The study adopts the virtual channel model, which assumes that the \ac{AoA} and \ac{AoD} are discretely distributed over a spatial grid utilizing the angular sparsity of the \ac{THz} channel, making it a sparse recovery problem suitable for \ac{ComS}-based algorithms. However, this discrete grid assumption reduces the estimation accuracy due to the grid resolution. To mitigate this limitation, the authors in \cite{Ref_Chen2018Super} proposed an iterative reweight-based super-resolution estimation scheme, which optimizes the on-grid estimation iteratively to converge to neighboring off-grid angles. Simulation results show that the off-grid solution has an improved accuracy and spectral efficiency compared to on-grid solutions. Similarly, the authors in \cite{Ref_Matilde2021Gridless} proposed a gridless \ac{ComS}-based algorithm to estimate the \ac{AoA} for arbitrary \ac{3D} antenna arrays, eliminating the quantization error of the grid assumption. 

The mentioned \ac{ComS}-based estimators are generally built on the assumption of angular sparsity of \ac{THz} channels, which holds in the far field but  is not valid in the near field. Therefore, the work in \cite{Ref_Mingyao2021Near} considers \ac{ComS}-based estimation in the near field where the angular sparsity assumption does not hold. The results show that the channel in the near field exhibits polar sparsity rather than angular sparsity, which was exploited by a \ac{ComS}-based polar-domain simultaneous orthogonal matching pursuit algorithm.

\subsubsection{Beamspace-based Estimation}
As we know, the \ac{MIMO} beamspace channel can be realized through the use of a \ac{DLA}. Such arrays function as passive phase shifters that steer beams towards specific directions based on the incident point to the lens aperture \cite {Ref_Yae2018RF}. The number of these directions is limited by the number of antenna elements, resulting in a beam-sparse channel. This artificially created sparsity reduces the pilot overhead required for channel estimation compared to conventional methods. In \cite{Ref_Xinyu2017Fast}, the authors adopted the \ac{DLA}-based \ac{MIMO} system architecture to create the \ac{MIMO} beamspace channel and utilized its sparsity for fast channel tracking. A \textit{priori} information-aided tracking scheme was proposed for \ac{MIMO} beamspace systems, where the channel was conventionally estimated in the first three time slots and the physical direction of the \ac{MS} was then derived based on a temporal variation law. The estimated physical direction was used to determine the support of the beamspace channel, which corresponds to the dominant beam directions. However, the estimation accuracy depends strongly on the localization accuracy. In \cite{Ref_Giorgos2019A}, the authors extended the work in \cite{Ref_Xinyu2017Fast} and proposed a cooperative localization-aided tracking algorithm with multiple \ac{BS}s, each equipped with a \ac{DLA}. These \ac{BS}s cooperate to accurately localize the \ac{MS} for improved channel tracking. While beamspace \ac{MIMO} solutions significantly reduce the overhead in comparison with that of conventional estimation methods, their accuracy may be limited by the discrete directions generated by the \ac{DLA} and restricted numbers of beams.

\subsubsection{Subspace-based Estimation}
When estimating continuously distributed \ac{AoA} and \ac{AoD}, another way referred to as subspace-based algorithms can be performed, with the aim of avoiding the estimation error caused by the sparse solutions or beam sparsity in beamspace solutions. In general, these algorithms collect channel measurement and identify the eigenvectors that correspond to the signal subspace. Two widely known algorithms, i.e., MUSIC (MUltiple SIgnal Classification) and ESPRIT (Estimation of Signal Parameters via Rotational Invariance Technique), belong to subspace-based estimation \cite{Ref_Anwen20172D, Ref_Yuhang2022Millidegree}.

In \cite{Ref_Yuhang2022Millidegree}, the authors adopted the MUSIC algorithm to achieve millidegree-level \ac{AoA} estimation. This study utilized a hybrid \ac{AoSA} architecture and collected measurement data by probing random steering vectors. The covariance matrix of the measurement data was then calculated and decomposed into signal and noise subspaces.  The \ac{AoA} was estimated by searching the MUSIC pseudospectrum function for vectors orthogonal to the noise eigenvectors. The estimation was further refined by collecting new measurement based on the coarse estimated angles. In \cite{Ref_Anwen20172D}, the ESPRIT algorithm was adopted for super-resolution channel estimation, which involved multiple steps such as spatial smoothing, forward-backward averaging techniques \cite{Ref_AJ1998Joint}, singular value decomposition (SVD), and joint diagonalization. While subspace solutions show improved performance compared to sparse solutions, their computational complexity is significantly higher.

\subsubsection{Deep Learning-based Estimation}
Recently, \ac{DL}-based techniques have emerged as a promising alternative to replace conventional estimators. \ac{DL}-based solutions are particularly effective for complex multi-user scenarios where the input and output of the channel are not directly related. In \cite{Ref_Yuhang2022Millidegree}, a branch of DL referred to as \ac{DCNN} is used to estimate the \ac{AoA} of a multipath channel. The measurement matrix is collected through random beamforming and combining matrices at the transmitter and receiver, respectively. The measured data is fed into the neural network. Three convolutional layers extract the spatial peculiarities of the channel, and two fully-connected layers capture the non-linear relationship between these peculiarities and \ac{AoA} estimation. The results show higher estimation accuracy compared to that of the subspace-based MUSIC algorithm at high \ac{SNR}. 

In \cite{Ref_Yuhang2020Deep}, a \ac{DCNN} architecture is used to estimate the near-field channel under the spherical wave propagation model. The proposed \ac{DCNN}-based approach addresses the spherical wave propagation model by considering the inter-subarray phase error as an output parameter of the network.
In \cite{Ref_Shuai2019Deep}, \ac{DKL} combined with \ac{GPR} is used to estimate the indoor \ac{THz} channel in a multi-user scenario. In particular, a \ac{DNN} is trained to capture the non-linear relationship between the input and output of the channel. The results show that this \ac{DNN}-based solution outperformed the \ac{MMSE} and least squares-based linear estimators. 

Prior studies have demonstrated the superiority of \ac{DL}-based solutions over the conventional solutions in complex scenarios. It also revealed that \ac{DL} networks need lots of computational and storage resources,  intensive offline training, and validation. Moreover, their efficiency in low \ac{SNR} scenarios requires further investigation. In order to achieve fast initial access in wireless networks, a \ac{DNN} framework called DeepAI has been proposed, which maps the \ac{RSS} to identify the optimal beam direction \cite{Ref_Tarun2022Deep}. The authors have introduced a sequential feature selection (SFS) algorithm that selects efficient and reliable beam pairs for DeepAI's inputs in \ac{LoS} \ac{mmWave} channels. However, the SFS algorithm fails to improve the accuracy and performance of DeepAI in \ac{NLoS} scenarios. The simulation results show that DeepAI outperforms the conventional beam-sweeping method. Another DL-based beam selection algorithm is suitable for \ac{5G} NR has been proposed by the authors in \cite{Ref_Min2020Deep}.


\section{\ac{THz} Systems and Networks Toward \ac{6G} and Beyond}
The ongoing research and development of the \ac{6G} system is set to revolutionize the way that various domains and layers of a mobile network interact and communicate with each other and with authorized third parties \cite{Ref_Habibi2021Towards, Ref_Habibi2022Enabling}. As we stated numerous times throughout this paper, one of the key enablers of \ac{6G} is \ac{THz} communications and sensing, which promise to deliver ultra-high data rates, ultra-low-latency connectivity, high-resolution sensing, and high-accuracy positioning in the coming decades. Nevertheless, the full potential of \ac{THz} communications and sensing can only be realized through its integration with other emerging technologies. 

This section explores \ac{THz} networks from a systematic point of view, with an emphasis on the synergy of \ac{THz} communications and sensing with a variance of potential \ac{6G} technologies, including \acl{MMIMO}, \acl{UMMIMO}, \acl{NOMA}, \acl{RIS}, non-terrestrial networks, digital twins, \acl{AI}m \acl{ML}. Moreover, we discuss security, localization, joint communications and sensing,  multi-connectivity, and channel awareness for \ac{THz} systems and networks. By examining these synergies, we hope to shed light on the most significant research challenges and opportunities facing the development and deployment of  \ac{THz} communications and sensing in \ac{6G} and beyond networks, as well as the potential benefits for future applications and services.

\subsection{\ac{THz}-\ac{MMIMO} Systems and Networks} 
Compared to lower frequencies at sub-6 \si{\giga\hertz} and \ac{mmWave}, the \ac{THz} bands have a much smaller signal wavelength, which leads to a tiny size of antenna (i.e. a larger number of antennas within the same surface area) and narrower beams. Both the factors are beneficial for \ac{MMIMO} and grant a great potential in the \ac{THz} band than at lower frequencies. For example, Akyildiz and Jornet demonstrated that an \ac{UMMIMO} system operating in 1 \ac{THz} with the dimension of $1024\times 1024$ is realized by the arrays that have a size of only \SI{1}{\milli\meter^2}~\cite{Ref_akyildiz2018combating}. 
However, the application of \ac{MMIMO} and/or \ac{UMMIMO} in practical \ac{6G} \ac{THz} communications and sensing are challenged in various aspects. In addition to the barriers in the fabrication of nano-antenna arrays, the complexity and sparsity of \ac{THz} channels are also limiting the exploitation of \ac{MMIMO} in this band. Accurate channel models, \ac{PHY} layer enabling technologies, as well as novel link layer design, are needed to release the full potential of \ac{THz} \ac{MMIMO}.

A lot of works in modeling \ac{THz} \ac{MMIMO} channels have been reported since the late 2010s, overwhelmingly with the ray-tracing methodology. Han \emph{et al.} proposed in \cite{Ref_han2018ultra} a model for \ac{UMMIMO} channels over a distance to \SI{20}{\meter} and in the frequency windows of \SIrange{0.3}{0.4}{\tera\hertz} and \SIrange{0.9}{1}{\tera\hertz}. Busari \emph{et al.} studied the \SI{0.1}{\tera\hertz} \ac{MMIMO} channel in \cite{Ref_busari2019terahertz} a specific outdoor street-side scenario, investigating the impacts of precoding scheme, carrier frequency, bandwidth, and antenna gain on the system regarding spectral and energy efficiencies. Sheikh \emph{et al.} focused on the critical features of rough surface reflection and diffuse scattering at \ac{THz} frequencies, and proposed in \cite{Bio_8859609} a \ac{3D} indoor model for \SI{0.3}{\tera\hertz} and \SI{0.35}{\tera\hertz} \ac{MMIMO} channels with different surface roughness levels, considering both \ac{LoS} and \ac{NLoS} scenarios. In the recent work \cite{Bio_9591285}, Tarboush \emph{et al.} reported their channel model for wideband \ac{UMMIMO} \ac{THz} communications and a simulator based thereon.

Efforts on the physical layer deal with the problem of beamforming and combining from the perspectives of beam training and beam tracking, i.e.,  finding the best beam pattern and online adjusting it, in order to obtain the best link quality and maintain it against the time variation of the channel. As outlined by Ning \emph{et al.} in their tutorial~\cite{Ref_ning2022prospective}, there are two basic principles of beamforming: the precoding/decoding that is executed in the digital domain, and the beam steering that works in the analog domain. Each of the principles, as well as their hybrid, when applied in the wideband \ac{THz} systems, must be carefully designed to address two main issues: the spatial-wideband effect that different antennas receive different symbols at the same time, and the frequency-wideband effect that the beam pattern of a phase array codeword changes with the frequency of the signal (a.k.a. the beam squint or beam split). While most existing methods leverage either the digital precoding approach \cite{Bio_9735144, Ref_gao2021wideband}, or the precoding/steering hybrid \cite{Ref_Xinyu2017Fast, Ref_dovelos2021channel, Bio_9398864, Ref_chen2022millidegree}, new research interests in the steering approach based on \ac{RIS} are arising \cite{Ref_wan2021terahertz, Ref_ning2021terahertz}. Whilst higher layer design has not been a major research focus of \ac{MMIMO} \ac{THz} systems so far, there is pioneering work on multi-access scheme reported in \cite{Bio_7913686}.

\subsection{\ac{THz}-\ac{NOMA} Systems and Networks} 
Another promising \ac{RAN} technology for enabling \ac{THz}-\ac{MIMO} systems and networks is \ac{NOMA}, which allows allocating of the same radio resources to more than one user simultaneously and invokes the so-called \ac{SIC} approach on the receiver side to decode the information for different users successively \cite{Ref_jiang2023orthogonal}. Compared to lower frequencies, the low-rank channels in the \ac{THz} bands can be much more correlated because of the limited-scattering transmission, which reduces the channel orthogonality between different users and makes \ac{NOMA} a promising technique to improve the spectral efficiency~\cite{Ref_liu2022developing}. 
Serghiou \emph{et al.} \cite{Ref_serghiou2022terahertz} believe that in \ac{LoS} scenarios where spatial processing approaches fail to separate the users from each other, a combination of \ac{NOMA} with \ac{UMMIMO} can provide more fair user access in terms of resource allocation, and therewith achieve better spectral efficiency of the overall network. Meanwhile, with proper resource allocation algorithm, \ac{NOMA} can also enhance the energy efficiency in \ac{THz} communications and sensing systems~\cite{Ref_zhang2021energy}.

To evaluate the feasibility of \ac{MIMO}-\ac{NOMA} systems in \ac{THz} bands, Sabuj \emph{et al.} proposed in \cite{Ref_sabuj2022machine} a \ac{FBL} channel model and therewith evaluated the system performance regarding \ac{CMTC} scenarios.
In contrast to its good performance in \ac{LoS} scenarios, \ac{THz}-\ac{NOMA} performs much more poorly when the connected devices are blocked by obstacles~\cite{Ref_chen2021towards}. To address this issue, \ac{RIS} appears as a promising solution. In \cite{Ref_xu2022graph}, Xu \emph{et al.} proposed a smart \ac{RIS} framework for \ac{THz}-\ac{NOMA}, which delivers significant enhancements in the system energy efficiency and the reliability of super-fast-experience users. The principle of \ac{NOMA} requires users to be paired/clustered for sharing radio resources, and relies on an appropriate clustering to achieve satisfactory system performance. Shahjala \emph{et al.} comparatively reviewed the user clustering techniques for \ac{MMIMO}-\ac{NOMA} \ac{THz} systems in \cite{Ref_shahjala2021user}, and proposed a fuzzy C-means-based clustering approach in \cite{Ref_shahjala2021fuzzy}. 

It should be noted that many popular clustering policies are tending to pair a user with good channel, called \ac{CCU}, to another with poor channel, called \ac{CEU}. Such policies lead to a gain in the spectral efficiency of the overall system, but a degradation at  the \ac{CEU} due to power splitting. To address this issue, Ding \emph{et al.} proposed in \cite{Ref_ding2015cooperative} a \ac{CNOMA} scheme where the \ac{CCU} always forwards the message for \ac{CEU} that it obtains during the \ac{SIC}, so that the performance loss at \ac{CEU} is compensated. However, this design is forcing the \ac{CCU} to work as a relay which drains its battery. Therefore, \ac{SWIPT} is often introduced into \ac{CNOMA} systems so that the \ac{CCU} is able to harvest energy from the radio signal to support the relaying \cite{Ref_parihar2021performance}. At \ac{THz} frequencies, due to the high spreading loss and atmospheric absorption, the power propagation loss is more critical than that at lower frequencies, and the \ac{SWIPT}-assisted \ac{CNOMA} solution can be more important. Oleiwi and Al-Raweshidy analyzed the performance of \ac{SWIPT} \ac{THz}-\ac{NOMA} in \cite{Ref_oleiwi2022cooperative}, and correspondingly designed a channel-aware pairing mechanism \cite{Ref_oleiwi2022cooperative}.

\subsection{\ac{THz}-\ac{RIS} Systems and Networks}\label{Subsec:RISforTHz} 
The \ac{6G} and beyond systems will be revolutionized by the tremendous potential of \ac{RIS} \cite{Ref_jiang2023performance, Ref_jiang2022intelligent, Ref_jiang2023capacity} and \ac{THz} \cite{Ref_Pan2021UAV}, the two cutting-edge enablers for the access domain of a futuristic communications and sensing network. The synergy between \ac{RIS} and \ac{THz} lies in the fact that \ac{RIS} can be utilized to improve the performance of \ac{THz} systems by providing a cost-effective solution to the propagation challenges associated with \ac{THz} frequencies \cite{chen2021terahertz}. By utilizing the reconfigurability and versatility features of RIS, it is possible to address the challenges of \ac{THz} wave propagation, especially the use of bypassing the blockage of \ac{THz} beams \cite{Ref_jiang2022dualbeam}, thereby improving the overall performance of \ac{THz} communications and sensing. 

By controlling the phase, amplitude, and polarization, \ac{RIS} can effectively steer, reflect, and amplify electromagnetic waves in \ac{THz} systems and networks. Consequently, it enables a vast array of applications and use-case scenarios, including beamforming, wireless power transfer, and indoor localization, among others. In addition, by employing RIS-assisted spatial modulation, the \ac{THz}-\ac{RIS} systems and networks have the potential to dramatically improve their spectral efficiency. More importantly, \ac{RIS} can be used to generate virtual channels that compensate for the propagation losses of \ac{THz} waves so that \ac{SNR} and the coverage area of \ac{THz} communications and sensing could be increased. 
By far, the intersection of the \ac{RIS} and \ac{THz} has been intensively studied in the literature. There exist a number of overview and survey papers that provide an insight into such synergies, including \cite{Ref_Hao2022Ultra, Ref_Yang2022Terahertz, Ref_Xu2022Reconfigurable, Ref_sarieddeen2021overview}. In addition to the aforementioned surveys, we discovered during our research that the intersection between these two technologies has been dramatic, including in the context of massive MIMO, millimeter wave, 3D beamforming, satellite networks, and many others. Moreover, a large number and various types of physical layer-related optimization problems have been jointly investigated. 

\subsection{\ac{THz}-Aided Non-Terrestrial Networks} 

With its ambition of ubiquitous \ac{3D} coverage, \ac{6G} and beyond are envisioned to include different non-terrestrial infrastructures, such as \ac{UAV}, \ac{HAP}, \ac{LEO} satellites, and \ac{GEO} satellites, as an indispensable part of its architecture. Since the air/space channels and air/space-to-ground channels are less subject to blockages w.r.t. terrestrial channels, the \ac{LoS} link availability is much higher, implying a vast potential for \ac{THz} communications and sensing~\cite{Ref_geraci2022what}. On the one hand, the tremendous amount of spectral resources offers the feasibility of efficient interconnection among these terrestrial, air, and space platforms through \ac{THz} communications links. On the other hand,  non-terrestrial infrastructures enable the flexible deployment of a variety of \ac{THz} sensing equipment at favorable altitudes and places.   

Nevertheless, the practical deployment of \ac{THz}-\ac{NTN} is still facing various technical challenges, which include but are not limited to the feasibility assessment of \ac{THz} frequencies for space-to-earth links, transceiver implementation, and accurate \ac{NTN} platform positioning~\cite{Ref_araniti2022toward}. Regarding the characterization of \ac{THz}-\ac{NTN} channels, the authors of \cite{Ref_you2022propagation} proposed an analytical propagation model for low-altitude \ac{NTN} platforms such as \acp{UAV} in the frequency range \SIrange{0.275}{3}{\tera\hertz}, while the authors of \cite{Ref_aliaga2023cross}  modeled the cross-link interference for \ac{LEO} satellites.  The use of satellites to serve air planes on the THz band and related channel models have been analyzed in \cite{Ref_kokkoniemi2021channel}. Concerning the \ac{THz} transceiver implementation, \ac{NTN} systems pose high antenna design requirements. For example, the antennas are supposed to produce multiple high-gain beams to support dynamic networking and realize long-range communications. There are various approaches towards this aim, which are well summarized in the survey by Guo \emph{et al.}~\cite{Ref_guo2021quasi}. For \ac{THz} CubeSat networks, the antennas are required to provide sufficient beamwidth angle to enable faster neighbor discovery, while simultaneously providing a high gain to overcome the path loss. To fulfill these requirements, Alqaraghuli \emph{et al.} designed a two-stage Origami horn antenna~\cite{Ref_alqaraghuli2022compact}.

On the \ac{PHY} layer, digital signal processing techniques are studied to overcome the limitations of the analog front end in \ac{THz} transceivers. Tamesue \emph{et al.} proposed to deploy digital predistortion in \ac{RF} power amplifiers of \ac{THz}-\ac{NTN} systems to compensate for nonlinear distortion \cite{Ref_tamesue2022digital}. In \cite{Ref_kumar2022dnn}, Kumar and Arnon reported a \ac{DNN} beamformer to replace the phase shifters in \ac{THz}-\ac{MMIMO} antenna arrays for wideband \ac{LEO} satellite communication.
It also creates additional benefits for \ac{NTN} by deploying \ac{THz} communications and sensing in conjunction with other novel enabling technologies. For example, \ac{RIS} can contribute to the deployment of \ac{THz} in future integrated terrestrial/non-terrestrial networks by means of enhancing the beamforming~\cite{Ref_ramezani2022toward}. By leveraging the \ac{ISAC} technology, the \acp{DAR}, which are traditionally used for weather sensing, can be granted an extra capability of communicating with \ac{LEO} satellites~\cite{Ref_aliaga2022joint}.

\subsection{Digital Twin-Aided \ac{THz} Systems and Networks}\label{Subsec:DTforTHz} 
The digital-twin technology \cite{Ref_Kuruvatti2022Empowering} is an emerging novel concept (which is also considered to be a key enabler of the \ac{6G} and beyond systems) in which a virtual replica of a physical system, object, process, network, or link is created through employing accurate data collected in real-time \cite{Ref_Han2023Digital}. It enables the autonomous control, intelligent monitoring, and accurate self-optimization of physical networks, processes, and systems in a fully virtualized environment. To our knowledge, there exists synergy between the digital-twin technology and \ac{THz} communications/sensing networks that can produce a combined effect on the \ac{6G} and beyond aimed at improving the overall performance in delivering data-driven services. This synergy stems from the fact that both rely on accurate and real-time data that is collected from their corresponding data nodes. On the one hand, \ac{THz} communications can support the transmission of large amounts of data generated by digital-twin nodes by facilitating high-speed communications links \cite{Ref_Kuruvatti2022Empowering}, while \ac{THz} sensing can help the acquisition of high-accuracy environment data for digital twin. On the other hand, a digital twin can improve the overall performance of \ac{THz} communications and sensing by offering a virtual testing, monitoring, decision-making, and optimization environment for the said \ac{THz} systems and networks \cite{Ref_Kuruvatti2022Empowering}. 

To be specific, digital twins can be utilized to generate and enable virtual replicas (also known as virtual models) for manufacturing processes and systems, such as the digital twins for the machines, links, services, materials, networks, and products contained in industry. By controlling and monitoring the virtual replicas in a real-time manner, it can be feasible to detect and subsequently address any maintenance issues and bottlenecks that may arise in the said industry, including device and machine complete failures, unsuccessful service delivery attempts, and material shortages, among many others. To enable the digital twinning of the manufacturing industry, \ac{THz} communications/sensing systems and networks can be deployed to acquire, transmit, and receive data (and at some points enrichment information) between the physical objects and virtual replicas of the manufacturing systems, enabling real-time control, accurate decision-making, and autonomous optimization \cite{Ref_Tao2022Optimal}.

In the literature, the relationship between digital-twin technologies and \ac{THz} systems has received scant attention. We uncovered three references addressing this intersection of the two technologies. First, the authors in \cite{Ref_Pengnoo2020Digital} proposed a \ac{THz} signal guidance system in which a digital twin is utilized to aim at modeling, controlling, and predicting the indoor signal propagation features and characteristics. The authors claim that their methodology achieved the ``best'' \ac{THz} signal path from a nearby base station to the targeted user equipment using a number of certain models. Second, in reference \cite{Ref_Zhang2022Digital}, the authors proposed a framework that is based on the \ac{THz} communications system and aimed at implementing the digital-twin prediction for enabling extremely security-sensitive systems and objects. Finally, reference \cite{Ref_Tao2022Optimal} studies the delay minimization optimization problem within the context of \ac{THz} communications system and visible light communications system. In their study, the authors claim that their approach reduces up to 33.2\% transmission delay in comparison with the traditional methods.

\subsection{AI/ML-Aided \ac{THz} Systems and Networks}\label{Subsec:AI/MLTHz} 
\ac{THz} communication/sensing systems and \ac{AI}/\ac{ML} \cite{Ref_jiang2019neural, Ref_jiang2020deep, Ref_jiang2019comparison} can benefit from each other synergistically. There are several facets of \ac{THz} systems that can benefit from the application of \ac{AI} techniques and \ac{ML} algorithms in \ac{6G} and beyond networks. For example, \ac{AI}/\ac{ML} can be employed  for (a) signal processing to enhance the quality of \ac{THz} signals and reduce noise; (b) \ac{THz} channel estimation to be maintained over long distances and not be affected by atmospheric effects; and (c) the optimization of error correction codes and modulation schemes. On the one hand, the performance, effectiveness, and dependability of \ac{THz} systems and networks can be improved through the utilization of \ac{AI}/\ac{ML} approaches. On the other hand, \ac{THz} systems can offer high-speed wireless data transfer and high-accuracy sensing capabilities that can be helpful for the deployment of AI/\ac{ML} services. 

In addition, AI/ML technologies can be utilized to create intelligent and data-driven \ac{THz} communications and sensing systems with being capable of adapting to quickly changing environmental conditions. For instance, with advanced \ac{AI}/ML algorithms, a self-healing, self-optimizing, and self-regulating \ac{THz} communication and sensing network can be built that can modify their parameters autonomously to maintain required performance and service levels. Moreover, AI/ML can be utilized in \ac{THz} imaging and sensing application use-case scenarios, including security screening, medical diagnosis, industrial inspection, and many others. \ac{THz} images can be processed using AI/ML algorithms to extract enrichment and/or useful information, resulting in more accurate and reliable results.

The synergistic relationship between \ac{AI}/ML and \ac{THz} systems has also been demonstrated by some recently published works. During our investigation, we found three papers \cite{Ref_Wang2021Key, Ref_Jiang2022Machine, helal2022signal}, which provide an overview of various aspects of the AI/ML applicability in \ac{THz} systems and future research directions in this domain. To be specific, the authors in \cite{Ref_Wang2021Key} provide a survey of the current state-of-the-art research in \ac{THz} communication, including signal processing, front-end chip design, channel modeling, modulation schemes, and resource management. The paper also highlights the challenges and opportunities in \ac{6G} \ac{THz} communications systems and discusses the potential applications of \ac{THz} communications in various fields. Reference \cite{Ref_Jiang2022Machine} provides a comprehensive review of the recent achievements and future challenges of ML in \ac{THz} communication. More specifically, the paper summarizes the state-of-the-art research on ML-based \ac{THz} imaging, sensing, and communications systems, including signal processing, feature extraction, classification, and optimization. The paper also discusses the potential applications of ML in \ac{THz} technology, such as medical diagnosis, security screening, and wireless communication, and outlines the future research directions and challenges in this field. The authors of \cite{helal2022signal} cover the fundamentals of \ac{THz} sensing, including sources of \ac{THz} radiation, detection techniques, and applications. This paper presents a comprehensive survey of signal processing techniques, including time-domain and frequency-domain methods, feature extraction, and classification. It also reviews recent developments in ML-based \ac{THz} sensing and highlights the challenges and future directions for signal processing and ML techniques in \ac{THz} sensing. 

Finally, and in addition to the above three overview papers related to \ac{THz} communications systems, there exist a number of papers that study ML techniques for time-domain spectroscopy and \ac{THz} imaging \cite{Ref_Park2021Machine}, the application of AI in \ac{THz} healthcare technologies \cite{Ref_Banerjee2020Terahertz}, two types of low-cost \ac{THz} tags using ML-assisted algorithms \cite{Ref_Mitsuhashi2020Terahertz}, and molecular screening for \ac{THz} detection using ML-assisted techniques \cite{Ref_Koczor2021Molecular}.

\subsection{Security in \ac{THz} Systems and Networks}\label{Subsec:PLSforTHz} 
Resiliency against eavesdropping and other security threats has become one of the key design priorities. Security measures are accessible at every layer of a wireless network and can be integrated across layers to ensure redundancy. These measures encompass various forms, including software (e.g., encryption and authentication at the upper layers), hardware (e.g., trusted platform modules), and the physical layer (e.g., wave-front engineering, near-field antenna modulation, and polarization multiplexing) \cite{Ref_ma2018security}. Physical-layer security offers notable advantages: they do not require a shared private key, demand minimal additional computing resources, and do not hinge on the assumption that the attacker possesses limited computational capacity, making it attractive, especially in \ac{THz} systems.

The intersection between physical-layer security and \ac{THz} communications and sensing is raised from the particular properties of \ac{THz} waves, such as channel sparsity and pencil-like beams, where an eavesdropper is difficult to intercept highly directional \ac{THz} signals. Additionally, the use of ultra-wide signal bandwidth and spread spectrum techniques can enhance the security of \ac{THz} transmission. Spread-spectrum techniques spread a signal over a wide range of frequencies, making it hard for signal detection and improving robustness against common jamming attacks \cite{Ref_akyildiza2014terahertz}. To the best of our knowledge, physical-layer security can be deployed in the following aspects of \ac{THz} communications and sensing: channel authentication, \ac{PHY} encryption, beamforming, and \ac{PHY} key generation. These security techniques can improve the security of \ac{THz} communications and sensing and make them suitable for a variety of \ac{6G} scenarios, such as hot spots, wireless backhaul, satellite interconnection, industrial networks, positioning, and imaging. 

As of the time of this writing, a number of studies have uncovered numerous facets of physical-layer security for \ac{THz} systems.  In \cite{Ref_Djordjevic2017OAM}, a hybrid physical and multi-dimensional coded modulation scheme for \ac{THz} communications systems was proposed. In \cite{Ref_Rahman2017Physical}, physical-layer authentication in \ac{THz} systems was presented. Following that, some works \cite{Ref_Fang2022Secure, Ref_Fang2021Physical, Ref_ma2018security} were conducted (both based on simulation and calculation) on resiliency against eavesdropping using a directional atmosphere-limited \ac{LoS} \ac{THz} link. Regarding the probability of eavesdropping, the authors of \cite{Ref_Petrov2019Exploiting} have also investigated about decreasing message detection using inherent multi-path \ac{THz} systems. Moreover, a physically secure \ac{THz} system operating in \SI{310}{\giga\hertz} using orbital angular momentum was presented in \cite{Ref_Woo2022Physcially}.  Last but not least, it is worth noting that \ac{THz} sensing and communications face unique security challenges that must be addressed. For example, \ac{THz} waves can penetrate some materials, including clothing and certain types of packaging, which could potentially be exploited by malicious actors \cite{Ref_ma2018security}. 

\subsection{Localization Services in \ac{THz} Systems and Networks} \label{Subse:LocalizationTHz} 
Localization services and \ac{THz} systems are two completely different research areas. Nevertheless, they can be integrated synergistically with being aimed at improving their capabilities and opening up new avenues for a variety of applications and services in \ac{6G}. The intersection of localization and \ac{THz} systems results from the unique propagation characteristics of \ac{THz} frequencies, which can be utilized for localization purposes in the access network domain. \ac{THz} frequencies are particularly sensitive to rapidly occurring environmental changes, such as the presence of obstacles (a.k.a. problematic objects) or changes in the refractive index of materials in an environment. This extreme sensitivity can be utilized to create \ac{THz}-based localization services and systems that can operate in a variety of environments, including indoor environments where GPS-based systems may not function optimally. 

The integration of localization services into \ac{THz} systems and network  has the potential to enable a variety of novel use cases and applications in the \ac{6G} era, such as intelligent factories, manufacturing, healthcare, and many others. \ac{THz}-based localization services, for instance, could be used to track assets within a factory or warehouse, while \ac{THz} communications could be used to enable high-speed data transfer between machines and objects in the said factory. \ac{THz}-based localization services could be deployed to monitor and control patient movement within a hospital, while \ac{THz} communications could be used to enable wireless video transmission for remote consultations and many other services.

The intersection between localization services and \ac{THz} communications systems has been studied to some extent in the literature. A tutorial  \cite{chen2022tutorial} provides a comprehensive review of \ac{THz}-band localization techniques for \ac{6G} systems. The authors discussed various aspects of \ac{THz} waves, including propagation characteristics, channel modeling, and antenna design. They also explore different localization methods such as the time of arrival, angle of arrival, and hybrid techniques. The paper concludes by highlighting some potential applications of \ac{THz}-band localization in \ac{6G} networks. In addition, two research articles discuss various aspects of this intersection. In \cite{Ref_Fan2020A}, the authors proposed a deep learning model for 3D \ac{THz} indoor localization using a structured bidirectional long short-term memory network. The authors claim that their proposed method achieves better localization accuracy than state-of-the-art methods, making it a promising solution for indoor localization in \ac{THz}-band communications systems. Finally, reference \cite{Ref_Fan2021A} proposes a new deep learning method for \ac{THz} indoor localization called SIABR, utilizing a structured intra-attention bidirectional recurrent neural network to learn features from the received signal and estimate the location of the target.

\subsection{Multi-Connectivity for \ac{THz} Systems and Networks} 
Due to the high atmospheric absorption and low penetration capability, \ac{THz} signals suffer from such strong propagation loss, fading, shadowing, and blockage, that they are hard to maintain with mobility even when beamforming and combining are ideally performed. To address this issue, \ac{THz} systems will need \ac{MC} as an essential feature so that a continuous and stable data connection between the users and the network can be ensured by means of radio link redundancy in case a single radio link fails. 
The basic principle of \ac{MC} is to keep multiple radio connections to different \acs{BS}s simultaneously, but only use one of them at a time for signal transmission. The effectiveness of multi-connectivity in addressing the issue of blockage in \ac{THz} band has been confirmed by evidence from various studies: a higher density of \ac{BS}s is proven to enhance the system performance from the perspectives of link probability \cite{Ref_liu2022performance}, capacity \cite{Ref_liu2022performance, Ref_shafie2020multi}, and session completion rate \cite{Ref_moltchanov2021uninterrupted, Ref_sopin2022user}. 

Especially, there are two different strategies for selecting the active radio link, namely the \ac{CMC} where the closest \ac{BS} with \ac{LoS} link is always selected for communications and the \ac{RMC} in which the active radio link is only re-selected when the current \ac{LoS} link is blocked. While the \ac{CMC} strategy is significantly outperforming the single-connectivity strategy, the \ac{RMC} strategy brings only a marginal - sometimes even negative - gain, and is therefore discouraged despite its low signaling overhead~\cite{Ref_shafie2020multi}. It should also be noted that the application of \ac{MC} has an influence on the handover mechanism since it links the status model. In \cite{Ref_oezkoc2021impact}, \"Ozko\c{c} \emph{et al.} established an analytical framework to assess the joint impact of the \ac{MC} degree and the handover constraints on system performance of \ac{THz} cellular networks.

An alternative approach to exploit \ac{MC} is to allow multiple \acp{BS} to simultaneously \emph{serve} multiple mobile stations, i.e., each user may be communicating with multiple \acp{BS} rather than one at a time. This is usually known as the network \ac{MIMO} or \ac{DMIMO}, which exploits the spatial diversity by means of intensifying the \acp{BS} instead of antenna units in each array like in classical \ac{MMIMO}/\ac{UMMIMO}. To minimize the cross-interference among adjacent \acp{BS} and maximize the throughput in \ac{DMIMO} networks, the \ac{CoMP} technologies shall be invoked. \ac{CoMP} allows different \acp{BS} to be clustered into small groups, and to coordinately optimize their user association and beamforming within each group. More specifically, there are two principles of \ac{CoMP}: the \ac{JT} where multiple \acp{BS} transmit the same signal simultaneously to the same \ac{UE}, and the \ac{CSCB} where each \ac{BS} sends a different signal and the signals are combined at the \ac{UE}. An example of applying \ac{CoMP} in the \ac{THz} band is presented in \cite{Ref_dowhuszko2017distance}, which combines joint power allocation and quantized co-phasing schemes to maximize the aggregated data rate.

Since the late 2010s, the concepts of \ac{DMIMO} and \ac{CoMP} have evolved into the \ac{CFN} paradigm, where all \acp{UE} in an area are jointly served by numerous single-antenna \acp{BS} in a \ac{CoMP} manner~\cite{Ref_jiang2021road}. Having been well studied at \ac{mmWave} frequencies, the applicability of \ac{CFN} in the \ac{THz} band still remains 
under-studied~\cite{Ref_faisal2020ultramassive}. Pioneering work was reported in 2022 by Abbasi and Yanikomeroglu~\cite{Ref_abbasi2022cell}, considering a \ac{NTN} scenario. In some research works \cite{Ref_suer2020multi}, multi-connectivity also refers to establishing connections in different communications bands, e.g., transmitting control signaling in the sub-6GHz band while delivering data in the \ac{THz} band (also deliver data when the \ac{THz} band is in an outage). 

\subsection{Channel Awareness for \ac{THz} Systems and Networks} 
While modern wireless data transmission technologies generally rely on the knowledge of channel state to achieve satisfactory performance, the acquisition of accurate \ac{CSI} can be a critical challenge for \ac{THz} systems and networks. First, the pilot symbols can be easily blocked due to the susceptibility of \ac{THz} channels, leading to a low efficiency of classical channel estimation methods. Second, \ac{THz} channels are selective regarding many different parameters, e.g., time, frequency, beam pattern, beam direction, polarization, etc. Therefore, it takes much effort to comprehensively measure the \ac{CSI} of a \ac{THz} channel, in addition to a significant overhead to encode the dimensional and sparse \ac{CSI}. Similar to the pilots, the \ac{CSI} reported from UE to the network can also be blocked if transmitted in the \ac{THz} band itself~\cite{Ref_shafie2022terahertz}.  

Regarding these challenges, the out-of-band channel estimation occurs as a promising solution. This involves acquiring the \ac{CSI} of \ac{THz} channels using the channel estimation at lower frequencies, leveraging the spatial correlation. To assess the feasibility of this approach, the authors of \cite{Ref_zhang2022out} and \cite{Ref_kyoesti2023feasibility} studied the spatial similarity among \ac{THz}, \ac{mmWave}, and sub-\SI{6}{\giga\hertz} bands based on point cloud ray-tracing simulation and field measurement. Their results support the use of out-of-band beam search strategy, not only in \ac{LoS} scenarios but also even in \ac{NLoS} ones, when using well-designed antenna patterns in specific frequency bands. Meanwhile, Peng \emph{et al.} demonstrated the feasibility of out-of-band channel estimation beam searching with both ray-tracing simulations \cite{Ref_Bile2014Fast} and real-world experiments \cite{Ref_peng2019power}.

However, the exploitation of channel similarity for \ac{THz} communications and sensing is still facing technical challenges. First, the difference in the size of the antenna array within the same aperture leads to a mismatch in the beamwidth between the lower frequencies and the \ac{THz} ones~\cite{Ref_shafie2022terahertz}, which is proven to have a strong impact on channel similarity than the frequency gap itself~\cite{Ref_kyoesti2023feasibility}. Second, the correlation matrix is difficult to estimate, considering its large size and the small dimension of antenna arrays measuring at lower frequencies~\cite{Ref_shafie2022terahertz}. Third, despite the feasibility of out-of-band estimation on static \ac{THz} channels, the dynamics such as user mobility, scatter mobility, and blockages, are lifting the difficulty of this task to the next level~\cite{Ref_zhang2022out}.

\section{Integrated \ac{THz} Communications and Sensing} \label{SEC:ISAC}
The novel concepts of ``network as a sensor'' and ``Internet of Senses'' are becoming unprecedented essentials in the upcoming \ac{6G} and beyond cellular networks so as to support a multitude of emerging use cases~\cite{Wymeersch2021}. The two major functionalities, i.e., communications and sensing (including positioning and imaging), will be merged, synergized, and integrated, benefiting from each other rather than competing for network resources. Future base stations are supposed to provide not only legacy communication services but also localization, sensing, and even electromagnetic imaging capabilities, acting as dual-functional \ac{ISAC} transceivers~\cite{yang2021intecom, li2021integrated}. 

\begin{figure}[t]
	\centering
\includegraphics[width=1 \linewidth]{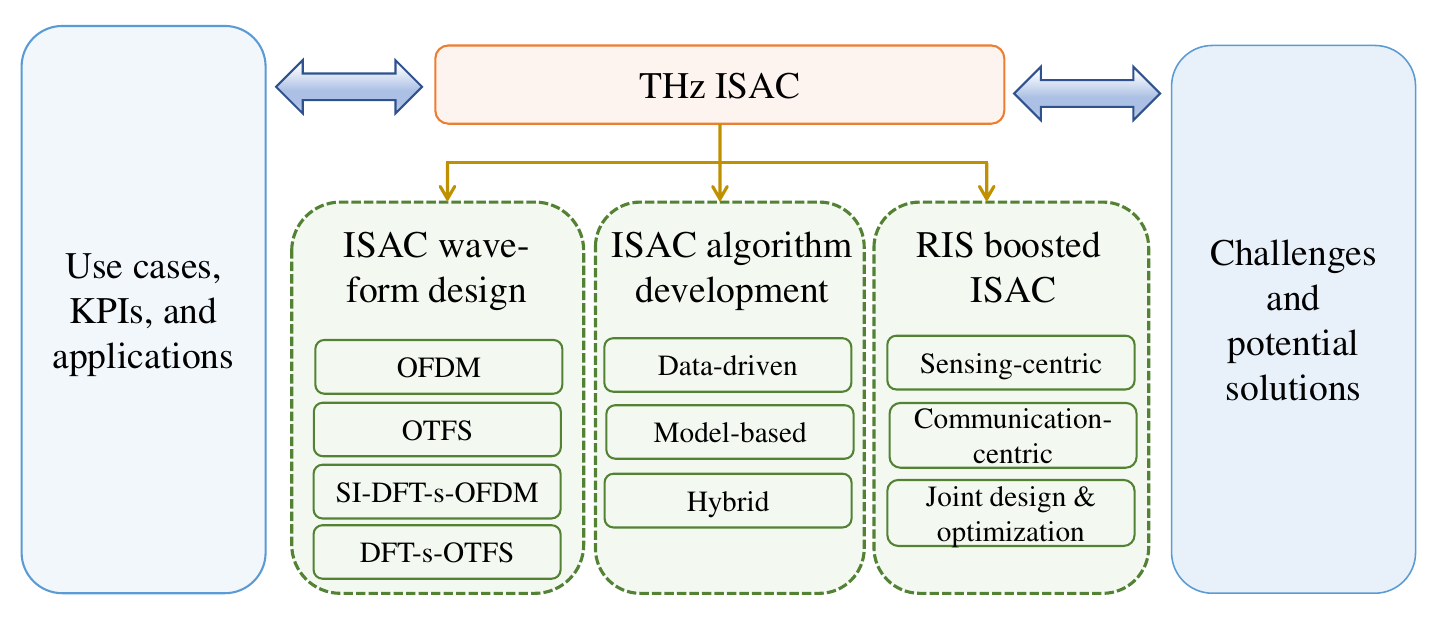}
	\caption{An architectural overview of \ac{THz} \ac{ISAC}.  
 }
		\label{THz_ISAC_outline}
		\vspace{-0.5cm}
\end{figure}

In general on the physical layer, \ac{ISAC} has two widely adopted terminologies, i.e., \ac{RCC} and \ac{DFRC}~\cite{elbir2022terahertz}. It aims for enhanced spectral and energy efficiency, reduced hardware cost, and decreased power consumption as well as deployment and computational complexity. In the literature, such as~\cite{chen2022tutorial,nie2017three, Ref_chen2019survey, Ref_sarieddeen2021overview,  Wymeersch2020, tan2021integrated}, \ac{THz} sensing, imaging, localization, and communications are treated separately. Different from these, we in this section provide a holistic survey on the recent activities in integrated \ac{THz} communications and sensing. Special attention is paid to use cases and \acp{KPI}, waveform design, algorithm development, \ac{RIS}-boosted \ac{ISAC}, and potential challenges and solutions. The architectural overview of \ac{THz} \ac{ISAC} is depicted in Fig.~\ref{THz_ISAC_outline}, and the selected major contributions related to \ac{THz} \ac{ISAC} are summarized in Table \ref{THz_ISAC_table}.

\begin{table*}[t]
{\color{black}
\renewcommand{\arraystretch}{1.5} 
\caption{\textsc{Summary of selected major contributions on \ac{THz}-\ac{ISAC}}}
\label{THz_ISAC_table}
\raggedleft
\begin{tabular}{|c| c | c | p{12cm} |}
\hline
\textbf{Category}&\textbf{Year} & \textbf{Authors} & \textbf{Contributions} \\
\hline \hline
\multirow{3}{4em}{  } & 2020 & Wymeersch \textit{et al.}~\cite{Wymeersch2020} & Listed a series of application examples of \ac{RIS}-based localization and mapping services.\\ 
\cline{2-4}
\multirow{3}{4em}{Use cases, \acp{KPI}, and applications} & 2021 & Wymeersch \textit{et al.}~\cite{Wymeersch2021} & Provided an overview of the vision of the European \ac{6G} flagship project Hexa-X  alongside
the envisioned use cases; Discussed the technical enablers and associated research challenges.\\ 
\cline{2-4}
\multirow{3}{4em}{ } & 2022 & Chen~\textit{et al.}~\cite{chen2022tutorial} & Identified the
prospects, challenges, and requirements of \ac{THz} localization techniques.\\ 
\cline{2-4}
\multirow{3}{4em}{} & 2022 & Moltchanov~\cite{Ref_moltchanov2022tutorial} & Evaluated user- and system-centric \acp{KPI} of \ac{mmWave}
and \ac{THz} communication systems.\\ 
\cline{2-4}
\multirow{3}{4em}{} & 2021 & Zhang~\cite{zhang2021enabling} &Set up \ac{JCAS} in the mobile network context
and envisage its potential applications.\\ 
\cline{2-4}
\multirow{3}{4em}{} & 2022 & Akyildiz~\cite{Akyildiz2022} &Pointed out \ac{6G} applications enabled by \ac{THz} communications and their corresponding performance objectives.\\ 

\hline\hline
\multirow{3}{4em}{Waveform design} & 2022 & Wu~\cite{wu2022sensing} & Proposed SI-DFT-s-OFDM system for \ac{THz}
\ac{ISAC} with \ac{FGI} scheme, which tremendously enhances the data transmission rate.\\ 
\cline{2-4}
\multirow{3}{4em}{ } & 2022 & Wu~\cite{wu2022dft} & Proposed \ac{DFT-s-OTFS} system to improve the robustness
to Doppler effects and reduce \ac{PAPR} for \ac{THz} \ac{ISAC}. \\ 
\hline\hline

\multirow{3}{4em}{ \ac{ISAC} algorithms} & 2021 & Yang~\cite{yang2021integrated} & Leveraged multi-domain cooperation to enhance the performance
of the \ac{ISAC} system through active and passive sensing, multi-user, and multi-frequency-band
networks.\\ 
\cline{2-4}
\multirow{3}{4em}{ } & 2022 & Helal~\cite{helal2022signal} & Addressed the effectiveness of deep learning techniques by
exploring their promising sensing and localization capabilities
at the \ac{THz} band.\\
\cline{2-4}
\multirow{3}{4em}{} & 2022 & Mateos-Ramos~\cite{mateos2022model} & Studied model-driven end-to-end learning for joint single
target sensing and \ac{MISO} communication.\\ 
\hline\hline
\multirow{3}{4em}{\ac{RIS}-boosted \ac{ISAC}} & 2021 & Jiang~\cite{jiang2021intelligent} & Investigated the joint optimization of
the \ac{IRS} passive \ac{PSM} and precoding matrix of
the radar-aided base station for the \ac{DRC} system.\\ 
\cline{2-4}
\multirow{3}{4em}{ } & 2021 & Wang~\cite{wang2021joint} &Investigated joint constant-modulus waveform and discrete \ac{RIS} phase shift design, with the aim of minimizing \ac{MUI} under the \ac{CRB} constraint for \ac{DoA} estimation. \\ 
\cline{2-4}
\multirow{3}{4em}{ } &2021  & Wang~\cite{wang2021jointwave}  & Studied the minimization of MUI under the strict beam pattern constraint by jointly optimizing \ac{DFRC} waveform and \ac{RIS} phase shift matrix.\\
\cline{2-4}
\multirow{3}{4em}{ } & 2022 & Wang~\cite{wang2022simultaneously} & Minimized the \ac{CRB} of the \ac{2D} \acp{DoA} estimation of the sensing target subject to the minimum communication requirement.\\
\hline\hline
\multirow{3}{4em}{Challenges and solutions} & 2022 & Elbir~\cite{elbir2022terahertz} &  Studied several design challenges such as beam
split, range-dependent bandwidth, near-field beamforming, and
distinct channel model for \ac{ISAC} at \ac{THz}-band, and provided research opportunities in developing
novel methodologies for channel estimation, near-field beam split,
waveform design and beam misalignment.\\ 
\cline{2-4}
\multirow{3}{4em}{ } & 2022 & Han~\cite{han2022THz} & Elaborated challenges from \ac{THz}
channel and transceiver perspectives, as well as the difficulties of
\ac{ISAC}.\\ 
\hline
\end{tabular}
}
\end{table*}

\subsection{\ac{THz}-\ac{ISAC} Use Cases and \acp{KPI}}
Localization and sensing, including determining the \ac{2D}/\ac{3D} location and \ac{EM} properties of objects, alongside multi-scale communications enable a multitude of emerging use cases, which may have different functional and non-functional \ac{QoS} requirements in terms of accuracy, range, latency, velocity, update rate, reliability, and availability. The use case families consist of various vertical applications, e.g., massive twinning, immersive telepresence, wireless \ac{XR}, cooperative robots, \ac{Tera-IoT}, local trust zones, vehicular communication
and radar sensing~\cite{petrov2019unified}, and \ac{THz} \ac{IAB}. In general, these fall into the category of data-demanding and delay-sensitive applications. However, the \ac{QoS} requirements differ from one use case to another. For instance, manufacturing and smart cities, as two sub-categories of massive twining, need different accuracy, data rates, latency, update rates, reliability, and availability ~\cite{Wymeersch2021}.  A full list of selected use cases and their corresponding performance metrics can be referred to~\cite{Wymeersch2021,wymeersch20226g,zhang20196g,Ref_sarieddeen2020next,Ref_moltchanov2022tutorial,chen2021terahertz,chaccour2021joint,zhang2021enabling, wild2021joint, Akyildiz2022}, closely tied to these of upcoming \ac{6G} and beyond systems, for which the ultimate goal would be realizing terabit-per-second links and millimeter-level sensing/localization accuracy~\cite{chen2021terahertz, chaccour2022seven}.

\subsection{\ac{THz}-\ac{ISAC} Waveform Design}
\ac{THz} channels usually have short delay spread and \ac{THz} transceivers suffer from low \ac{PA} efficiency (rapid decline on its saturated
output power). Besides, frequency selectivity becomes less severe. As a result, coherence bandwidth increases. Therefore, the single-carrier waveform is more preferred compared to the \ac{OFDM} waveform~\cite{Yongzhi2021}. Sensing integrated DFT-spread-orthogonal frequency-division multiplexing (SI-DFT-s-OFDM) can yield lower \ac{PAPR} than \ac{OFDM} while maintaining single-carrier characteristic, as reported in~\cite{wu2021sensing,wu2022dft,wu2022sensing}. In addition, it brings a ten-fold improvement in velocity estimation of the moving target and significant enhancement on data rate, beneficial from the \ac{FGI} approach, which is capable of reducing the \ac{CP} overhead. In \ac{ISAC} systems, communication and sensing channels may possess different properties. For instance, a sensing channel can expose a significant delay spread. Thus, different design criteria for communications and sensing waveforms should be considered in terms of cyclic prefix lengths and pilots~\cite{wu2021sensing}. 

Another type of intensively studied waveform, so-called \ac{OTFS}, can well handle the Doppler effect and accommodate the channel dynamics in the delay-Doppler domain. Nevertheless, it still can not meet the strict requirements of power amplifier efficiency and signal processing complexity. Similar to SI-DFT-s-OFDM, \ac{DFT-s-OTFS} was proposed to address the high \ac{PAPR} issue faced in the original \ac{OTFS}~\cite{wu2022dft}. In contrast to \ac{OTFS}, \ac{DFT-s-OTFS} can achieve strong robustness to Doppler spread compared to \ac{OFDM} and \ac{DFT-s-OFDM} and reach lower \ac{PAPR} compared to its \ac{OTFS} counterpart~\cite{wu2021dft_s_otfs,wu2022dft, Akyildiz2022}. By imposing superimposed pilots in the delay-Doppler domain alongside the \ac{DFT} precoding operation, \ac{THz} \ac{ISAC} system facilitates low-complexity iterative
channel estimation and data detection. Because of the aforementioned advantages, \ac{DFT-s-OTFS} is the most promising candidate waveform for \ac{THz} \ac{ISAC} systems.

\subsection{\ac{THz}-\ac{ISAC} Algorithm Development}
The \ac{ISAC} algorithms typically fall into three categories, i.e., the data-driven AI-based approaches, model-based approaches, and hybrid approaches (a combination of the former two). \ac{AI} techniques rely on large-volume date sets for training customized \ac{NN} models for sensing, localization, and signal detection, while tackling the mathematically intractable non-linearity issues from, e.g., phase noise and offset, power amplifier, and mutual coupling. On the contrary, for the model-based \ac{ISAC} algorithms, the majority of them need to harness the well-justified domain knowledge and modeling, such as geometric relationship among the transceivers and the environmental objects, and take full advantage of channel sparsity in the form of rank deficiency of the channel matrix or a limited number of resolvable paths, as to obtain satisfactory performance.   

Under the framework of \ac{THz} \ac{ISAC}, joint data detection (signal recovery) and sensing parameter estimation are conducted with multi-task \acp{NN} in~\cite{Yongzhi2021}. In a broad sense, the \ac{ML} roles on \ac{ISAC} can be classified into three categories: 
\begin{itemize}
    \item  \ac{JSAC}
    \item sensing-aided communications~\cite{Anzhong2022,Jiguang2021}
    \item communication-aided sensing~\cite{demirhan2022integrated}.
\end{itemize}

To be specific, the first category includes the following activities: \ac{JSAC} waveform design, spatial beam pattern design,  inter- and self-interference cancellation, resource allocation, etc. Without any doubt, communications and sensing can be mutually beneficial for each other. In the category of sensing-aided communication, sensing information (treated as prior information), e.g., the location of the transmitter, receiver, and environmental objects, can be leveraged for enhancing the beam prediction/alignment and reducing the overhead of beam training as well as channel sounding~\cite{yang2021integrated, tan2021integrated}. In dynamic scenarios where the user is under mobility, regardless of low or high velocity, such sensing information can be utilized for predicting potential blockages and enabling smooth handovers~\cite{demirhan2022radar}. Similarly, communication signals can also be exploited to boost the sensing performance during the data transmission phase. The back-scattered data signals can gradually refine/improve the sensing parameter estimation, similar to data-aided channel estimation in the literature~\cite{Junjie2014}.    

The \ac{DL} algorithms alongside other counterparts, e.g., deep reinforcement learning and transfer learning, pave the way for the integrated detectors and estimators for both communications and sensing, in terms of e.g., sensing parameters estimation, interference mitigation/cancellation, beam tracking/prediction, and network resource allocation/management~\cite{helal2022signal,wu2022sensing}. Meanwhile, it can successfully tackle the mathematically intractable non-linearity issues and hardware impairments in \ac{ISAC} systems~\cite{wu2022sensing}. Furthermore, the latent features related to sensing parameters can be more readily learned and extracted by the adoption of \ac{DL} algorithms. For instance, two sensing neural network
(SensingNet) models, one for range estimation and the other for velocity estimation, were proposed in~\cite{wu2022sensing}, composed of an input layer, a flatting
layer, five dense layers for feature extraction and nonlinear
mapping, and an average output layer. Meanwhile, a concatenated two-level communication neural network
(ComNet) model for data detection was developed, where the first level is leveraged for channel information acquisition and its output is further utilized in the second level. Combining both the SensingNet and ComNet models bring us a promising \ac{DL}-empowered solution for \ac{THz} \ac{ISAC} systems. It should be noted that other \ac{DL} variants can also be adopted for both sensing and communication applications.   

In the \ac{DFT-s-OTFS} system~\cite{wu2022dft}, a two-stage sensing parameter estimation approach was proposed, i.e., coarse on-grid search in the first stage followed by refined off-grid search in the second stage for extracting the sensing parameters. Under the framework of \ac{ISAC}, data detection and sensing parameter estimation can be performed in an iterative manner by considering the conjugate gradient method until a certain preset stopping criterion is reached. Besides, \ac{ISAC} performance can be further enhanced by multi-domain cooperation through joint active and passive sensing, and multi-user and multi-frequency operations~\cite{yang2021integrated}. The tensor decomposition approach is capable of leveraging the channel sparsity and guaranteeing a unique solution for each environmental sensing parameter without any ambiguity~\cite{chaccour2021joint}. Such sensed information can be then utilized to reconstruct a high-resolution indoor mapping to further boost the prediction of blockages and the availability of \ac{LoS} path, and reduce the beam tracking frequency. Thus, higher spectrum efficiency in data transmission can be achieved accordingly.   

The traditional model-based algorithms adopt \acl{ComS} techniques, on-the-grid, off-the-grid, and the combination of the former two (e.g., in~\cite{wu2022dft}), for extracting channel and sensing parameters by taking advantage of channel sparsity~\cite{cui2019low,rahman2020joint,JiguangIEEETWC2021}. Model-driven end-to-end learning, falling into the category of hybrid approaches, for joint single target sensing and precoder design in \ac{MISO} communication networks was studied in \cite{mateos2022model}. In particular, the authors jointly consider precoder design at the transmitter and target \ac{AoA} estimation at the receiver by applying an \ac{AE} while accounting for the hardware impairment. During the model-driven end-to-end learning, the matrix composed of steering vectors at discretized grid points is considered as trainable parameters to simplify the transmit precoder design and optimized by only maximizing the sensing-related performance metric~\cite{mateos2022model}.  

\subsection{\ac{RIS}-Boosted \ac{THz}-\ac{ISAC}}
With the newly-introduced capability of manipulating the radio propagation environment, \ac{RIS} is able to expand the communication coverage and enhance the sensing performance~\cite{elbir2022terahertz,he2022joint}. The potential roles that can be played by an \ac{RIS} are multi-fold: scattering, reflecting, refraction, absorption, polarization, and diffraction. With all the preceding \ac{DoF}, intelligent, programmable wireless propagation environments can be established for different tasks, e.g., communications, sensing, localization, and imaging. The \ac{RIS} can enrich the \ac{LoS} availability by establishing a virtual one when the real one is suffering from temporary blockage, which frequently occurs at \ac{THz} frequencies.

The various benefits of integrating \ac{RIS} into \ac{ISAC} were discussed in~\cite{chepuri2022integrated}. The gains against the \ac{RIS}-free counterpart heavily rely on the cross-correlation between the sensing and communication channels. The more the mutual coupling, the more gain can be accomplished in terms of \ac{ISAC} performance. By introducing the \ac{RIS}, enhanced flexibility and adaptation to channel dynamics is seen in altering the coupling level of these channels~\cite{chepuri2022integrated}. The importance of tight coupling of communications and localization was also emphasized in~\cite{he2022beyond} for the purpose of harnessing the full potential of \acp{RIS}. That is to say, the \ac{SLAC} requires smart \ac{RIS} control, co-design of communications and localization, and the flexible trade-off and reinforcement between the two functionalities. 

In the rich \ac{RIS}-boosted \ac{ISAC} literature, various optimization problems are formulated with different objectives along with different constraints. These works can be cast into three different classes: 
\begin{itemize}
    \item sensing-centric design~\cite{jiang2021intelligent}
    \item communication-centric design~\cite{wang2021joint}
    \item joint design and optimization~\cite{wang2021jointwave, zhang2021overview}.
\end{itemize}

For the first category, the objective is sensing-oriented, while the communication metrics are taken as constraints. For instance, the authors of~\cite{jiang2021intelligent} maximized the \ac{SNR} at the radar while considering a communication \ac{SNR} constraint. By addressing this optimization problem, \ac{SDR} along with bisection search was considered for transmit beamforming design while majority-minimization is considered for \ac{RIS} design. With respect to the second category, the reference~\cite{wang2021joint} takes interference among the communication users as the objective while treating the desired \ac{MSE} of \ac{DoA} estimation as a constraint. To be specific, the authors developed an alternating optimization algorithm for finding the optimal design of constant-modulus waveform and discrete \ac{RIS} phase shifts for the \ac{RIS}-assisted \ac{ISAC} systems. For the former design, manifold optimization algorithm was considered, while for the later design, two schemes, one based on manifold optimization and the other based on successive optimization, were proposed. In terms of the last category, the weighted sum of two objectives, one for communications and the other for sensing, are usually considered~\cite{wang2021jointwave}. Similar to that in~\cite{wang2021joint}, an alternating optimization algorithm based on oblique manifold optimization alongside Riemannian steepest descent was introduced to jointly optimize the \ac{DFRC} waveform and \ac{RIS} phase shift matrix. All the above categories share some common constraints, e.g., individual transmit power, sum transmit power, hardware (especially for \ac{RIS}, e.g., phase quantization, constant modulus of amplitude), etc. A holistic comparison among the three classes can be referred to~\cite{zhang2021enabling,wang2022integrated}.     

Recently, a more promising type of \ac{RIS}, termed as \ac{STAR-RIS}, was introduced, which is able to offer additional benefits thanks to its inherent dual-mode operation and full-dimensional coverage~\cite{Jiguangstarris2022,he2023star}. The \ac{STAR-RIS} can concurrently reflect and refract the incident signals towards multiple desired \acp{MS}. Because of this, the \ac{STAR-RIS} can further boost the \ac{ISAC} performance compared to the sole-reflection-type \ac{RIS}~\cite{wang2022simultaneously,gao2022joint, wang2022stars} with extended flexibility. Either an outdoor or indoor BS is capable of providing both communications and sensing services to the users located indoors and outdoors by installing a \ac{STAR-RIS} on a transparent glass window~\cite{Jiguangstarris2022,he2023star}.

\subsection{Challenges and Solutions for \ac{THz}-\ac{ISAC}}
The open problems for \ac{THz}-\ac{ISAC} are listed and discussed in~\cite{Wymeersch2021, Yongzhi2021}. For example, waveform design should be customized depending on sensing applications. Dynamic beamforming control faces great challenges since the beamwidth is narrow and highly directional~\cite{Ref_sarieddeen2020next}. As a consequence, the probability of beam misalignment can be inevitably high. A robust design of candidate beams for communication purposes requires a wider beamwidth. However, to enhance the sensing resolution and accuracy, narrow beams are preferred. Multiple concurrent beams comprise one fixed sub-beam for point-to-point communications and multiple time-varying sub-beams for sensing purposes can achieve a well-balanced performance between communications and sensing~\cite{chen2021terahertz}. However, multiple simultaneous beams suffer from degraded beamforming gains. 

Imperfections, resulting from \ac{IQ} imbalance, \ac{PA} nonlinear distortions, and phase noise at the local oscillator, need to be compensated for when designing robust \ac{THz}-\ac{ISAC} algorithms. The wide-band channel becomes highly selective with high Doppler spread, which may break the orthogonality of \ac{OFDM} transmission and incur inter-carrier interference~\cite{han2022THz,elbir2022terahertz}. Besides, the near-filed propagation, where channel sparsity vanishes in the angular domain, makes the beamforming design intractable. However, a recent study~\cite{Cui2022_TCOM} shows that sparse representations of the near-field channel from the polar domain are still available, making efficient \ac{CSI} acquisition feasible with the aid of advanced compressive sensing techniques. 

The beam squint effect makes the designed beam deviate from the exact one, resulting in reduced array gain and performance degradation~\cite{han2022THz}. The beam squint and split effect will become more obvious as the increase of carrier frequency and bandwidth, causing significant performance degradation on sensing and communication. As examined in~\cite{elbir2022terahertz} for a broadside target, beam split can reach as much as $4^\circ$ for 0.3 \ac{THz} with \SI{30}{\giga\hertz} bandwidth while it is only $1.4^\circ$ for \SI{60}{\giga\hertz} with \SI{2}{\giga\hertz} bandwidth. This effect should be mitigated and compensated for when designing the beamforming patterns, such as \ac{DPP}~\cite{dai2022delay}. By introducing a time delay  network between the \ac{RF} chains and
frequency-independent phase shifters in the hybrid precoding architecture, DPP is capable of performing frequency-dependent delay-phase controlled beamforming for reducing the array gain loss introduced by the beam split effect. The newly-introduced time delays can make beams aligned with the target physical directions across the entire bandwidth. Due to the user mobility and frequent blockage, beam misalignment occurs when \ac{LoS} path is unavailable between the \ac{BS} and \ac{MS}. Provided that the user can be tracked and blockage can be predicted in advance, beam misalignment can be avoided. However, this requires high-precision sensing information. In line with the \ac{5G} \ac{CSI} acquisition signals, the authors in~\cite{chen2022isac} adopt \ac{SSB} for blockage detection and \ac{RS} for user tracking. 

Until all the above-mentioned challenges are thoroughly addressed in the forthcoming years, the vision that everything will be sensed, connected, and intelligent can be fulfilled.

\section{\ac{THz} Trials and Experiments}

\begin{table*}[!t]
\renewcommand{\arraystretch}{1.3}
\caption{Summary of experimental and demonstration systems for \ac{THz} Communications and Sensing.}
\label{table_THzExpe}
\centering
\begin{tabular}{|c|c|c|c|c|c|m{8cm}|}
\hline \hline
\textbf{Ref.} & \textbf{Year} & \textbf{Freq.[\si{\giga\hertz}}] & \textbf{BW[\si{\giga\hertz}}] & \textbf{Rate[\si{\giga\bps}}] & \textbf{Distance [\si{\meter}}] & \textbf{Contributions}    \\  \hline \hline
 \cite{Bio_moeller20112} &  2011 & 625 & narrow N/A & 2.5 & 0.2 &  A novel approach is reported for \SI{2.5}{\giga\bps} signalling at a carrier frequency of \SI{625}{\giga\hertz}. Duobinary baseband modulation on the transmitter side generates a signal with a sufficiently narrow spectral bandwidth to pass an upconverting frequency multiplier chain.\\ \hline
 \cite{Bio_koenig2013wireless} &  2013 & 237.5 & 35 & 100 & 20 &  Present for the first time, a single-input and single-output wireless communications system at \SI{237.5}{\giga\hertz} for transmitting data over 20 m at a data rate of \SI{100}{\giga\bps} from combining terahertz photonics and electronics.  \\ \hline 
 \cite{Bio_kallfass201564} &  2015 & 240 & 32 & 64 & 850 & A directive fixed wireless link operating at a center frequency of \SI{240}{\giga\hertz} achieves a data rate of \SI{64}{\giga\bps} over a transmission distance of 850 m using QPSK and 8PSK modulation, in a single-channel approach without the use of spatial diversity concepts. \\ \hline 
 \cite{Bio_qiuyu2017design} &  2017 & 140 & N/A & 5 & 21000 & The 16QAM modulation scheme is used in baseband processing, and mixers are used for cascading frequency up and down-converting. Cascading power amplification technique is adopted with a solid-state power amplifier and a vacuum electronic device. A 21 km wireless communications testing is carried out, by means of two Cassegrain antennas with \SI{50}{\dB i} gain each.  \\ \hline 
 \cite{Bio_8424023} &  2019 & 375-500 & 30 & 120 & 1.42 &  The first experimental demonstration of 2 × 2 MIMO wireless transmission of multi-channel THz-wave signal, which realizes 6 x \SI{20}{\giga\bps} six-channel polarization division multiplexing QPSK THz-wave signal delivery over 10 km wireline single-mode fiber-link and 142 cm wireless 2 x 2 MIMO link. \\ \hline 
 \cite{Bio_9096572} &  2020 & 335-365 & 30 & 600 & 2.8 & Demonstration of a hybrid THz photonic-wireless transmission based on a THz orthogonal polarization dual-antenna scheme. Probabilistic shaped 64QAM-OFDM modulation format is used to realize a high transmission rate. A potential total system throughput of \SI{612.65}{\giga\bps} is successfully achieved. \\ \hline 
 \cite{Bio_harter2020generalized} &  2020 & 300 & 40 & 115 & 110 &  A local-oscillator tone is transmitted along with the signal, and the amplitude and phase of the complex signal envelope are digitally reconstructed from the photocurrent by exploiting their Kramers–Kronig-type relation. Using Schottky-barrier diode as a nonlinear receiver element and 16-state QAM, a net data rate of \SI{115}{\giga\bps} at a carrier frequency of 0.3 THz over a distance of 110 m is achieved.  \\ \hline 
 \cite{Bio_li202154} &  2021 & 340 & 30 & 44.8 & 104 & Demonstrates the capability of over 54/104 meters wireless transmission with a record-breaking net data rate of 128/\SI{44.8}{\giga\bps} at THz-band by utilizing both suitable dielectric lenses and DSP algorithms, without THz amplifier. \\ \hline 
 \cite{Bio_9490004} &  2021 & 231 & 79 & 240 & 115 & The first transparent optical-\ac{THz}-optical link providing record-high line-rates up to 240 and \SI{190}{\giga\bps} over distances from 5 to 115 meters is demonstrated.\\ \hline
 \cite{Bio_9881873}  &  2022 & 340-510 & 37.7 & 103 & 3 & Demonstrates a real-time fiber-\ac{THz}-fiber 2 × 2 \ac{MIMO} seamless integration system at 340–\SI{510}{\giga\hertz} using commercial DCO modules for baseband signals processing, which realizes a record net rate of \SI{103.125}{\giga\bps} DP-QPSK signals delivery over two spans of 20 km wireline single-mode fiber-link and 3 m wireless 2 × 2 MIMO link without using THz power amplifier. \\ \hline
 \cite{Bio_zhu2023ultra}  & 2023 & 360-430 & 37.7 & 206 & 1 &  A novel UWB fiber-\ac{THz}-fiber seamlessly converged real-time architecture, which utilizes the commercially mature digital coherent optical module to realize ultrahigh-capacity \ac{THz} real-time wireless communication, is proposed.\\ \hline \hline
\end{tabular}
\end{table*}

In order to give readers an insightful view of the current status of the practical use of \ac{THz} communications and sensing towards \ac{6G} and beyond, we summarize state-of-the-art \ac{THz} trials and experiments worldwide in this section, where the achieved data rates at the certain \ac{THz} bands with specific features are surveyed.

In the past decade, the electronic mixing technology was widely applied to generate  high-frequency \ac{THz} signals by up-converting a low-frequency microwave signal, as the traditional way to realize \ac{THz} transmission as listed in Table \ref{table_THzExpe} \cite{Bio_moeller20112,Bio_kallfass201564,Bio_qiuyu2017design}. One of the most remarkable approaches was done by Bell Labs in 2011, where \ac{THz} radiation at \SI{625}{\giga\hertz} was generated by using an all-solid-state electric mixer. It achieved a data rate of \SI{2.5}{\giga\bps} at a distance of \SI{0.2}{\meter} under the transmission power of \SI{1}{\milli\watt}  \cite{Bio_moeller20112}. In 2015, the researchers at the University of Stuttgart in Germany successfully transmitted \SI{240}{\giga\hertz} \ac{THz} signals to the receiver at a distance of \SI{850}{\meter}.  The trial achieved a peak data rate of \SI{64}{\giga\bps} using \ac{QPSK} and \ac{8PSK} modulation in a single-channel approach without the use of spatial diversity \cite{Bio_kallfass201564}. In the year 2017, a research team from the China Academy of Engineering Physics achieved ultra-long-distance \ac{THz} wireless communications over up to \SI{21}{\kilo\meter} and realized single-channel transmission speed up to \SI{5}{\giga\bps}, taking advantage of two Cassegrain antennas with \SI{50}{\decibel}i gain each \cite{Bio_qiuyu2017design}. In the same year, the first demonstration of photonically-enabled independent side-bands D-Band wireless transmission up to \SI{352}{\giga\bps} by means of $2\times2$ antenna polarization multiplexing was achieved \cite{Ref_puerta2017demostration}.  

Because of the inherent properties of electronic devices, the parameters of high-frequency electronic devices gradually approach the theoretical limit, with relatively lower bandwidth and a limited transmission rate. 
Recently, much attention was shifted to the photonics-assisted heterodyne beating technique for higher data rate and better signal quality, where the rates of \ac{THz} transmission is able to reach hundreds of \SI{}{\giga\bps} or even \SI{}{\tera\bps} \cite{Bio_koenig2013wireless,Bio_8424023,Bio_9096572,Bio_harter2020generalized,Bio_li202154,Bio_9490004}. It should be pointed out that  the \ac{THz} signal power generated by the photonics-assisted heterodyne beating method is usually limited to the \SI{}{\milli\watt} level because of the lower responsivity of the uni-traveling carrier photodiode (UTC-PD), resulting in the limited transmission distance. Therefore, some researchers utilized high-gain \ac{THz} amplifiers or high-gain lens antennas to extend distances to \SI{100}{\meter}. In early 2013, the researchers \cite{Bio_koenig2013wireless} utilized the large frequency range in the \ac{THz} window between \SI{200}{\giga\hertz} and \SI{300}{\giga\hertz} to implement a \ac{SISO} wireless \SI{100}{\giga\bps} link with a carrier frequency of \SI{237.5}{\giga\hertz} over a distance of \SI{20}{\meter}. Several years later, a team from Fudan University, China successfully applied $2\times2$ \ac{MIMO} and wavelength division multiplexing (WDM) technologies at \ac{THz} signal transmission, achieving a data rate of \SI{120}{\giga\bps} by using \ac{QPSK} modulation \cite{Bio_8424023}. Meanwhile, some researchers at Zhejiang University in China achieved \ac{THz} signal transmission of \SI{600}{\giga\bps} using 64QAM multi-carrier modulation \cite{Bio_9096572}. However, the distances of \ac{THz} signal transmission of the above two approaches are only \SI{1.42}{\meter} and \SI{2.8}{\meter}, respectively. 

In the past three years, some research teams have presented prominent improvements in \ac{THz} communications. The wireless transmission distances were effectively extended to more than \SI{100}{\meter} with the assistance of high-gain \ac{THz} ampliﬁers or high-gain lens antenna. In 2020, a team at Karlsruhe Institute of Technology (KIT), Germany took advantage of \ac{THz} amplifiers and the Kramers-Kronig method for simplifying the design of the receiver and launched an offline multi-carrier \ac{THz} system. It offers a peak data rate of \SI{115}{\giga\bps} at a carrier frequency of \SI{300}{\giga\hertz} over a distance of \SI{110}{\meter} \cite{Bio_harter2020generalized}. One year later, the Fudan University in China successfully transmitted a \SI{44.8}{\giga\bps} 64QAM-modulated signal over a distance of \SI{104}{\meter} without using the \ac{THz} amplifier but utilizing both suitable dielectric lenses and \ac{DSP} algorithms \cite{Bio_li202154}. In the same year, Yannik Horst \emph{et al.} \cite{Bio_9490004} from Switzerland demonstrated the transparent optical-\ac{THz}-optical link, providing a transmission rate of \SI{240}{\giga\bps} over a distance up to \SI{115}{\meter}. 

With the objective of achieving full-coverage and low-cost deployment towards future \ac{6G} mobile communications,  the priority of the hybrid optoelectronic down-conversation solution was presented \cite{Bio_9881873} and \cite{Bio_zhu2023ultra}, where a novel fiber-\ac{THz}-fiber seamlessly converged real-time architecture was successfully demonstrated. It adopts both dual-polarization photonic up-conversion for \ac{THz} signal generation and hybrid optoelectronic down-conversion for \ac{THz} reception, by thoroughly reusing commercial digital coherent optical modules. In the case of hybrid channel transmission with two hops consisting of a 20km-long fiber and 1m-long \ac{THz} wireless link, a \ac{THz} signal with a net rate of \SI{206.25}{\giga\bps} was successfully transmitted real-timely \cite{Bio_zhu2023ultra}. It is also pointed out that zhe \ac{THz} phased array techniques are key to realizing \ac{6G} \ac{THz} mobile communications and sensing, which meets the needs of application scenarios, such as multiple users and beam tracking.

In addition to the excellent demonstrations and validations that have been achieved by research teams around the world, some equipment suppliers and organizations have also presented great advances in \ac{THz} commercialization. The NYU WIRELESS is currently focusing on sub-\ac{THz} bands at \SI{140}{\giga\hertz}, \SI{220}{\giga\hertz}, and higher. The radio-frequency integrated circuit (RFIC) probe stations working up to \SI{220}{\giga\hertz}, and channel sounders for propagation measurement at \SI{140}{\giga\hertz} \cite{Bio_9685929} are provided by the Keysight Technologies. Keysight has also closely corporated with Nokia Bell labs on the sub-\ac{THz} testbed, which was chosen to verify the performance of transceiver modules, power amplifiers, and antennas under both linear and nonlinear conditions.  Recently, the Huawei \ac{6G} research team has developed and demonstrated \ac{THz} integrated sensing and communications (\ac{THz}-\ac{ISAC}) prototype. Using wireless electromagnetic waves, the prototype can sense and produce images of blocked objects with millimeter-level resolution and communicates at an ultra-high rate of \SI{240}{\giga\bps}, opening up new service possibilities for \ac{6G} and beyond systems \cite{Ref_li2021integrated}.

\section{CONCLUSIONS}
In summary, the upcoming \ac{6G} and beyond cellular systems are envisioned to exploit the \ac{THz} band beyond \SI{100}{\giga\hertz}, which not only offers an abundant amount of spectral resources for globally ubiquitous, ultra-high-rate, super-reliable, hyper-low-latency, massive-density telecommunications services but also empowers high-resolution cognition through \ac{THz} sensing, positioning, and imaging. The use of \ac{THz} frequencies will bring novel applications such as tera-bits-per-second/Tbps hot spots or links, and, in addition, disruptive uses like nano-scale networks and on-chip communications.   
Despite its high potential, we do not expect that the \ac{THz} band can replace the sub-6GHz and \ac{mmWave} bands, which have been employed as the basis of previous generations of cellular communications networks. Instead, the \ac{THz} band is highly probably being used as the complementary resource to aid the success of low-frequency bands in future generations of cellular systems. Meanwhile, there is still tremendous work to be done in terms of characterizing and modeling \ac{THz} channels, developing affordable, usable \ac{THz} antennas and devices, designing novel algorithms for long-range \ac{THz} signal transmission, proposing efficient protocols for flexible \ac{THz} networking, and elaborately considering its synergy with other 6G-enabling technologies. It is hoped that this survey could be able to provide the researchers with a holistic view of all technical aspects and issues required to design and build \ac{THz} communications and sensing for \ac{6G} and beyond from an application and implementation perspective. Although there is a long journey to go before the success of \ac{THz} communications and sensing in \ac{6G} and beyond cellular systems, this survey might be able to speed up a bit the research endeavors. 



\section{List of Acronyms}
\addcontentsline{lof}{section}{List of acronyms}
\input{acronyms.tex}
\input{glossary.tex}

\bibliographystyle{IEEEtran}
\bibliography{IEEEabrv,Ref_Wei.bib,Ref_Qiuheng.bib, Ref_Jiguang,Ref_Mohammed.bib,Ref_Sergiy.bib, Ref_Asif.bib,Ref_Bin.bib}

\begin{IEEEbiography}
[{\includegraphics[width=1in,height=1.25in,clip,keepaspectratio]{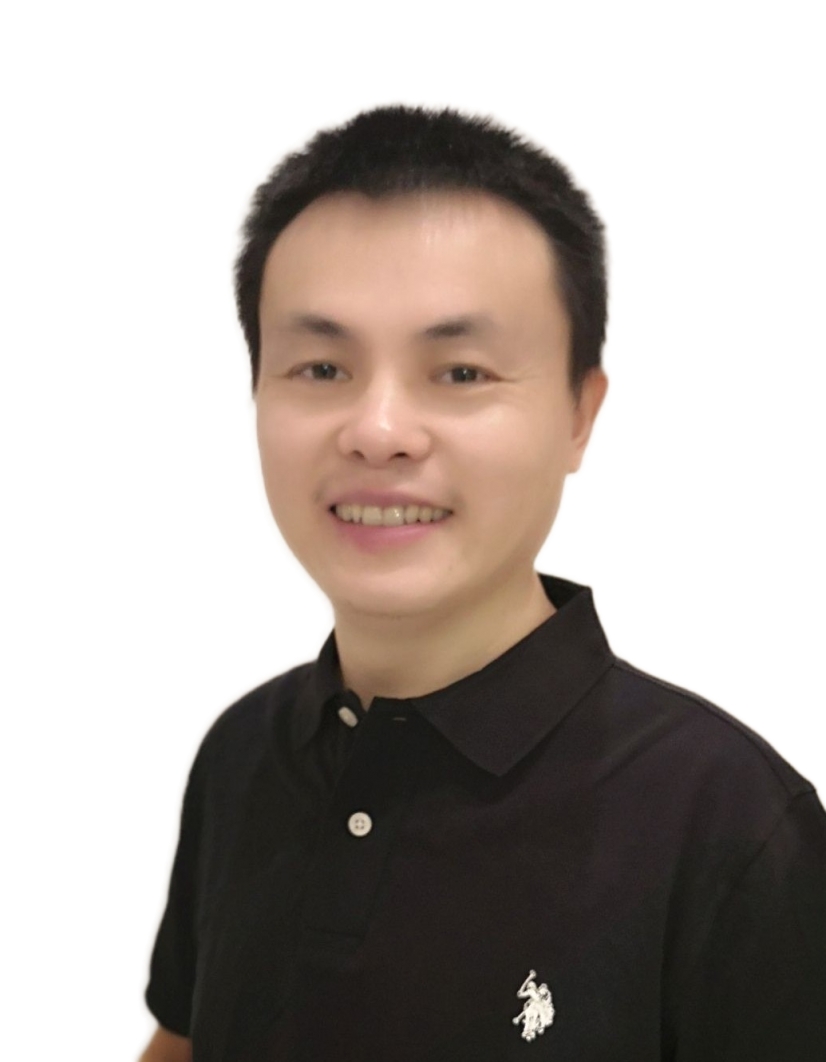}}]
{Wei Jiang} (M'09-SM'19) received a Ph.D. degree in Computer Science from Beijing University of Posts and Telecommunications (BUPT) in 2008. From 2008 to 2012, he was with the 2012 Laboratory, HUAWEI Technologies. From 2012 to 2015, he was with the Institute of Digital Signal Processing, University of Duisburg-Essen, Germany. Since 2015, he has been a Senior Researcher with German Research Center for Artificial Intelligence (DFKI), the biggest European AI research institution and the birthplace of the ``Industry 4.0" strategy. Meanwhile, he was a Senior Lecturer at the University of Kaiserslautern, Germany, from 2016 to 2018. He has published two monographs -- \textit{6G Key Technologies: A Comprehensive Guide (Wiley \& IEEE Press, 2023)} and \textit{Cellular Communication Networks and Standards: The Evolution from 1G to 6G (Springer, 2024)}, and contributed three book chapters in 5G and machine-learning-based communications. He has over 100 conference and journal papers, holds around 30 granted patents, and participated in a number of EU and German research projects: \textit{ABSOLUTE}, \textit{5G COHERENT}, \textit{5G SELFNET}, \textit{5G-ACIA}, \textit{AI@EDGE}, \textit{TACNET4.0}, \textit{KICK}, \textit{AI-NET-ANTELLAS}, and \textit{Open6GHub}. He received the best paper award in IEEE CQR 2022 and the best presentation award in IEEE CCAI 2023. He was the Guest Editor for the Special Issue on ``Computational Radio Intelligence: A Key for 6G Wireless" in \textit{ZTE Communications} (December 2019). He serves as an Associate Editor for \textit{IEEE Access} (2019-2023), an Editor for \textit{IEEE Communications Letters} and \textit{IEEE Open Journal of the Communications Society}, and a Moderator for \textit{IEEE TechRxiv}. He served as a member of the organizing committee or technical committee for many conferences such as IEEE ICASSP 2022, CCS 2014, PIMRC 2015/2020, GLOBECOM 2022/2023/2024, ICC 2023/2024, WCNC 2024, ICCC 2017/2021/2022/2023, HP3C 2022, HSPR 2024, and icWCSN 2022/2023. He was the founding member and vice chair of the special interest group (SIG) “Cognitive Radio in 5G” under the IEEE Technical Committee on Cognitive Networks (TCCN). 
\end{IEEEbiography}

\begin{IEEEbiography}
[{\includegraphics[width=1in,height=1.25in,clip,keepaspectratio]{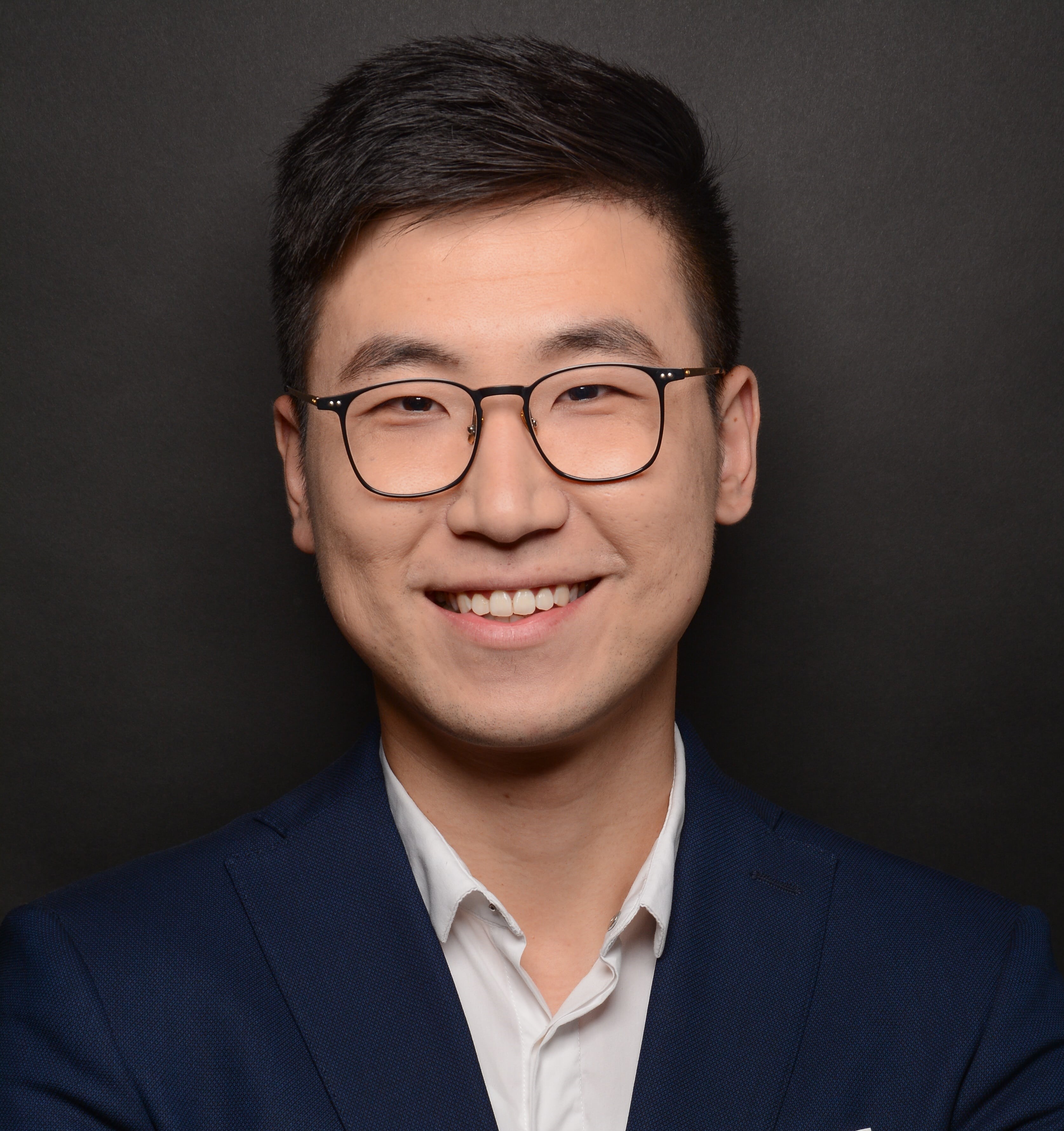}}]
{Qiuheng Zhou} received his B.Sc degree in Electronics and Information Engineering from Tianjin University of Technology, China, in 2015. He obtained his M.Sc. degree in autonomous and networked driving from the University of Stuttgart, Germany, in 2020. Since April 2020, he has been working as a researcher in the intelligent networks of German Research Center for Artificial Intelligence (DFKI).  His main research interests include channel measurement, software-defined radio networks, machine learning, channel prediction, and resource allocation. 
\end{IEEEbiography}

\begin{IEEEbiography}[{\includegraphics[height=1.15in,clip,keepaspectratio]{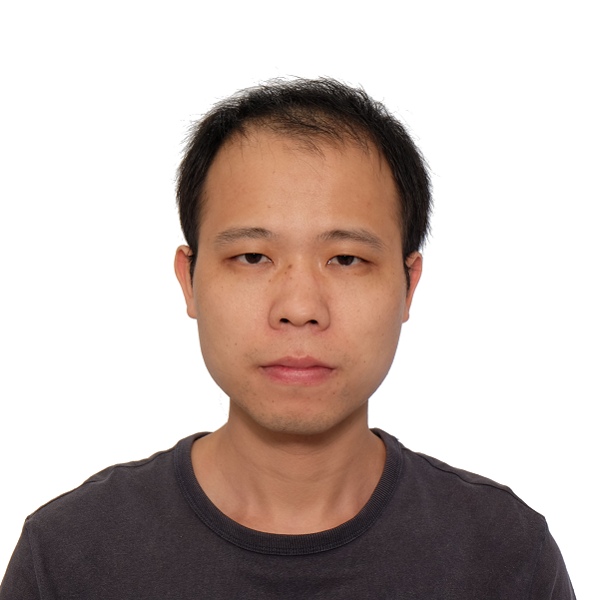}}]{Jiguang He} (M'20-SM'22)
received the Ph.D. degree from the University of Oulu, Finland, in 2018 on communications engineering. He is now a senior researcher at Technology Innovation Institute, Abu Dhabi, United Arab Emirates, and holds Docentship (adjunct professor) at the University of Oulu. From September 2013 to March 2015, he was with the State Key Laboratory of Terahertz and Millimeter Waves at the City University of Hong Kong, working on beam tracking over millimeter wave MIMO systems. From June 2015 to August 2021, he has been with the Centre for Wireless Communications (CWC), University of Oulu, Finland, first as a doctoral candidate, and then a postdoctoral researcher. He was an assistant professor at Macau University of Science and Technology from August 2021 to March 2022. He has participated in many international and national projects, e.g., EU FP7 RESCUE, EU H2020 ARIADNE, 6G Flagship, and received one FDCT-GDST joint research project from Macau Science and Technology Development Fund. He is an Exemplary Reviewer for IEEE Transactions on Communications as well as IEEE Communications Letters and a TPC member for various prestigious IEEE conferences. His research interests span millimeter wave MIMO communications, reconfigurable intelligent surfaces for simultaneous localization and communications (SLAC), and advanced signal processing techniques.
 \end{IEEEbiography}

\begin{IEEEbiography}[{\includegraphics[width=1in,height=1.25in,clip,keepaspectratio]{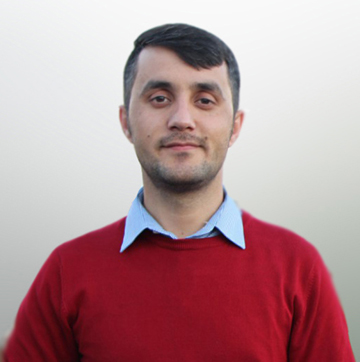}}]{Mohammad Asif Habibi} received his B.Sc. degree in Telecommunications Engineering from Kabul University, Afghanistan, in 2011. He obtained his M.Sc. degree in Systems Engineering and Informatics from the Czech University of Life Sciences, Czech Republic, in 2016. Since January 2017, he has been working as a research fellow and Ph.D. candidate at the Division of Wireless Communications and Radio Navigation, Rheinland-Pf\"alzische Technische Universit\"at (previously known as Technische Universit\"at Kaiserslautern), Germany.  From 2011 to 2014, he worked as a radio access network engineer for HUAWEI. His main research interests include network slicing, network function virtualization, resource allocation, machine learning, and radio access network architecture.
\end{IEEEbiography} 

\begin{IEEEbiography}
[{\includegraphics[width=1in,height=1.25in,clip,keepaspectratio]{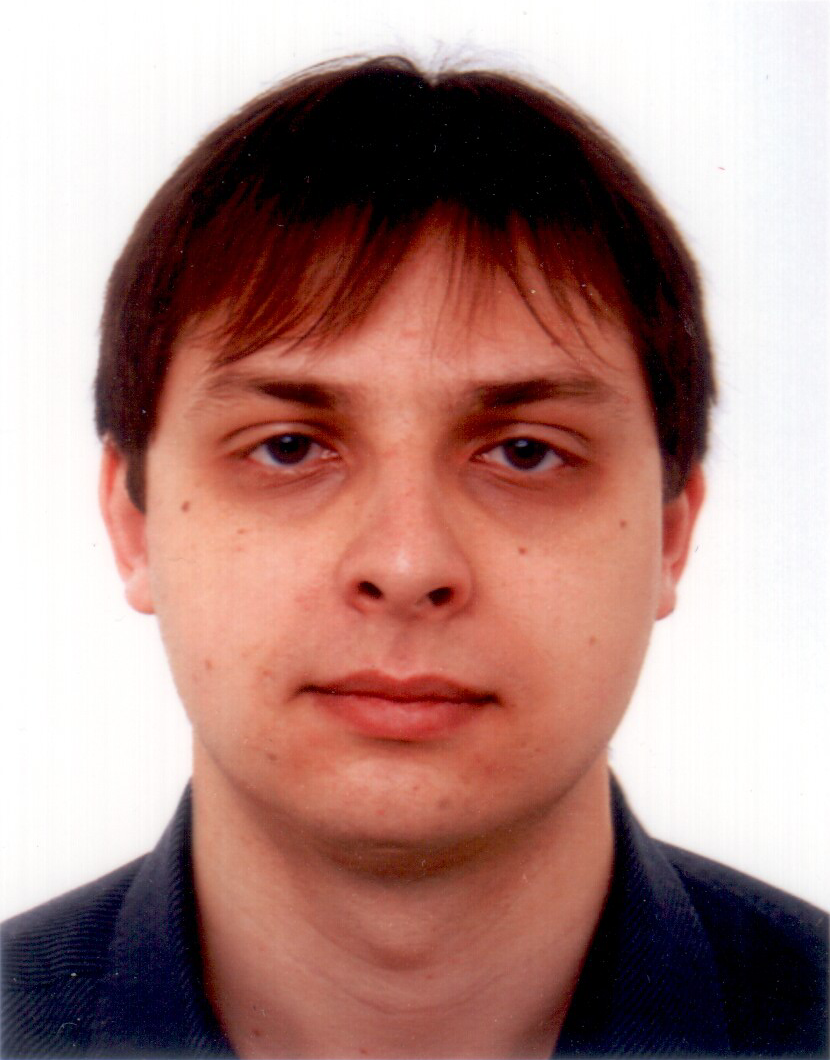}}]
{Sergiy Melnyk}  received his Dipl.-Ing degree in Electronics and Information Engineering from Technische University Munich in 2012. Since April 2015, he has been working as a researcher in the Intelligent Networks Group of German Research Center for Artificial Intelligence (DFKI).  His main research interests include industrial communications systems,  software-defined radio networks, and lower-layer protocol design.
\end{IEEEbiography}

\begin{IEEEbiography}[{\includegraphics[width=1in,height=1.25in,clip,keepaspectratio]{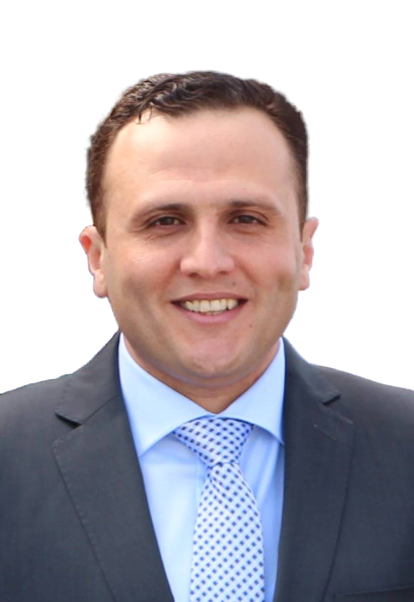}}]{Mohammed El-Absi} is currently working as a Senior Researcher at Digital Signal Processing Institute at University of Duisburg-Essen, Duisburg, Germany, where he received his Ph.D. degree (Summa Cum Laude) in electrical engineering in 2015. He received the M.S. degree in electrical engineering in 2008 from Jordan University of Science and Technology and the B.E. degree in electrical engineering in 2005 from Islamic University of Gaza, Gaza, Palestine. He received a Mercator fellow at the Collaborative Research Center "Mobile Material Characterization and Localization by Electromagnetic Sensing" (MARIE) in the period of 2017-2018. He received the German Academic Exchange Service Fellowship in 2006 and 2011. He is currently contributing in 6G research hub for open, efficient and secure mobile radio systems (6GEM) and Collaborative Research Center "Mobile Material Characterization and Localization by Electromagnetic Sensing" (MARIE). He is a principal investigator in the excellent terahertz research for communication, localization, material characterization, medical technology and environmental monitoring (terahertz.NRW). His research interests are in the area of communications and signal processing.
\end{IEEEbiography}

\begin{IEEEbiography}
[{\includegraphics[width=1in,height=1.25in,clip,keepaspectratio]{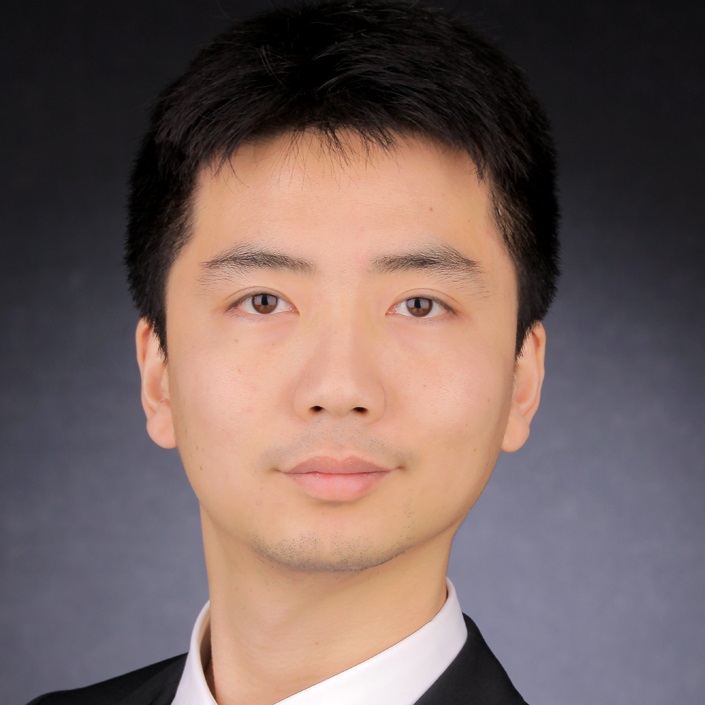}}]
{Bin Han} (M'15-SM'21) received his B.E. degree in 2009 from Shanghai Jiao Tong University, M.Sc. in 2012 from the Technical University of Darmstadt, and a Ph.D. degree in 2016 from Karlsruhe Institute of Technology. Since July 2016 he has been with the Division of Wireless Communications and Radio Positioning, RPTU Kaiserslautern-Landau (formerly: Technical University of Kaiserslautern) as a Postdoctoral Researcher and Senior Lecturer. His research interests are in the broad area of wireless communications and networking, with the current focus on B5G/6G and MEC. He is the author of one book, five book chapters, and over 50 research papers. He has participated in multiple EU research projects. He is Editorial Board Member for Network, Guest Editor for Electronics, and has served as Organizing Committee Member and/or TPC Member for GLOBECOM, ICC, EuCNC, EW, and ITC. He is a voting member of the IEEE Standards Association Working Groups P2303 and P3106.
\end{IEEEbiography}

\begin{IEEEbiography}
[{\includegraphics[width=1in,height=1.25in,clip,keepaspectratio]{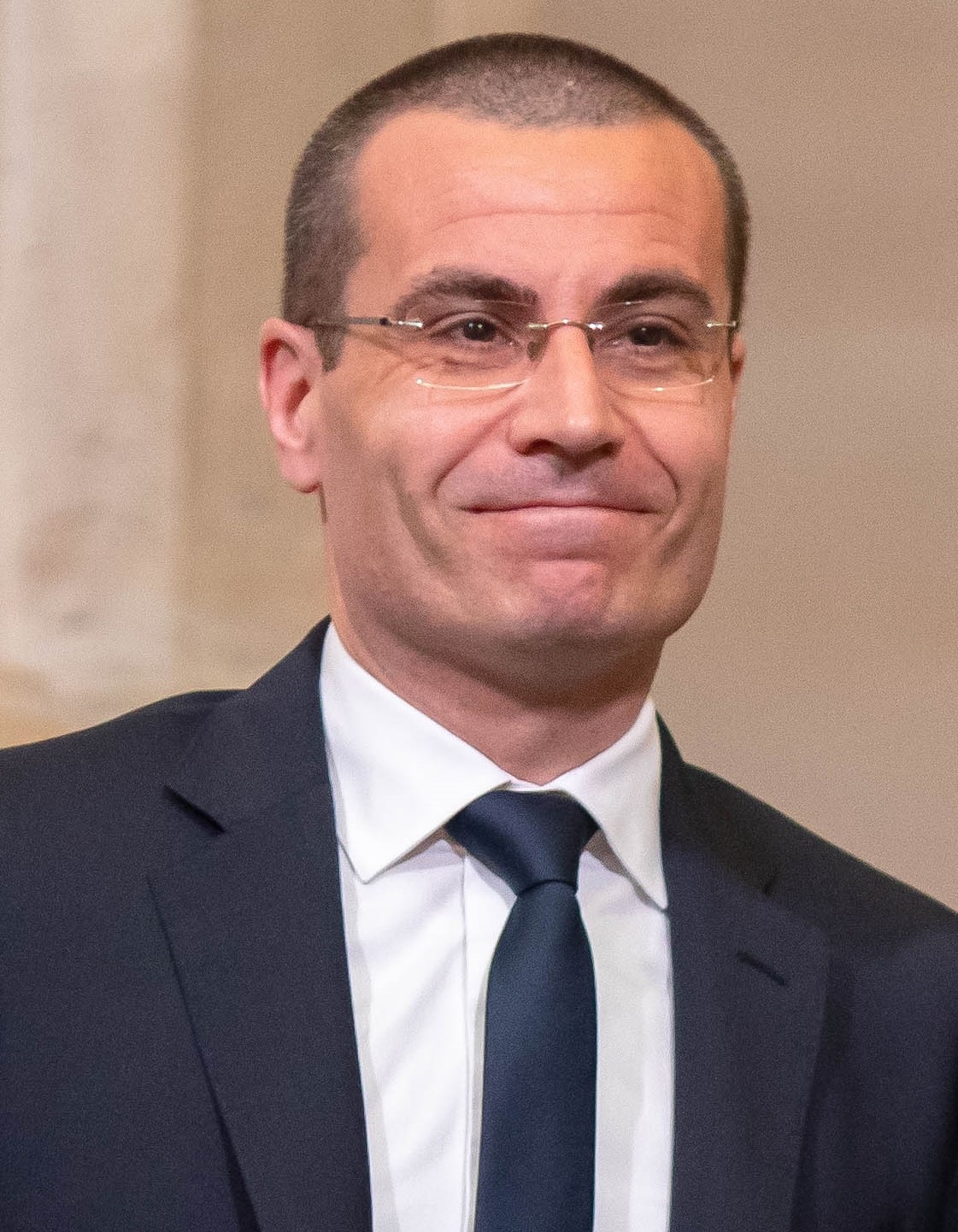}}]
{Marco Di Renzo} (Fellow, IEEE) received the Laurea (cum laude) and Ph.D. degrees in electrical engineering from the University of L’Aquila, Italy, in 2003 and 2007, respectively, and the Habilitation à Diriger des Recherches (Doctor of Science) degree from University Paris-Sud (currently Paris-Saclay University), France, in 2013. Currently, he is a CNRS Research Director (Professor) and the Head of the Intelligent Physical Communications group in the Laboratory of Signals and Systems (L2S) at Paris-Saclay University – CNRS and CentraleSupelec, Paris, France. Also, he is an elected member of the L2S Board Council and a member of the L2S Management Committee, and is a Member of the Admission and Evaluation Committee of the Ph.D. School on Information and Communication Technologies, Paris-Saclay University. He is a Founding Member and the Academic Vice Chair of the Industry Specification Group (ISG) on Reconfigurable Intelligent Surfaces (RIS) within the European Telecommunications Standards Institute (ETSI), where he served as the Rapporteur for the work item on communication models, channel models, and evaluation methodologies. He is a Fellow of the IEEE, IET, and AAIA; an Academician of AIIA; an Ordinary Member of the European Academy of Sciences and Arts, an Ordinary Member of the Academia Europaea; and a Highly Cited Researcher. Also, he holds the 2023 France-Nokia Chair of Excellence in ICT, and was a Fulbright Fellow at the City University of New York (USA), a Nokia Foundation Visiting Professor (Finland), and a Royal Academy of Engineering Distinguished Visiting Fellow (UK). His recent research awards include the 2021 EURASIP Best Paper Award, the 2022 IEEE COMSOC Outstanding Paper Award, the 2022 Michel Monpetit Prize conferred by the French Academy of Sciences, the 2023 EURASIP Best Paper Award, the 2023 IEEE ICC Best Paper Award, the 2023 IEEE COMSOC Fred W. Ellersick Prize, the 2023 IEEE COMSOC Heinrich Hertz Award, the 2023 IEEE VTS James Evans Avant Garde Award, and the 2023 IEEE COMSOC Technical Recognition Award from the Signal Processing and Computing for Communications Technical Committee. He served as the Editor-in-Chief of IEEE Communications Letters during the period 2019-2023, and he is now serving on the Advisory Board. He currently serves as a Voting Member of the Fellow Evaluation Standing Committee and as the Director of Journals of the IEEE Communications Society.
\end{IEEEbiography}

\begin{IEEEbiography}
[{\includegraphics[width=1in,height=1.25in,clip,keepaspectratio]{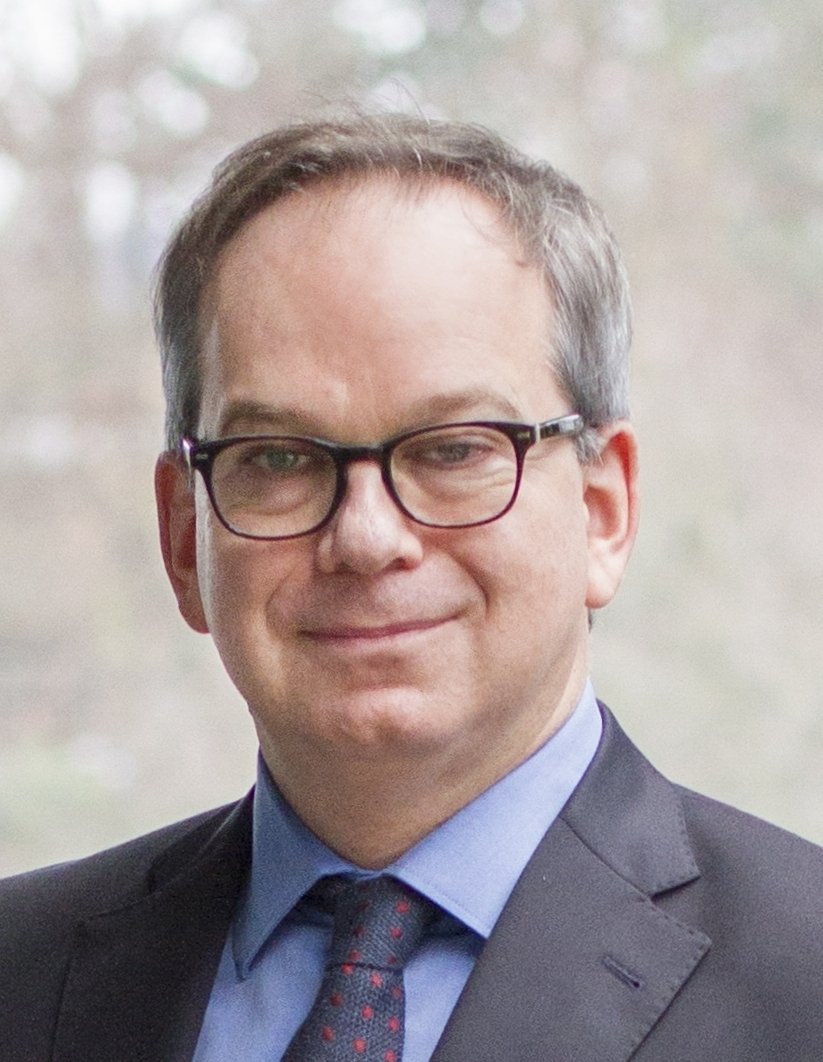}}]
{Hans D. Schotten} (S'93-M'97) received the Ph.D. degree from the RWTH Aachen University of Technology, Germany, in 1997. From 1999 to 2003, he worked for Ericsson. From 2003 to 2007, he worked for Qualcomm. He became manager of a R\&D group, Research Coordinator for Qualcomm Europe, and Director for Technical Standards. In 2007, he accepted the offer to become a full professor at the University of Kaiserslautern. In 2012, he - in addition - became the scientific director of the German Research Center for Artificial Intelligence (DFKI) and head of the Department for Intelligent Networks. Professor Schotten served as dean of the Department of Electrical Engineering of the University of Kaiserslautern from 2013 until 2017. Since 2018, he is chairman of the German Society for Information Technology and a member of the Supervisory Board of the VDE. He is the author of more than 200 papers and participated in 30+ European and national collaborative research projects.
\end{IEEEbiography}

\begin{IEEEbiography}
[{\includegraphics[width=1in,height=1.25in,clip,keepaspectratio]{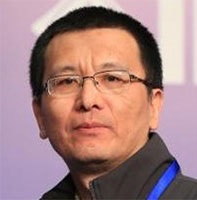}}]
{Fa-Long Luo} (SM'95-F'16) has served as a Board Member of both the Conference Board and the Membership Board of the IEEE Signal Processing Society (SPS) as well as an IEEE Fellow Committee Member. Dr. Luo served as the Society Representative of SPS in the IEEE TAB Committee on Standards.  He was the Chairman of the Industry DSP Technology Standing Committee (IDSP-SC) and a Technical Directions Board Member of IEEE SPS as well as a founding member of the IoT SIG of SPS.  He was the founding Editor-in-Chief of the International Journal of Digital Multimedia Broadcasting. 
He has 39 years of academic, industry, and research experience with an emphasis on translating adaptive signal processing techniques to practical applications and commercial products related to multimedia, wireless communications, and digital broadcasting.  Including his well-received books: “Applied Neural Networks for Signal Processing” (1997, Cambridge University Press) and “Signal Processing for 5G: Algorithms and Implementations” (2016, Wiley-IEEE), Dr. Luo has published 7 books and more than 100 technical papers in the related fields.  Dr. Luo has also contributed 115 USA patents (issued or pending) which have successfully resulted in a number of new or improved commercial products in mass production. 
\end{IEEEbiography}

\begin{IEEEbiography}
[{\includegraphics[width=1in,height=1.25in,clip,keepaspectratio]{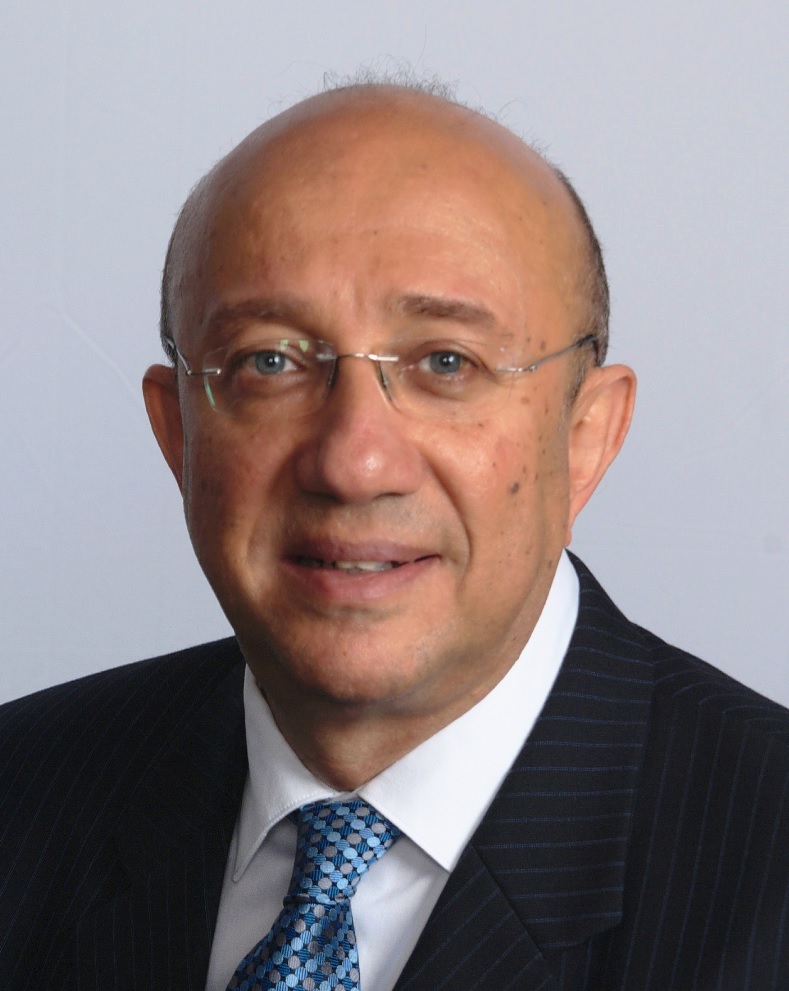}}]
{Tarek S. El-Bawab} (Fellow, IEEE) is the Dean of the School of Engineering and Professor of Electrical and Computer Engineering at the American University of Nigeria. Before this, he was Professor and Dean of Engineering and Applied Sciences at Nile University (Egypt), Professor of Electrical and Computer Engineering at Jackson State University (USA), and Project Manager with the Network Strategy Group of Alcatel-Lucent USA (now Nokia). Earlier, he assumed research roles with Alcatel-Lucent, Colorado State University (USA), and the University of Essex (UK). Before this he led large-scale international telecommunication projects in the Middle East and Africa for 10 years.

Dr. El-Bawab research interests include telecommunications, network architectures, performance analysis, and Discipline Based Education Research (DBER). He has more than 80 scholarly journal/conference papers and patents. His book Optical Switching is one of the most comprehensive references in its subject. He is the Series Editor of Springer’s Textbooks in Telecommunication Engineering, the Editor in Chief of the IEEE Communications Magazine (2017-2021), IEEE Distinguished Lecturer (2016-2019), Eta Kappa Nu (HKN) member, and NSF Review Panelist. Tarek led the Telecommunication Engineering Education (TEE) initiative and movement (2008-2014), which resulted in recognition of network/telecommunication engineering as distinct ABET-accreditable education discipline. He is the first recipient of the IEEE Communications Society’s (ComSoc) Education Award due to this work (2015). 

He has served IEEE and ComSoc in numerous capacities. He serves/served as Board Member of the IEEE Educational Activities Board, EAB (2016-2017), as Board Member of the IEEE PSPB’s Thesaurus Editorial Board (2021-2023), as Board Member of the ComSoc Board of Governors (2014-2015, 2018-2019, and 2020-2021), and Board Member of the ComSoc Educational Services Board (2012-2019). He served as the ComSoc Director of Industry Communities (2020-2021), Director for Standards Development (2018-2019) and Director of Conference Operations (2014-2015). He was elected as Chair of the Transmission, Access, and Optical Systems (TAOS) Technical Committee for two terms, and as chair in several ICC/Globecom Conferences. Dr. El-Bawab has B.Sc. in electrical engineering from Ain Shams University (Egypt), M.Sc. in solid state science from the American University in Cairo, M.Sc. in telecommunications and information systems from the University of Essex (UK), and Ph.D. in electrical engineering from Colorado State University (USA).
\end{IEEEbiography}

\begin{IEEEbiography}
[{\includegraphics[width=1in,height=1.25in,clip,keepaspectratio]{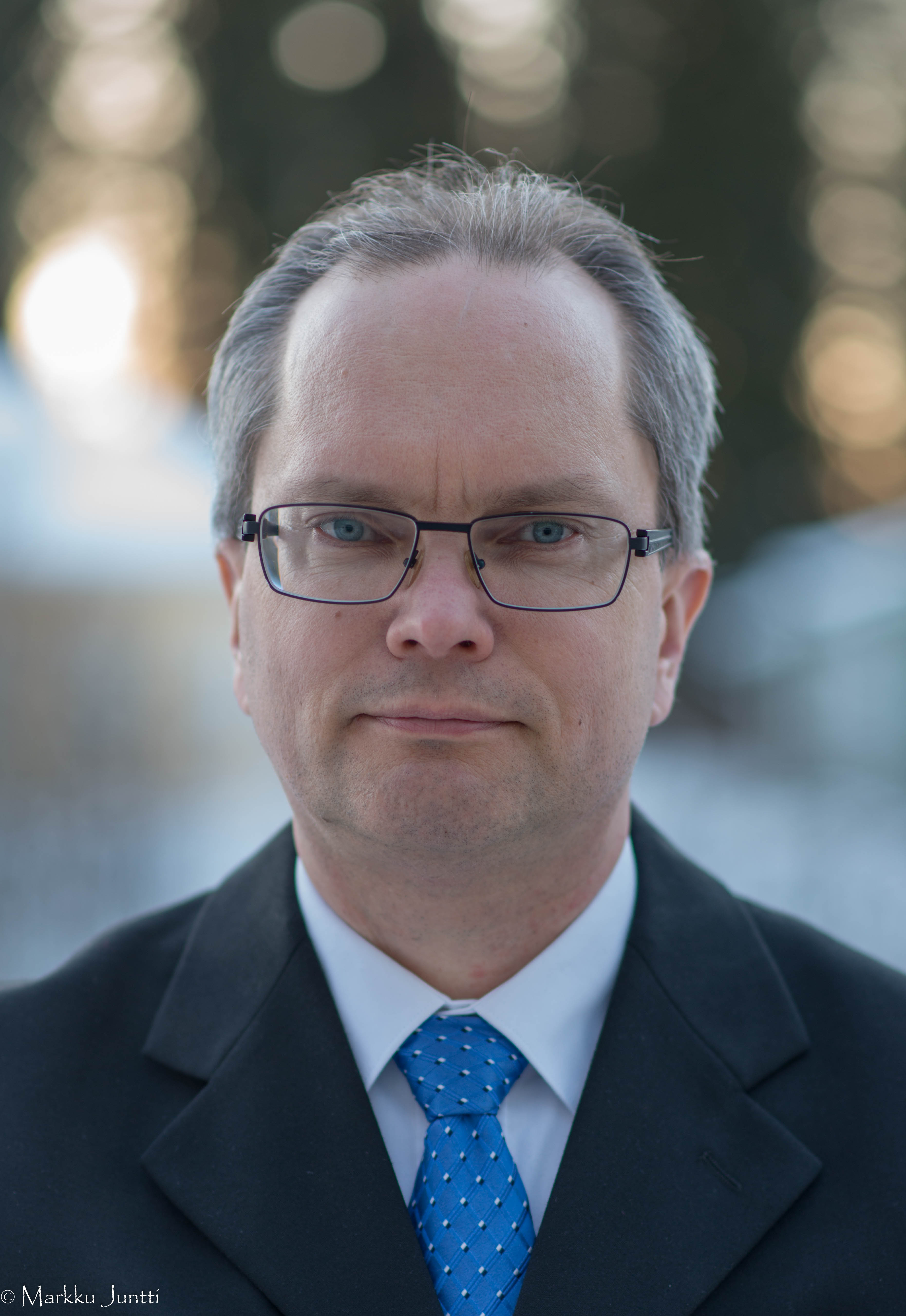}}]{Markku Juntti} (Fellow, IEEE) received his M.Sc.\ (EE) and Dr.Sc.\ (EE) degrees from University of Oulu, Oulu, Finland in 1993 and 1997, respectively.
Dr.\ Juntti was with University of Oulu in 1992--98. In academic year 1994--95, he was a Visiting Scholar at Rice University, Houston, Texas. In 1999--2000, he was a Senior Specialist with Nokia Networks in Oulu, Finland. Dr.\ Juntti has been a professor of communications engineering since 2000 at University of Oulu, Centre for Wireless Communications (CWC), where he leads the Communications Signal Processing (CSP) Research Group. He also serves as Leader of CWC -- Radio Technologies (RT) Research Unit. His research interests include signal processing for wireless networks as well as communications and information theory. He is an author or co-author in almost 500 papers published in international journals and conference records as well as in books {\it Wideband CDMA for UMTS} in 2000--2010, {\it Handbook of Signal Processing Systems} in 2013 and 2018 and {\it 5G Wireless Technologies} in 2017. Dr.\ Juntti is also an Adjunct Professor at Department of Electrical and Computer Engineering, Rice University, Houston, Texas, USA.

Dr.\ Juntti is an Editor of \textsc{IEEE Transactions on Wireless Communications}, and served previously in a similar role in \textsc{IEEE Transactions on Communications} and \textsc{IEEE Transactions on Vehicular Technology}. He was Secretary of IEEE Communications Society Finland Chapter in 1996--97 and the Chairman for years 2000--01. He has been Secretary of the Technical Program Committee (TPC) of the 2001 IEEE International Conference on Communications (ICC), and the Chair or Co-Chair of the Technical Program Committee of several conferences including 2006 and 2021 IEEE International Symposium on Personal, Indoor and Mobile Radio Communications (PIMRC), the Signal Processing for Communications Symposium of IEEE Globecom 2014, Symposium on Transceivers and Signal Processing for 5G Wireless and mm-Wave Systems of IEEE GlobalSIP 2016, ACM NanoCom 2018, and 2019 International Symposium on Wireless communications Systems (ISWCS). He has also served as the General Chair of 2011 IEEE communications Theory Workshop (CTW 2011) and 2022 IEEE Workshop on Signal Processing Advances in Wireless Communications (SPAWC).
\end{IEEEbiography}

\begin{IEEEbiography}
[{\includegraphics[width=1in,height=1.25in,clip,keepaspectratio]{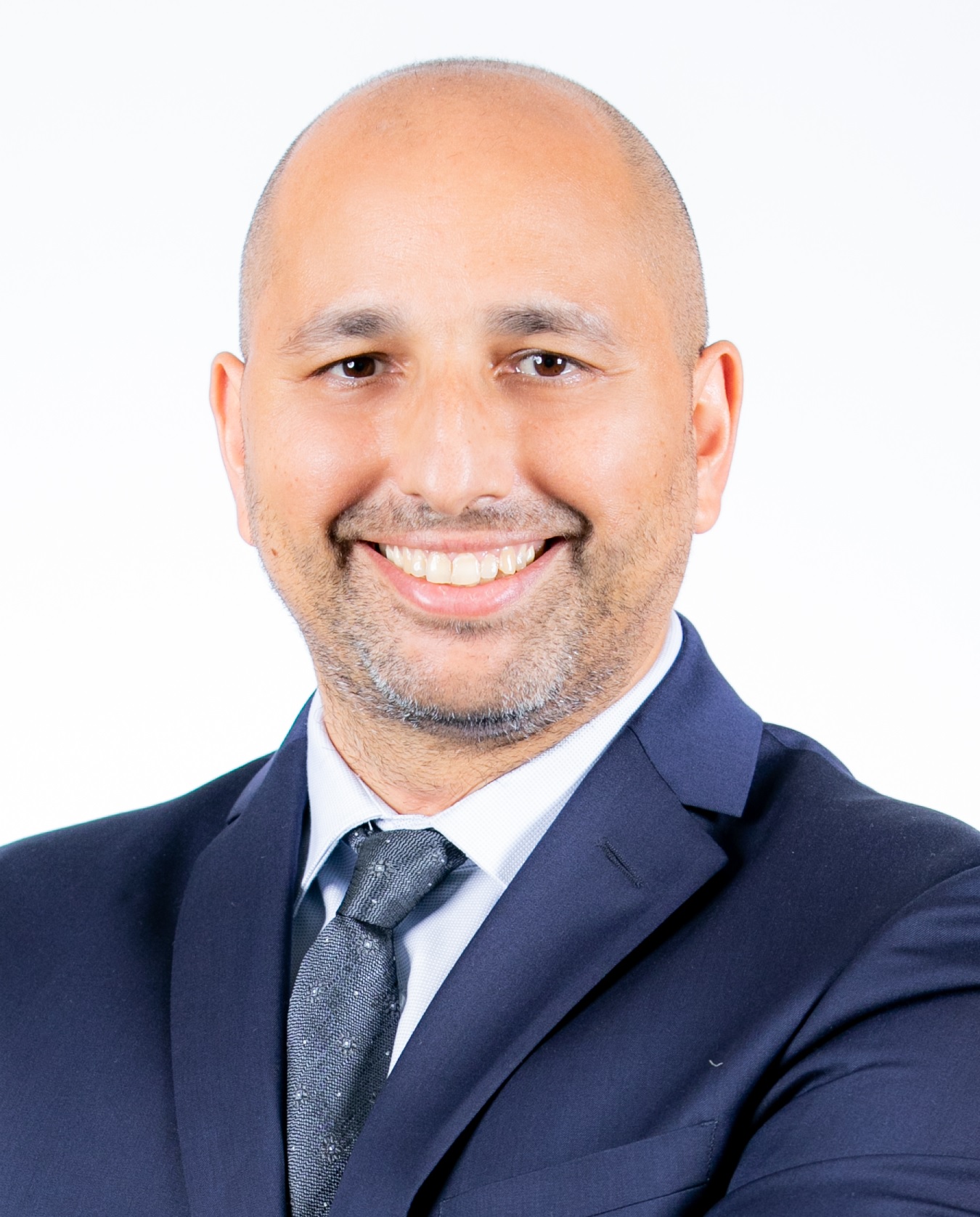}}]{Mérouane Debbah} (Fellow, IEEE) is a professor at  Khalifa University of Science and Technology in Abu Dhabi and founding Director of the KU 6G Research Center. He is a frequent keynote speaker at international events in the field of telecommunication and AI. His research has been lying at the interface of fundamental mathematics, algorithms, statistics, information, and communication sciences with a special focus on random matrix theory and learning algorithms. In the Communication field, he has been at the heart of the development of small cells (4G), Massive MIMO (5G), and Large Intelligent Surfaces (6G) technologies. In the AI field, he is known for his work on Large Language Models, distributed AI systems for networks, and semantic communications. He received multiple prestigious distinctions, prizes, and best paper awards for his contributions to both fields. He is an IEEE Fellow, a WWRF Fellow, an EURASIP Fellow, an AAIA Fellow, an Institut Louis Bachelier Fellow, and a Membre émérite SEE.
\end{IEEEbiography}

\begin{IEEEbiography}
[{\includegraphics[width=1in,height=1.25in,clip,keepaspectratio]{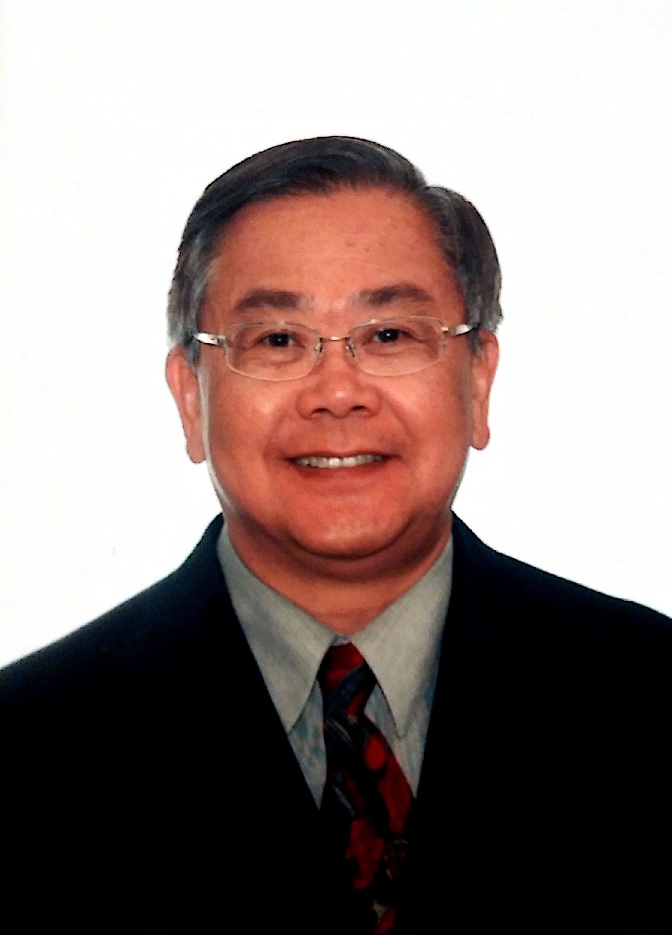}}]
{VICTOR C. M. LEUNG} (Life Fellow, IEEE) is a Distinguished Professor of Computer Science and Software Engineering at Shenzhen University, China. He is also an Emeritus Professor of Electrical and Computer Engineering and Director of the Laboratory for Wireless Networks and Mobile Systems at the University of British Columbia (UBC), Canada.  His research is in the broad areas of wireless networks and mobile systems, and he has published widely in these areas. His published works have together attracted more than 60,000 citations. He is named in the current Clarivate Analytics list of “Highly Cited Researchers”. Dr. Leung is serving on the editorial boards of the IEEE Transactions on Green Communications and Networking, IEEE Transactions on Cloud Computing, IEEE Transactions on Computational Social Systems, IEEE Access, and several other journals. He received the 1977 APEBC Gold Medal, 1977-1981 NSERC Postgraduate Scholarships, IEEE Vancouver Section Centennial Award, 2011 UBC Killam Research Prize, 2017 Canadian Award for Telecommunications Research, 2018 IEEE TCGCC Distinguished Technical Achievement Recognition Award, and 2018 ACM MSWiM Reginald Fessenden Award. He co-authored papers that won the 2017 IEEE ComSoc Fred W. Ellersick Prize, 2017 IEEE Systems Journal Best Paper Award, 2018 IEEE CSIM Best Journal Paper Award, and 2019 IEEE TCGCC Best Journal Paper Award.  He is a Life Fellow of IEEE, and a Fellow of the Royal Society of Canada (Academy of Science), Canadian Academy of Engineering, and Engineering Institute of Canada. 
\end{IEEEbiography}

\end{document}

%% file: config.tex
\usepackage{datatool}

\newcommand*{\addacronym}[2]{%
  \DTLnewrow{acronyms}%
  \DTLnewdbentry{acronyms}{Acronym}{#1}%
  \DTLnewdbentry{acronyms}{Description}{#2}%
}

\newenvironment{abbreviations}{%
                \begin{list}{failure!}{%
                        \setlength{\labelwidth}{6.2em}%
                        \setlength{\leftmargin}{6.2em}%
                        \setlength{\itemindent}{0cm}%
                        \setlength{\labelsep}{0pt}%
                }%
        }{%
                \end{list}%
        }

%% file: acronyms.tex
\begin{acronym}
\let\NewAcronym\acro
\NewAcronym{RAN}{radio access network}
\NewAcronym{ORAN}{open radio access network}
\NewAcronym{6G}{sixth generation}
\NewAcronym{1G}{first generation}
\NewAcronym{2D}{two-dimensional}
\NewAcronym{3D}{three-dimensional}
\NewAcronym{3G}{third generation}
\NewAcronym{3GPP}{Third Generation Partnership Project}
\NewAcronym{4G}{fourth generation}
\NewAcronym{5G}{fifth generation}
\NewAcronym{THz}{terahertz}
\NewAcronym{UAV}{unmanned aerial vehicle}
\NewAcronym{UE}{user equipment}
\NewAcronym{ULA}{uniform linear array}
\NewAcronym{UPAs}{uniform planar arrays}
\NewAcronym{ULBC}{ultra-reliable low-latency broadband communication}
\NewAcronym{uMBB}{ubiquitous mobile broadband}
\NewAcronym{UMMIMO}{ultra-massive multi-input multi-output}

\NewAcronym{URLLC}{ultra-reliable low-latency communications}
\NewAcronym{VLC}{visible light communication}
\NewAcronym{VNF}{virtual network function}
\NewAcronym{VR}{virtual reality}
\NewAcronym{WCDMA}{Wideband Code-Division Multiple Access}
\NewAcronym{WSN}{wireless sensor network}
\NewAcronym{WRC}{World Radiocommunication Conference}
\NewAcronym{ITU-T}{International Telecommunication Union - Telecommunication}
\NewAcronym{ITU-R}{International Telecommunication Union - Radiocommunication}
\NewAcronym{NGMN}{Next Generation Mobile Networks}
\NewAcronym{LEO}{low Earth orbit}
\NewAcronym{5GPPP}{Fifth Generation Private Public Partnership}
\NewAcronym{AI}{artificial intelligence}
\NewAcronym{AoI}{age of information}
\NewAcronym{AR}{augmented reality}
\NewAcronym{AoS}{age of synchronization}
\NewAcronym{AoSA}{array of subarrays}
\NewAcronym{AP}{access point}
\NewAcronym{AMPS}{Advanced Mobile Phone System}
\NewAcronym{GSM}{Global System for Mobile Communications}
\NewAcronym{LTE-Advanced}{Long-Term Evolution Advanced}
\NewAcronym{LoS}{line-of-sight}
\NewAcronym{MIMO}{multi-input multi-output}
\NewAcronym{MMIMO}{massive multi-input multi-output}
\NewAcronym{KPI}{key performance indicator}
\NewAcronym{NLoS}{non-line-of-sight}
\NewAcronym{BDCM}{beam-domain channel model}
\NewAcronym{CSI}{channel state information}
\NewAcronym{NOMA}{non-orthogonal multiple access}
\NewAcronym{ML}{machine learning}
\NewAcronym{PHY}{physical}
\NewAcronym{MC}{multi-connectivity}
\NewAcronym{CMC}{closest line of sight multi-connectivity}
\NewAcronym{RMC}{reactive multi-connectivity}
\NewAcronym{DMIMO}{distributed MIMO}
\NewAcronym{CoMP}{coordinated multi-point}
\NewAcronym{JT}{joint transmission}
\NewAcronym{CSCB}{coordinated scheduling and beamforming}
\NewAcronym{NTN}{non-terrestrial network}
\NewAcronym{CFN}{cell-free network}
\NewAcronym{SIC}{successive interference cancellation}
\NewAcronym{FBL}{finite blocklength}
\NewAcronym{CMTC}{critical machine-type communications}
\NewAcronym{RIS}{reconfigurable intelligent surfaces}
\NewAcronym{CCU}{cell center user}
\NewAcronym{CEU}{cell edge user}
\NewAcronym{CNOMA}{cooperative non-orthogonal multiple access}
\NewAcronym{SWIPT}{simultaneous wireless information and power transfer}
\NewAcronym{GEO}{geostationary Earth orbit}
\NewAcronym{RF}{radio frequency}
\NewAcronym{HAP}{high-altitude platform}
\NewAcronym{DNN}{deep neural network}
\NewAcronym{ISAC}{integrated sensing and communications}
\NewAcronym{DAR}{differential absorption radar}
\NewAcronym{FC}{fully-connected}
\NewAcronym{DAoSA}{dynamic array-of-subarrays}
\NewAcronym{FSPL}{free-space path loss}
\NewAcronym{SPP}{surface plasmon polariton}
\NewAcronym{WSMS}{widely-spaced multi-subarray}
\NewAcronym{TTD}{true-time-delay}
\NewAcronym{mmWave}{millimeter wave}
\NewAcronym{DKL}{deep kernel learning}
\NewAcronym{GPR}{Gaussian process regression}
\NewAcronym{TDMA}{time division multiple access}
\NewAcronym{EC}{European Commission}
\NewAcronym{NR}{new radio}
\NewAcronym{QAM}{quadrature amplitude modulation}
\NewAcronym{1024QAM}{1024-ary quadrature amplitude modulation}
\NewAcronym{FCC}{Federal Communications Commission}
\NewAcronym{IR}{infrared }
\NewAcronym{IMT}{International Mobile Telecommunications}

\NewAcronym{QCL}{quantum cascade laser}
\NewAcronym{GaN}{Gallium Nitride}
\NewAcronym{InP}{Indium Phosphide}
\NewAcronym{SiGe}{Silicon Germanium}
\NewAcronym{CMOS}{complementary metal-oxide-semiconductor}
\NewAcronym{HBT}{heterojunction bipolar transistor}
\NewAcronym{LO}{local oscillator}
\NewAcronym{FET}{field-effect transistor}
\NewAcronym{HEMT}{high-electron-mobility transistor}
\NewAcronym{GaAs}{Gallium Arsenide}
\NewAcronym{LTCC}{low-temperature co-fired ceramic}
\NewAcronym{SIW}{substrate-integrated waveguide}
\NewAcronym{UTC-PD}{uni-traveling-carrier photodiode}

\NewAcronym{B6G}{beyond sixth generation}
\NewAcronym{RCC}{radar-communications coexistence}
\NewAcronym{DFRC}{dual-functional radar-communications}
\NewAcronym{EM}{electromagnetic}
\NewAcronym{QoS}{quality of service}
\NewAcronym{XR}{extended reality}
\NewAcronym{Tera-IoT}{THz internet-of-things}
\NewAcronym{IAB}{integrated access and backhaul}
\NewAcronym{PA}{power amplifier}
\NewAcronym{OFDM}{orthogonal frequency-division multiplexing}
\NewAcronym{DFT}{discrete Fourier transform}
\NewAcronym{DFT-s-OFDM}{DFT-spread-OFDM} 
\NewAcronym{PAPR}{peak-to-average power ratio}
\NewAcronym{OTFS}{orthogonal time frequency space}
\NewAcronym{DFT-s-OTFS}{discrete Fourier transform spread OTFS} 
\NewAcronym{NN}{neural network}
\NewAcronym{JCAS}{joint communications and sensing}
\NewAcronym{MISO}{multiple-input single-output}
\NewAcronym{IRS}{intelligent reflecting surfaces}
\NewAcronym{PSM}{phase-shift matrix}
\NewAcronym{CRB}{Cram\'er-Rao bound}
\NewAcronym{MUI}{multi-user interference}
\NewAcronym{JSAC}{joint sensing and communication}
\NewAcronym{DL}{deep learning}
\NewAcronym{CS}{channel sounder}
\NewAcronym{AoA}{angle-of-arrival}
\NewAcronym{AoD}{angle-of-departure}
\NewAcronym{DoF}{degree of freedom}
\NewAcronym{SLAC}{simultaneous localization and communications}
\NewAcronym{SNR}{signal-to-noise ratio}
\NewAcronym{SDR}{semidefinite relaxation}
\NewAcronym{DoA}{direction of arrival}
\NewAcronym{STAR-RIS}{simultaneously transmitting (refracting) and reflecting reconfigurable intelligent surface}
\NewAcronym{NF}{near field}
\NewAcronym{DPP}{delay-phase precoding}
\NewAcronym{SSB}{synchronization signal block}
\NewAcronym{RS}{reference signal}
\NewAcronym{MSE}{mean square error}
\NewAcronym{BS}{base station}
\NewAcronym{GHz}{gigahertz}
\NewAcronym{DRC}{dual-function radar and communication}
\NewAcronym{FGI}{flexible guard interval}
\NewAcronym{CP}{cyclic prefix}
\NewAcronym{AE}{autoencoder}

\NewAcronym{LWA}{leaky-wave antennas}
\NewAcronym{DLA}{discrete lens array}

\NewAcronym{OWC}{optical wireless communications}
\NewAcronym{MS}{mobile station}
\NewAcronym{IQ}{in-phase and quadrature}
\NewAcronym{TDD}{time-division multiplexing}
\NewAcronym{EESS}{Earth Exploration Satellite Service}
\NewAcronym{AGV}{automated guided vehicle}
\NewAcronym{D2D}{device-to-device}
\NewAcronym{IoNT}{Internet of Nano-Things} 
\NewAcronym{SLAM}{simultaneous localization and mapping}
\NewAcronym{GNSS}{Global Navigation Satellite System}
\NewAcronym{EIRP}{effective isotropic radiated power}
\NewAcronym{VNA}{vector network analyzer}
\NewAcronym{UMi}{urban microcell}
\NewAcronym{2D}{two-dimensional}
\NewAcronym{PPP}{Poisson point process}
\NewAcronym{DSSS}{direct-sequence spread spectrum}
\NewAcronym{CTF}{channel transfer function}
\NewAcronym{RFoF}{radio frequency over fiber}
\NewAcronym{TDS}{time-domain spectroscopy}
\NewAcronym{SC}{sliding correlation}
\NewAcronym{FDTD}{finite-difference time-domain}
\NewAcronym{ToA}{time of arrival}
\NewAcronym{UWB}{ultra wideband}
\NewAcronym{BDMA}{beam division multiple access}
\NewAcronym{ComS}{compressive sensing}
\NewAcronym{ADC}{analogue-to-digital converter}
\NewAcronym{DCNN}{deep convolutional neural network}
\NewAcronym{MMSE}{minimum-mean square error}
\NewAcronym{RSS}{received signal strength}
\NewAcronym{QPSK}{quadrature phase-shift keying}
\NewAcronym{8PSK}{8-ary phase-shift keying}
\NewAcronym{SISO}{single-input single-output}
\NewAcronym{DSP}{digital signal processing}

\end{acronym}

%% file: glossary.tex
\DTLnewdb{acronyms}
\let\acro\addacronym
\input{acronyms}
\DTLsort*{Acronym}{acronyms}

 \begin{abbreviations}
 \DTLforeach*{acronyms}{\thisAcronym=Acronym,\thisDesc=Description}%
 {\item[\thisAcronym]\thisDesc}%
 \end{abbreviations}